\documentclass[aps,prb,preprint,floats,amsmath,amssymb,superscriptaddress]{revtex4-2}
\usepackage[utf8]{inputenc}
\usepackage{graphicx}
\usepackage{epsfig}
\usepackage{amsmath}
\usepackage{amssymb}
\usepackage{longtable}
\usepackage{natbib}
\usepackage{multirow}
\usepackage{bm}
\usepackage{epstopdf}
\usepackage{ulem}
\usepackage{color}
\usepackage{listings}
\usepackage{appendix}
\usepackage{multirow}
\usepackage{hyperref}
\usepackage{breakurl}

\definecolor{green}{rgb}{0.0, 0.72, 0.92}

\newcommand{\black}[1]{\color{black}#1}
\newcommand{\ronaldo}[1]{\black #1}

\definecolor{gray}{rgb}{0.4,0.4,0.4}
\definecolor{darkblue}{rgb}{0.0,0.0,0.6}
\definecolor{cyan}{rgb}{0.0,0.6,0.6}

\newcommand{\exciting}{{\usefont{T1}{lmtt}{b}{n}exciting}}
\newcommand{\FHIaims}{FHI-aims}
\newcommand{\ie}{{\it i.e.}, }
\newcommand{\eg}{{\it e.g.}, }
\newcommand{\rgkmax}{\texttt{rgkmax}}

\newcommand{\kpts}{\textbf{k}-points}
\newcommand{\kgrid}{\textbf{k}-grid}
\newcommand{\GW}{$G_0W_0$}

\lstdefinelanguage{XML}
{
	morestring=[b]",
	morestring=[s]{>}{<},
	morecomment=[s]{<?}{?>},
	stringstyle=\color{black},
	identifierstyle=\color{blue},
	keywordstyle=\color{cyan},
	morekeywords={xmlns,version,type}
}
\begin{document}
	\title{\ronaldo{Critical assessment of \GW{} calculations for 2D materials: the example of monolayer MoS$_2$
	}}
	\author{Ronaldo Rodrigues Pela}
	\address{Supercomputing Department, Zuse Institute Berlin (ZIB), Berlin, Takustraße 7, 14195 Berlin, Germany}
	\author{Cecilia Vona}
	\affiliation{Physics Department and CSMB, Humboldt-Universit\"{a}t zu Berlin, Zum Gro\ss en Windkanal 2, 12489 Berlin, Germany}
	\author{Sven Lubeck}
	\affiliation{Physics Department and CSMB, Humboldt-Universit\"{a}t zu Berlin, Zum Gro\ss en Windkanal 2, 12489 Berlin, Germany}
	\author{Ben Alex}
	\affiliation{Physics Department and CSMB, Humboldt-Universit\"{a}t zu Berlin, Zum Gro\ss en Windkanal 2, 12489 Berlin, Germany}\author{Ignacio Gonzalez Oliva}
	\affiliation{Physics Department and CSMB, Humboldt-Universit\"{a}t zu Berlin, Zum Gro\ss en Windkanal 2, 12489 Berlin, Germany}
	\author{Claudia Draxl}
	\affiliation{Physics Department and CSMB, Humboldt-Universit\"{a}t zu Berlin, Zum Gro\ss en Windkanal 2, 12489 Berlin, Germany}
	\affiliation{European Theoretical Spectroscopic Facility (ETSF)}	
	\date{\today}
	\begin{abstract}
		Two-dimensional (2D) {\ronaldo materials combine many fascinating properties that make them more interesting than their three-dimensional counterparts for a variety of applications. For example, 2D materials exhibit stronger electron-phonon and electron-hole interactions, and their energy gaps and effective carrier masses can be easily tuned.} Surprisingly, published band gaps of {\ronaldo several 2D materials} obtained with the $GW$ approach, the state-of-the-art in electronic-structure calculations, are quite scattered. The details of these calculations, such as the underlying geometry, the starting point, the inclusion of spin-orbit coupling, and the treatment of the Coulomb potential can critically determine how accurate the results are. {\ronaldo Taking monolayer MoS$_2$ as a representative material,} we employ the linearized augmented planewave + local orbital method to systematically investigate how all these aspects affect the quality of \GW{} calculations, and also provide a summary of literature data. We conclude that the best overall agreement with experiments and coupled-cluster calculations is found for \GW{} results with HSE06 as a starting point including spin-orbit coupling, a truncated Coulomb potential, and an analytical treatment of the singularity at $q=0$.
		
	\end{abstract}
	\maketitle
	
	\section{Introduction}
	The isolation of graphene in 2004 can be regarded as a milestone in materials science that triggered a novel field of research, namely atomically thin 2D materials \cite{Geim_2007}. {\ronaldo Compared to their 3D counterparts, 2D materials have a higher surface-to-volume ratio, making them ideal candidates for catalysts and sensors \cite{Shanmugam_2022,Kumbhakar_2023}. Due to the confinement of electrons, holes, phonons, and photons in the 2D plane, the electronic, thermal, and optical properties of 2D materials present unusual features not found in their 3D counterparts \cite{Mir_2020,Zeng_2018,Zhang_2018}. For instance, their electronic structure -- especially band gaps -- can be easily adjusted by acting on the vertical quantum confinement through \eg the number of atomic layers, or external perturbations, such as an external electric field, and strain \cite{Chaves_2020,Guinea_2014}. The sensitivity to strain, \ie to structural details, implies that 2D materials also exhibit strong electron-phonon coupling \cite{Guinea_2014}. In addition, exciton binding energies are significantly larger than in 3D materials, and they can be tuned by the dielectric environment, \eg{} by encapsulation or deposition on substrates \cite{Xiao_2017,Mueller_2018,Thygesen_2017}. All these characteristics make them outstanding components in novel applications for electronics and optoelectronics \cite{Mir_2019,Bolotin_2008,Radisavljevic_2011,Liu_2016,Jariwala_2014,Wang_2012}. 
	}
	
\begin{figure}[hb]
		\centering
		\includegraphics[scale=1]{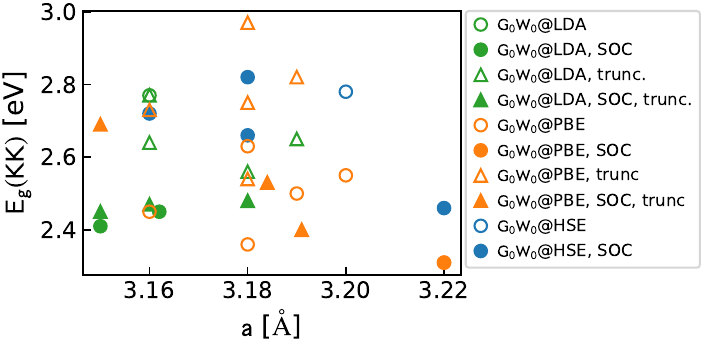}
		\caption{Literature results for the \GW{} energy gap of {\ronaldo MoS$_2$} at the $\mathrm{K}$ point of the Brillouin zone as a function of the in-plane lattice parameter $a$. They are obtained with and without SOC (filled and open symbols, respectively), as well as with and without truncation of the Coulomb potential ({\ronaldo triangles} and circles, respectively), using LDA, PBE, and HSE as starting points (green, orange and blue, respectively). Table \ref{tab:gw_gaps_literature} summarizes these data.}
		\label{fig:gaps-literature}
	\end{figure}

	For a deep understanding of {\ronaldo 2D materials}, an accurate description of {\ronaldo their} band structure is a must. Many-body perturbation theory within the \textit{GW} approach has become the state-of-the-art for \textit{ab initio} electronic-structure calculations of materials. 
	In this sense, many studies have employed \textit{GW} to investigate the electronic properties of {\ronaldo 2D materials 
		\cite{MolinaSanchez2015, Conley_2013, Ataca_2011, Shi_2013, Ramasubramaniam_2012, Rasmussen_2015, Rasmussen_2016, MolinaSanchez_2013, Qiu_2013, Cheiwchanchamnangij_2012, Hueser_2013, Echeverry_2016, Schmidt_2017, Komsa_2012, Liang_2013, Jiang_2021, Zibouche_2021, Soklaski_2014, Xia_2020, Gao_2016, Qiu_2016,Gillen_2016,Zhuang_2013,Haastrup_2018,Gjerding_2021,Kim_2021,Smart_2018,Horzum_2013,Lee_2017,Elliott_2020,Kirchhoff_2022,Hueser_2013_b,Ferreira_2019,Attaccalite_2011,Galvani_2016,Wu_2017,Blase_1995,Wang_2020,Mengle_2019,Fu_2016,Berseneva_2013,Cudazzo_2016,Ferreira_2017,Steinkasserer_2016,Rudenko_2014,Tran_2014,Cakir_2014,Tran_2015}. Surprisingly, as illustrated in Fig.~\ref{fig:gaps-literature} for monolayer MoS$_2$, they show a wide dispersion in the fundamental band gap. The same has been found for a number of 2D materials that have been extensively studied in the last years. Results for MoS$_2$, MoSe$_2$, MoTe$_2$, WS$_2$, WSe$_2$, BN, and phosphorene are summarized in the Appendix.
		In the extreme cases of MoS$_2$, WS$_2$, WSe$_2$, and BN, the calculated band gaps are scattered between 2.31-2.97, 2.43-3.19, 1.70-2.89, and 6.00-7.74 eV, respectively; in the worst case, the deviation (ratio between largest and smallest values) is as much as 61\% .}
	Moreover, {\ronaldo for some materials, such as for MoS$_2$, MoTe$_2$, WS$_2$, WSe$_2$, and BN, not even the gap character} is uniquely obtained -- being direct or indirect, depending on the details of the calculation.
	
	Many factors contribute to this confusing and unsatisfactory situation:
	\begin{itemize}
		\item 
		In various works, different geometries have been adopted. In this context, it must be said that the properties of {\ronaldo 2D materials} are highly sensitive to structural parameters \cite{Pisarra_2021, Conley_2013, MolinaSanchez2015}. {\ronaldo Small changes in the lattice constant $a$ already have a large impact on the energy gap, as seen in Fig. \ref{fig:gaps-literature} and Tables \ref{tab:gw_gaps_literature}-\ref{tab:gw_gaps_literature_phosphorene}. Moreover, often the lattice parameter alone is not sufficient to unambiguously determine the structure of a 2D material. For instance, phosphorene is characterized by four structural parameters; transition metal dichalcogenides require, besides $a$, the distance between two chalcogens (for MoS$_2$, $d_{\mathrm{S}\mathrm{S}}$ as depicted in Fig. \ref{fig:structure}). These ``other'' structural parameters have a notable effect on the electronic properties as well \cite{Ortenzi_2018,Pisarra_2021}.} Unfortunately, in several studies, only the lattice parameter is reported, which prevents not only a fair comparison between published results but also reproducibility. 
		
		\item
		A second reason can be attributed to the various ways of performing \GW{} calculations. First, there is the well-known starting problem \cite{Bechstedt,Onida_2002,Li_2012,Koerzdoerfer_2012,Chen_2014,Pela_2016,McKeon_2022,Knight_2016,Gant_2022}. Then, especially for 2D materials, \GW{} energy gaps converge very slowly with respect to the vacuum thickness and the number of \kpts{}. Even slabs with a vacuum layer of 60~\AA{} together with a $33\times 33\times 1$ \kgrid{} have been shown to be insufficient to obtain fully converged results \cite{Hueser_2013}. However, by truncating the Coulomb potential, convergence can be achieved with a reasonable amount of vacuum \cite{Hueser_2013_b}. The number of \kpts{} can be drastically reduced by an analytic treatment of the $q=0$ singularity of the dielectric screening or by using nonuniform \kgrid{s} \cite{Rasmussen_2016, Qiu_2016}. 
		
		\item 
		Last but not least, also {\ronaldo spin-orbit coupling (SOC)} plays an important role {\ronaldo in many cases}. Besides decreasing the size of the fundamental gap {\ronaldo mainly} through a splitting of the valence band \cite{Wang_2014}, {\ronaldo in some 2D materials and} for certain geometries and methods, disregarding or including this effect may change its character from indirect to direct or vice versa \cite{Pisarra_2021}.
	\end{itemize}
	
	In this manuscript, we address all these issues and provide a benchmark data set of DFT and \GW{} calculations, {\ronaldo taking monolayer MoS$_2$ as a representative 2D material. Due to its unique properties, it can be considered the most important 2D material after graphene. MoS$_2$ exhibits high electron mobility \cite{Radisavljevic_2011,Lembke_2012}; moderate SOC that can be exploited in spin- and valleytronics \cite{Xiao_2012,Cao_2012,Zeng_2012,Yang_2015,Berghauser_2018,Caruso_2022,Ji_2023}; a direct fundamental band gap with intermediately strong exciton binding, which is suitable for (opto)electronic devices operating at room temperature \cite{Klots_2014,Komsa_2012,Qiu_2016,Echeverry_2016,Radisavljevic_2011,Lembke_2012}. For these reasons, there are many experimental and theoretical works in the literature that investigate MoS$_2$, allowing for a better comparison with our results.
		
		We employ} the linearized augmented planewave + local orbital (in short LAPW+LO) method as implemented in the \exciting\ code. LAPW+LO is known to achieve ultimate precision for solving the Kohn-Sham (KS) equations of density functional theory (DFT) \cite{Gulans_2018} and high-level $GW$ results \cite{Nabok_2016}. Besides the local and semilocal DFT functionals LDA~\cite{Perdew_1992} and PBE~\cite{Perdew_1996,Perdew_1997_E} respectively, we include HSE06~\cite{Heyd_2003,Heyd_2006_E,Krukau_2006} both for geometry optimization and as a starting point for \GW. So far, HSE06 has not often been used for such calculations of MoS$_2$ \cite{Ataca_2011,Jiang_2021,Echeverry_2016,Ramasubramaniam_2012}, and to the best of our knowledge, neither a Coulomb truncation nor an adequate treatment of the singularity at $q=0$ was applied. For brevity, hereafter, we will refer to HSE06 as HSE. In our \GW{} calculations, we truncate the Coulomb potential \cite{Ismail-Beigi_2006,Hueser_2013,Hueser_2013_b}, and apply a special analytical treatment for the $q=0$ singularity \cite{Rasmussen_2016}. Moreover, we investigate the role of SOC at all levels. We carefully evaluate the impact of all these elements and conclude what leads to the most reliable electronic structure of this important material. Besides a detailed analysis of energy gaps, we address effective masses and spin-orbit splittings. 
	
	\section{Methods}
	
	\subsection{Linearized augmented planewave + local orbital methods}
	
	The full-potential all-electron code \exciting{} \cite{Gulans_2014} implements the family of LAPW+LO methods. 
	In this framework, the unit cell is divided into non-overlapping muffin-tin (MT) spheres, centered at the atomic positions, and the interstitial region in between the spheres. \exciting{} treats all electrons in a calculation by distinguishing between core and valence/semi-core states. For core electrons, assumed to be confined inside the respective MT sphere, the KS potential is employed to solve the Dirac equation which captures relativistic effects including SOC. The KS wavefunctions $|\Psi_{n\mathbf{k}} \rangle$ for valence and semicore states, characterized by band index $n$ and wavevector $\mathbf{k}$, are expanded in terms of augmented planewaves, $|\phi_{\mathbf{G}+\mathbf{k}} \rangle$, and local orbitals, $|\phi_{\gamma} \rangle$, as 
	\begin{equation}
		|\Psi_{n\mathbf{k}} \rangle = 
		\sum_{\mathbf{G}} C_{\mathbf{G}n,\mathbf{k}} |\phi_{\mathbf{G}+\mathbf{k}} \rangle + 
		\sum_{\gamma} C_{\gamma n,\mathbf{k}} |\phi_{\gamma} \rangle .
	\end{equation}
	$|\phi_{\mathbf{G}+\mathbf{k}} \rangle$ are constructed by augmenting each planewave with reciprocal lattice vector $\mathbf{G}$, living in the interstitial region, by a linear combination of atomic-like functions inside the MT spheres. In contrast, the LOs $|\phi_{\gamma} \rangle$ are non-zero only inside a specific MT sphere. They are used for reducing the linearization error \cite{sjostedt2000alternative,Gulans_2014}, for the description of semicore states, as well as for improving the basis set for unoccupied states \cite{Gulans_2018,Nabok_2016}. The quality of the basis set can be systematically improved by increasing the number of augmented planewaves (controlled in \exciting{} by the dimensionless parameter \rgkmax{}) and by introducing more LOs \cite{Gulans_2018,Gulans_2014}. With all these features, \exciting{} can be regarded as a reference code not only for solving the KS equation \cite{Lejaeghere_2016}, where it is capable of reaching micro-Hartree precision \cite{Gulans_2018}, but also for \GW{} calculations \cite{Nabok_2016}.
	
	\subsection{\GW{} approximation}\label{Sec:GW} In the \GW{} approximation, one takes a set of KS eigenvalues~$\{\varepsilon_{n\mathbf{k}}\}$ and eigenfunctions~$\{\Psi_{n\mathbf{k}}\}$ as a reference, and evaluates first-order quasi-particle (QP) corrections to the KS eigenvalues in first-order perturbation theory as
	\begin{equation}
		\varepsilon^{QP}_{n\mathbf{k}} = \varepsilon_{n\mathbf{k}} + Z_{n\mathbf{k}} \langle \Psi_{n\mathbf{k}} | \Sigma(\varepsilon_{n\mathbf{k}}) - V_{\text{xc}} | \Psi_{n\mathbf{k}} \rangle,
	\end{equation}
	where $Z_{n\mathbf{k}}$ is the renormalization factor, $V_{\text{xc}}$ is the exchange-correlation  (xc) potential, and $\Sigma$ is the self-energy. The latter is given as the convolution
	\begin{equation}
		\Sigma(\mathbf{r},\mathbf{r}',\omega)= \frac{\mathrm{i}}{2\pi}\int G(\mathbf{r},\mathbf{r}',\omega+\omega')W(\mathbf{r},\mathbf{r}',\omega') \mathrm{d}\omega',
	\end{equation}
	with $G$ being the single-particle Green function and $W$ the screened Coulomb potential. 
	
	In this non-selfconsistent method, the quality of the QP eigenvalues may depend critically on the starting point. 
	In many cases, LDA and PBE have been proven to be good starting points for \GW{}, leading to QP energies that agree well with experiments {\ronaldo \cite{Aryasetiawan_1998,Kotani_2007,Pela_2016}. However, \eg for materials containing $d$ electrons, such as Mo, hybrid functionals, like HSE, usually provide an improved reference for QP corrections compared to semilocal functionals \cite{Chen_2014,Leppert_2019,Rinke_2005,Atalla_2013,Bruneval_2013,Marom_2012,Koerzdoerfer_2012,Ren_2009}.} Here, we evaluate the quality of each of these three as a starting point.

	\subsection{Coulomb truncation}
	In calculations with periodic boundary conditions, for treating 2D systems, a sufficient amount of vacuum is required to avoid spurious interaction between the replica along the out-of-plane direction. Local and semilocal density functionals have a (nonphysical) asymptotic decay much faster than $1/r$, facilitating convergence of unoccupied states --and thus KS gaps-- with respect to the vacuum size. In \GW, the $1/r$ decay of the self-energy complicates this task. Specifically for MoS$_2$, even a vacuum layer with 60~\AA\ thickness turned out not being sufficient to obtain a fully converged band gap \cite{Hueser_2013,Hueser_2013_b}. Truncating the Coulomb potential along the out-of-plane direction $z$, however, leads to well-converged \GW{} band gaps with a considerably smaller vacuum size \cite{Hueser_2013,Hueser_2013_b}. 
	
	Here, we truncate the Coulomb potential with a step function along $z$. Setting the cutoff length to $L/2$, where $L$ is the size of the supercell along $z$ (Fig. \ref{fig:structure}), the truncated Coulomb potential can be written in a planewave basis as \cite{Ismail-Beigi_2006}:
	\begin{equation}
		v_{\mathbf{G}\mathbf{G}'}(\mathbf{q})=\delta_{\mathbf{G}\mathbf{G}'}
		\frac{4\pi}{Q^2}\left[
		1-\mathrm{e}^{-Q_{xy}L/2}
		\cos(G_z L/2)
		\right],
	\end{equation}
	where $\mathbf{Q}=\mathbf{q}+\mathbf{G}$, and $\mathbf{q}$ is a vector in the first Brillouin zone (BZ).
	
	\subsection{Treatment of the $\mathbf{q=0}$ singularity}
	\label{sec:singularity}
	
	On the down-side, truncating the Coulomb interaction slows down the convergence in terms of \kpts{} because of the non-smooth behavior of the dielectric function around the singularity at $\mathbf{q} = \mathbf{0}$ \cite{Haastrup_2018,Rasmussen_2015,Rasmussen_2016}. To bypass this problem, we follow an analytical treatment of $W^c$, the correlation part of $W$, close to the singularity as proposed in Ref. \cite{Rasmussen_2016}. Without this treatment, the correlation part of the self-energy $\Sigma^c(\omega)$ can be written as \cite{Jiang_2013}
	
	\begin{equation}\label{eq:sigmac}
		\langle \Psi_{n\mathbf{k}} |\Sigma^c(\omega)| \Psi_{n\mathbf{k}} \rangle = \frac{\mathrm{i}}{2\pi}\sum_{mij}
		\int_{-\infty}^{\infty} d\omega^{\prime} \frac{1}{N_\mathbf{q}}\sum_{\mathbf{q}}\frac{1}{\omega +\omega^{\prime}-\tilde{\epsilon}_{m\mathbf{k-q}}}
		[{M}^i_{nm}(\mathbf{k},\mathbf{q})]^*
		{M}^j_{nm}(\mathbf{k},\mathbf{q})W^c_{ij}(\mathbf{q},\omega^{\prime})
		,
	\end{equation}
	where $\tilde{\epsilon}_{n\mathbf{k}}=\epsilon_{n\mathbf{k}}+\mathrm{i}\,\eta \,\mathrm{sign}(E_F-\epsilon_{n\mathbf{k}})$ and $E_F$ the Fermi energy. ${M}^i_{nm}(\mathbf{k},\mathbf{q})$ are the expansion coefficients of the mixed-product basis, an auxiliary basis to represent products of KS wavefunctions. Like LAPWs and LOs, they have distinct characteristics in the MT spheres and interstitial region~\cite{Aguilera_2013,Jiang_2013}. To treat the $\mathbf{q} = \mathbf{0}$ case separately, the corresponding term in Eq. (\ref{eq:sigmac}) is replaced by
	\begin{equation}\label{eq:q_equal_zero}
		\frac{1}{\omega +\omega^{\prime}-\tilde{\epsilon}_{m\mathbf{k}}}
		[{M}^i_{nm}(\mathbf{k},\mathbf{0})]^*
		{M}^j_{nm}(\mathbf{k},\mathbf{0})
		\frac{1}{\Omega_0}\int_{\Omega_0} \mathrm{d}\mathbf{q} \ \!
		W^c_{ij}(\mathbf{q},\omega^{\prime}),
	\end{equation}
	where $\Omega_0$ is a small region around $\mathbf{q}=\mathbf{0}$. Analytical expressions for $W^c_{ij}(\mathbf{q},\omega^{\prime})$ in the limit $\mathbf{q} \to \mathbf{0}$ \cite{Rasmussen_2016} are then employed to calculate the integral in Eq. (\ref{eq:q_equal_zero}).
	
	\subsection{Spin-orbit coupling}
	In this study, SOC is treated via the second-variational (SV) scheme~\cite{singh_1994}. In LDA and PBE calculations, the conventional SV approach is utilized \cite{macdonald1980linearised,li1990magnetic,Gulans_2014,vona_2023}: First, the scalar-relativistic problem, \ie omitting SOC, is solved. A subset of the resulting eigenvectors is then used as basis set for addressing the full problem. The number of eigenvectors is a convergence parameter. In this work, the SOC term is evaluated with the zero-order regular approximation (ZORA)~\cite{Lenthe_1993,Lenthe_1994}. 
	
	A ground-state calculation with HSE is performed via a nested loop \cite{betzinger2010hybrid}: In the outer loop, the nonlocal exchange is computed for a subset of KS wavefunctions, using a mixed-product basis. The inner loop solves the generalized KS equations self-consistently, where only the local part of the effective potential is updated in each step. Within this inner loop, the SV scheme is applied self-consistently to incorporate SOC. The corresponding term, evaluated within the ZORA, is based on PBE, which is justified by the minimal contribution due to the gradient of the nonlocal potential \cite{Huhn_2017,Wang_2018,Vona_2022}. 
	
	\GW{} calculations with SOC are performed in two steps on top of ground-state calculations that include SOC. First, the QP energies are computed as explained in Sec.~\ref{Sec:GW}, using the scalar-relativistic KS eigenvalues and eigenvectors. In the second step, the obtained QP energies are used, together with the SV KS eigenvectors, to evaluate SOC through the diagonalization of the SV Hamiltonian. This approach is sufficient in the case of MoS$_2$ since SOC does not cause band inversion~\cite{Aguilera_2013,Vona_2022}. 
	
	\section{Computational details}
	In our calculations, we employ the all-electron full-potential code \exciting\ \cite{Gulans_2014}. The only exception is for obtaining the HSE equilibrium geometry, where \FHIaims\ \cite{Blum2009, Ren2012} is used, since so far \exciting\ lacks geometry relaxation with hybrid functionals. Even though \exciting{} and \FHIaims{} implement very different basis sets to expand the KS wavefunctions, the two codes have been shown to be among those with the best mutual agreement \cite{Lejaeghere_2016}. Moreover, a comparison of energy gaps and geometry relaxations for MoS$_2$ confirms this finding (see Appendix \ref{sec:fhi_vs_exciting}).
	
	In all calculations, the unit-cell size $L$ along the out-of-plane direction (Fig. \ref{fig:structure}) is set to 30 bohr. Different flavors of xc functionals are applied, namely LDA, PBE, and HSE. In the latter, we use the typical parameters ~\cite{Krukau_2006}, \ie a mixing factor of $\alpha= 0.25$ and a screening range of $ \omega = 0.11$~bohr$^{-1}$. To determine the respective equilibrium geometries, we relax the atomic positions until the total force on each ion is smaller than 10 $\mu$Ha/bohr. For these geometries, the electronic structure is calculated with all three functionals, with and without SOC, giving rise to a set of 18 calculations. These calculations are followed by \GW{} calculations, taking the respective DFT solutions as starting points.
	
	The dimensionless parameter \rgkmax{} that controls the \exciting\ basis-set size is set to 8. In the LDA and PBE calculations, we use a $30\times 30 \times 1$ \kgrid{}. In HSE and \GW{} calculations, we employ 400 empty states and an $18\times 18 \times 1$ \kgrid{}. In \GW{}, the correlation part of the self-energy is evaluated with 32 frequency points along the imaginary axis, and then analytic continuation to the real axis is carried out by means of Pade's approximant. For the bare Coulomb potential, we use a 2D cutoff \cite{Ismail-Beigi_2006} combined with a special treatment of the $\mathbf{q=0}$ singularity as described in Section \ref{sec:singularity}. Furthermore, we carefully determine the minimal set of LOs, sufficient to converge at least the lowest 400 KS states. This is achieved with 2 and 6 LOs for sulfur $s$ and $p$ states, respectively, as well as 3, 6, and 10 LOs for molybdenum $s$, $p$, and $d$ states, respectively. Overall, we estimate a numerical precision of 50-100 meV in the energy gaps obtained with our \GW{} calculations. To determine effective masses, we use parabolic fits within a range of 0.05 \AA$^{-1}$ around the valence-band maximum (VBM) and the conduction-band minimum (CBm) at the $\mathrm{K}$ point of the BZ. As will be shown further below, depending on the respective case, $\mathrm{K}$ can host either global or local extrema.
	
	\section{Results}
	\subsection{Ground-state geometries}
	
	\begin{figure}[bht]
		\centering
		\includegraphics[width=14cm]{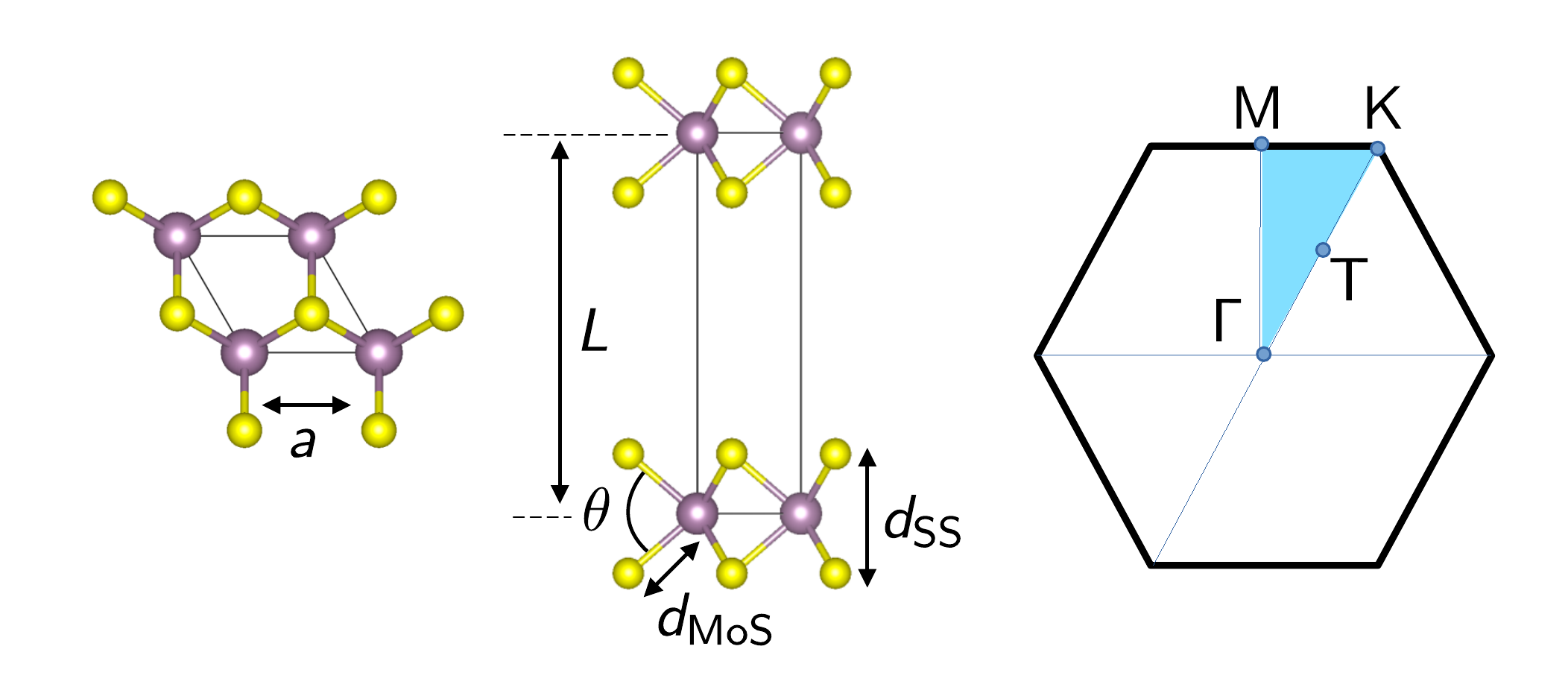}
		\caption{Top view (left) and side view (middle) of the MoS$_2$ slab geometry, determined by the in-plane lattice constant $a$ and the distance between sulfur atoms, $d_{\mathrm{SS}}$. The Mo-S bond length $d_{\mathrm{MoS}}$ and the S-Mo-S angle $\theta$ are shown as well. $L$ is the unit-cell size along the out-of-plane direction $z$, including a vacuum layer. High-symmetry points and paths used in the band-structure plots (Fig. \ref{fig:bandstructure}) are highlighted in the BZ (right panel).}
		\label{fig:structure}
	\end{figure}
	
	The geometry of MoS$_2$, depicted in Fig. \ref{fig:structure}, is determined by the in-plane lattice parameter $a$ and the distance between sulfur atoms, $d_{\mathrm{SS}}$. In Table \ref{tab:equilibrium_geometry}, we list these structural parameters as obtained with LDA, PBE, and HSE, and include the Mo-S bond length $d_{\mathrm{MoS}}$ and the angle $\theta$ between Mo and S atoms as well. As expected, LDA underestimates the lattice spacing, PBE slightly overestimates it, and HSE shows the best performance with respect to experiment. All three xc functionals underestimate the S-S bond length, PBE being closest to its measured counterpart. Comparison with computed literature data reveals good agreement. 
	
	\begin{table}[htb]
		\centering
		\caption{Equilibrium geometry of MoS$_2$ obtained with LDA, PBE, and HSE, compared with literature values (a \cite{Pisarra_2021}; b \cite{Ellis_2011}; c \cite{Boeker_2001}; d \cite{Ramasubramaniam_2012}; e \cite{Ataca_2011}; f \cite{Shi_2013}; g \cite{Komsa_2012}; h \cite{Kadantsev_2012}; i \cite{Rasmussen_2015}; j \cite{Rasmussen_2016}, k \cite{Kang2013}). Experimental results for bulk MoS$_2$ are also provided (c \cite{Boeker_2001}). For the definition of $d_{\mathrm{SS}}$, $d_{\mathrm{MoS}}$ and $\theta$, see Fig. \ref{fig:structure}. }\label{tab:equilibrium_geometry}
		\setlength{\tabcolsep}{5pt}
		\begin{tabular}{ccccc}
			{\bf This work} \\
			\hline 
			xc functional & $a$ [\AA] & $d_{\mathrm{SS}}$ [\AA] & $d_{\mathrm{MoS}}$ [\AA] & $\theta$ [°]\\ \hline
			LDA      &   {\ronaldo 3.121} &   {\ronaldo 3.106} &   {\ronaldo 2.379} &    81.51 \\
			PBE      &   {\ronaldo 3.186} &   {\ronaldo 3.125} &   {\ronaldo 2.414} &    80.70 \\
			HSE      &   {\ronaldo 3.160} &   {\ronaldo 3.101} &   {\ronaldo 2.394} &    80.73 \\ 
			\hline\hline
			{\bf Literature} \\
			\hline
			LDA & 3.11$^e$, 3.12$^a$, 3.122$^h$ & 3.11$^{a,e}$, 3.116$^h$ & 2.37$^e$, 2.383$^h$ & 81.62$^e$ \\
			PBE & 3.18$^{d,g,i,k}$,3.184$^j$, 3.19$^{a,f}$, 3.20$^e$& 3.12$^{a,d}$, 3.127$^j$, 3.13$^{e,i}$ & 2.41$^k$, 2.42$^e$ & 80.69$^e$\\
			HSE& 3.16$^{b,k}$ & & 2.40$^k$\\
			Experiment     &   3.160$^c$  &   3.172$^c$\\
			\hline
		\end{tabular}    
	\end{table} 	
	
	\subsection{Electronic structure}
\begin{figure}[h]
		\centering
		\includegraphics{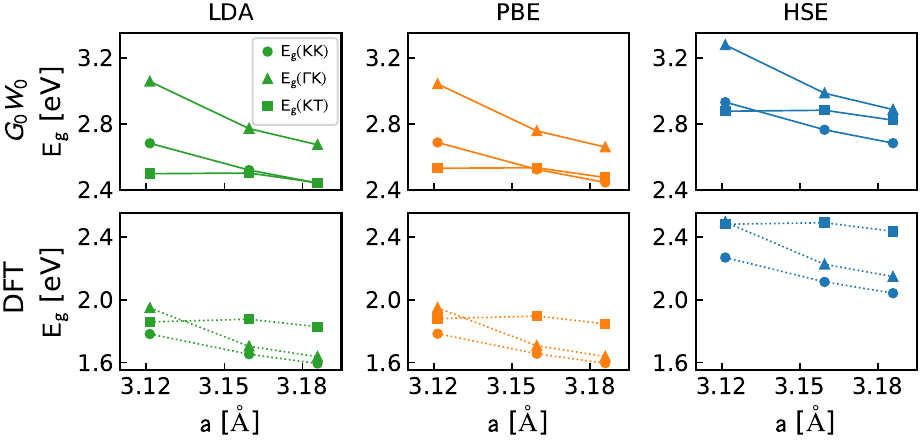}
		\caption{\ronaldo Calculated energy gaps of MoS$_2$ as a function of the lattice parameter $a$. The values for KK, $\Gamma$K, and KT are represented as circles, triangles, and squares, respectively. Dotted (solid) lines stand for DFT (\GW) results; all include SOC. Note that these values only indirectly reflect the S-S distance $d_{\mathrm{SS}}$. }
		\label{fig:gaps-vs-lattice}
	\end{figure}

	Table \ref{tab:bandgaps} summarizes the energy gaps obtained with different functionals for the different geometries. We consider here the direct gap at the $\mathrm{K}$ point ($E_g(\mathrm{KK})$) as well as the indirect gaps between $\Gamma$ and $\mathrm{K}$ ($E_g(\mathrm{\Gamma K})$) and between $\mathrm{K}$ and $\mathrm{T}$ ($E_g(\mathrm{KT})$). For the definition of the $\mathrm{T}$ point, see Fig. \ref{fig:structure}. For each geometry and methodology, the bold font highlights the fundamental gap. {\ronaldo For the calculations that include SOC, Fig. \ref{fig:gaps-vs-lattice} displays the energy gaps given in Table \ref{tab:bandgaps} with respect to the lattice parameter. In the DFT calculations, regardless of the calculation method, $E_g(\mathrm{KT})$ (squares) shows a weak dependence on the geometry. The fundamental gap obtained with LDA, PBE, and HSE, is always direct at K. In \GW, the fundamental gap is $E_g(\mathrm{KT})$ for the structures with smaller lattice parameter and$E_g(\mathrm{KK})$  for larger lattice constants.}
	\begin{table}[hbt]
		\centering
		\caption{Energy gaps (in eV) obtained for the 18 different cases considered. For each case, the fundamental gap is highlighted in bold. The second column indicates whether SOC is included (Y) or not (N). {\ronaldo The experimental gap is 2.6 eV (direct at K) \cite{Klots_2014}. Note that this value is corrected for the zero-point renormalization energy of 75 meV~\cite{MolinaSanchez_2016}.}}\label{tab:bandgaps}
		\begin{tabular}{l|c|c|ccc|ccc}
			Geometry    &  SOC & Gap & LDA & PBE & HSE & \GW{@LDA} & \GW{@PBE} & \GW{@HSE}\\ \hline
			\multirow{6}{*}{LDA} & \multirow{3}{*}{N} & $E_g(\mathrm{KK})$     & {\ronaldo\textbf{1.86}}& {\ronaldo\textbf{1.86}} & {\ronaldo\textbf{2.35}} & {\ronaldo 2.76} & {\ronaldo 2.77} & {\ronaldo 3.01} \\
			& & $E_g(\Gamma \mathrm{K})$& {\ronaldo 1.96}& {\ronaldo 1.96} & {\ronaldo 2.50}& {\ronaldo 3.07} & {\ronaldo 3.05}& {\ronaldo 3.29}\\
			& &  $E_g(\mathrm{KT})$     & {\ronaldo 1.97}& {\ronaldo 1.99} & {\ronaldo 2.58} & {\ronaldo\textbf{2.60}} & {\ronaldo\textbf{2.64}}& {\ronaldo\textbf{2.98}} 
			\\
			\cline{2-9}
			& \multirow{3}{*}{Y} & $E_g(\mathrm{KK})$ & {\ronaldo\textbf{1.78}} & {\ronaldo\textbf{1.78}}& {\ronaldo\textbf{2.27}} & {\ronaldo 2.68} & {\ronaldo 2.69} & {\ronaldo 2.93}\\
			& & $E_g(\mathrm{\Gamma K})$ & {\ronaldo 1.95} & {\ronaldo 1.95} & {\ronaldo 2.50} & {\ronaldo 3.06} & {\ronaldo 3.04} & {\ronaldo 3.28}\\ 
			& & $E_g(\mathrm{KT})$ & {\ronaldo 1.86} & {\ronaldo 1.88} & {\ronaldo 2.48} & {\ronaldo \textbf{2.50}} & {\ronaldo \textbf{2.53}} & {\ronaldo \textbf{2.88}}\\
			\hline
			\multirow{6}{*}{PBE} & \multirow{3}{*}{N} & $E_g(\mathrm{KK})$ & {\ronaldo 1.67} & {\ronaldo 1.67}& {\ronaldo \textbf{2.12}}& {\ronaldo \textbf{2.52}}& {\ronaldo \textbf{2.51}}& {\ronaldo \textbf{2.76}}\\
			& & $E_g(\mathrm{\Gamma K})$ & {\ronaldo \textbf{1.65}}& {\ronaldo \textbf{1.65}} & {\ronaldo 2.16} & {\ronaldo 2.68}& {\ronaldo 2.67} & {\ronaldo 2.89}
			\\
			& & $E_g(\mathrm{KT})$ & {\ronaldo 1.94} & {\ronaldo 1.94} & {\ronaldo 2.55} & {\ronaldo 2.56}& {\ronaldo 2.55}& {\ronaldo 2.93}\\
			\cline{2-9}
			& \multirow{3}{*}{Y} & $E_g(\mathrm{KK})$ &{\ronaldo \textbf{1.59}} & {\ronaldo \textbf{1.60}} & {\ronaldo \textbf{2.04}} & {\ronaldo \textbf{2.44}} & {\ronaldo \textbf{2.45}} & {\ronaldo \textbf{2.68}} \\
			& & $E_g(\mathrm{\Gamma K})$ & {\ronaldo 1.64} & {\ronaldo 1.64} & {\ronaldo 2.15} & {\ronaldo 2.67} &  {\ronaldo 2.66} &  {\ronaldo 2.89}\\ 
			&           & $E_g(\mathrm{KT})$ & {\ronaldo 1.83}& {\ronaldo 1.85} & {\ronaldo 2.44} & {\ronaldo 2.44} & {\ronaldo 2.48}&  {\ronaldo 2.82}\\
			\hline
			\multirow{6}{*}{HSE} & \multirow{3}{*}{N} & $E_g(\mathrm{KK})$ & {\ronaldo 1.73} & {\ronaldo 1.74} & {\ronaldo \textbf{2.19}} & {\ronaldo \textbf{2.60}} & {\ronaldo \textbf{2.60}}& {\ronaldo \textbf{2.84}} \\
			& & $E_g(\mathrm{\Gamma K})$ & {\ronaldo \textbf{1.71}}& {\ronaldo \textbf{1.72}}& {\ronaldo 2.23} & {\ronaldo 2.78}& {\ronaldo 2.77}& {\ronaldo 2.99}
			\\
			& &        $E_g(\mathrm{KT})$ & {\ronaldo 1.99} & {\ronaldo 2.01} & {\ronaldo 2.60} & {\ronaldo 2.61} & {\ronaldo 2.65} & {\ronaldo 2.99}
			\\
			\cline{2-9}
			& \multirow{3}{*}{Y} & $E_g(\mathrm{KK})$ & {\ronaldo \textbf{1.65}} & {\ronaldo \textbf{1.66}} & {\ronaldo \textbf{2.11}} & {\ronaldo 2.52}& {\ronaldo \textbf{2.52}}&  {\ronaldo \textbf{2.76}}\\
			& & $E_g(\mathrm{\Gamma K})$ & {\ronaldo 1.70} & {\ronaldo 1.71} & {\ronaldo 2.23} & {\ronaldo 2.77} & {\ronaldo 2.76} & {\ronaldo 2.99}\\ 
			& & $E_g(\mathrm{KT})$ & {\ronaldo 1.88} & {\ronaldo 1.90} & {\ronaldo 2.49} & {\ronaldo \textbf{2.50}} & {\ronaldo 2.53} & {\ronaldo 2.88}\\
		\end{tabular}
	\end{table}
	
	As to be expected and also observed in Ref. \cite{Qiu_2016}, for a fixed geometry, the energy gaps obtained by LDA and PBE are quite similar, with the largest difference being {\ronaldo 0.02 eV}. The two functionals also agree on the location of the VBM and the CBm. For both, the band gap is direct if SOC is included, and {\ronaldo indirect otherwise} for the PBE and HSE geometries. In contrast, HSE{\ronaldo gives a direct gap}, regardless of whether SOC is considered.
	
	Also \GW{@LDA} and \GW{@PBE} are very close to each other, the largest difference being {\ronaldo 0.04 eV. This can be attributed to the similarity between LDA and PBE when the same geometry is adopted. However, the similarity between \GW{@LDA} and \GW{@PBE} results is material dependent, as observed in other works \cite{Mansouri_2023,Sun_2020,Janssen_2015,Klimes_2014}.
		
		For} a given geometry, the locations of the VBM and the CBm are independent of the starting point, with the only exception being the HSE geometry when SOC is included. When comparing the three geometries, we encounter three different scenarios. First, for the LDA geometry, the fundamental gap changes from a direct KS gap at $\mathrm{K}$ to an indirect QP gap (between $\mathrm{K}$ and $\mathrm{T}$), independent of the starting point. Second, for the PBE and HSE geometries, when SOC is disregarded, the indirect gap $\mathrm{\Gamma K}$ obtained with LDA and PBE becomes direct and located at $\mathrm{K}$ upon applying \GW{}. Third, for the HSE geometry, and SOC being included we observe an indirect band gap for \GW{@LDA} while it is direct for PBE and HSE as starting points. This can be understood in terms of the small differences between the $\mathrm{KK}$ and $\mathrm{KT}$ gaps, $\Delta_{\mathrm{KT}}$, which are {\ronaldo 0.01 eV, 0.05 eV, and 0.15} eV for \GW{@LDA}, \GW{@PBE} and \GW{@HSE}, respectively. Including SOC, splits the conduction band state at T (K) by {\ronaldo $\sim  0.07$~eV} ($\sim 3$~meV), decreasing $\Delta_{\mathrm{KT}}$ by {\ronaldo $\sim 0.03$~eV}. This is enough to make $\Delta_{\mathrm{KT}}$ negative for \GW{@LDA}, but not for \GW{@PBE}, and \GW{@HSE}. A more detailed discussion about $\Delta_{\mathrm{KT}}$ can be found in Section \ref{subsubsec:experimental_data}.
	
\begin{figure}[htb]
		\centering
		\includegraphics[scale=1]{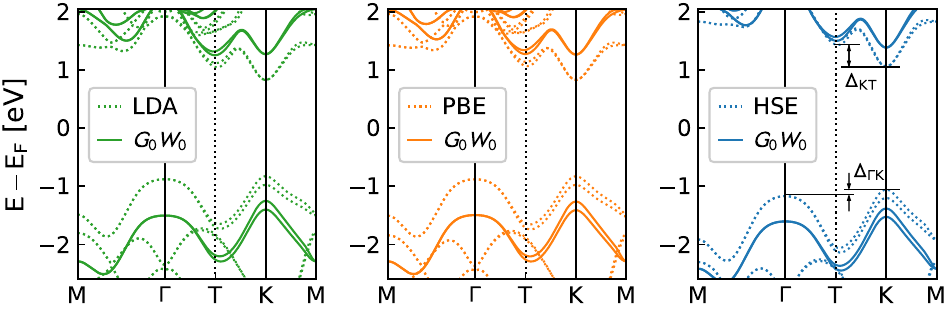}
		\caption{Band structures including SOC obtained for the HSE geometry. The parameters $\Delta_{\mathrm{KT}}$ and $\Delta_{\mathrm{\Gamma K}}$ are discussed in Section \ref{subsubsec:experimental_data}. }
		\label{fig:bandstructure}
	\end{figure}

	Figure \ref{fig:bandstructure} shows the band structures obtained for the HSE geometry including SOC. As expected, the differences between \GW{@HSE} and HSE bands are less pronounced than those between \GW{@LDA} (\GW{@PBE}) and LDA (PBE) bands. For all three starting points, the \GW{} corrections are not uniform over all \kpts{}, \ie a simple scissors approximation is, strictly speaking, not applicable. We will explore this in a more quantitative fashion further below. The SOC splitting in the valence band is zero at the $\Gamma$ point and increases toward the $\mathrm{K}$ point. The SOC effect on the conduction bands is much less pronounced. These observations are in agreement with other theoretical  \cite{Salehi_2018,Kormanyos_2015,Dou_2016,Zhu_2011,Echeverry_2016,MolinaSanchez_2013} and experimental works \cite{Miwa_2014,Klots_2014}.
	
\begin{figure}[htb]
		\centering
		\includegraphics[scale=0.8]{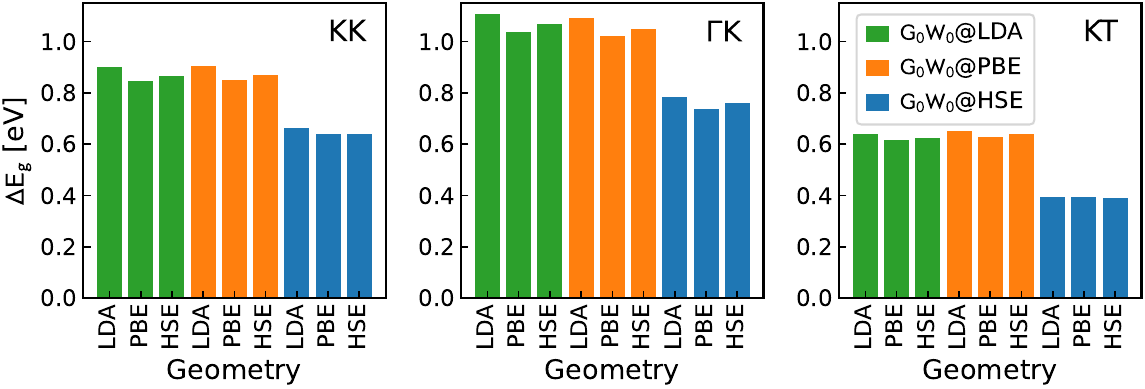}
		\caption{\GW{} self-energy correction, $\Delta E_g$, to the band gaps obtained for different starting points (LDA green, PBE orange, HSE blue) and geometries, including SOC effects. Left, middle, and right panels refer to $\Delta E_g$ evaluated at the $\mathrm{K}$ point, between $\Gamma$ and $\mathrm{K}$, and between $\mathrm{K}$ and $\mathrm{T}$, respectively.}
		\label{fig:delta-gaps}
	\end{figure}

	The impact of the self-energy correction on the energy gaps, $\Delta E_g = E_g^{G_0W_0} - E_g^{DFT}$, is shown in Fig. \ref{fig:delta-gaps} for the case when SOC is included. Clearly, $\Delta E_g$ is more significant for (semi)local DFT starting points than for HSE. Interestingly, for a given starting point, $\Delta E_g$ hardly depends on the geometry. For LDA and PBE as the starting points, the ranges of $\Delta E_g(\mathrm{KK})$, $\Delta E_g(\mathrm{\Gamma K})$, and $\Delta E_g(\mathrm{KT})$ are 0.84-0.90, 1.0-1.1, and 0.61-0.65 eV, respectively. The dependence is even weaker for \GW{@}HSE with values of 0.64-0.66, 0.74-0.78, and 0.39-0.40 eV, respectively. Very similar results are observed when SOC is disregarded. 
	
	\subsection{Spin-orbit splittings and effective masses}
	\label{section:SOC_meff}
	
	In Table \ref{tab:SOC_and_meff}, we report for the HSE structure the spin-orbit splitting $\Delta_{val}$ ($\Delta_{cond}$) at the $\mathrm{K}$ point for the highest occupied (lowest unoccupied) band. Effective electron (hole) masses $m_e$ ($m_h$) calculated at the $\mathrm{K}$ point along different directions are shown as well. We observe that neither the spin-orbit splittings nor the effective masses are very sensitive to the geometry (see Table \ref{tab:SOC_and_meff_all_geometries} for more details).
	
	\begin{table}[htb]
		\centering
		\caption{Spin-orbit splittings in valence ($\Delta_{val}$) and conduction ($\Delta_{cond}$) band (in meV) and effective hole ($m_h$) and electron ($m_e$) masses, in units of the free electron mass $m_0$, at the $\mathrm{K}$ point along different directions, obtained for the HSE geometry. {\ronaldo Experimental results are taken from Refs. (a \cite{Zhang_2015}; b \cite{Splendiani_2010}; c \cite{Li_2014}; d \cite{Dou_2016}; e \cite{Miwa_2014}; f \cite{Shen_2013}; g \cite{Klots_2014}; h \cite{Schmidt_2015}; i \cite{Eknapakul_2014}; j \cite{Jin_2015}).}}
		\label{tab:SOC_and_meff}
		\begin{tabular}{c|ccc|ccc|c}
			& LDA & PBE & HSE & \GW{@LDA} & \GW{@PBE} & \GW{@HSE} & {\ronaldo Experiment}\\ \hline
			$\Delta_{val}$ & 149 & 148 & 144& 149 & 148 & 143 & {\ronaldo 130$^a$, 130$^b$, 140$^c$, 141$^d$} \\
			& & & & & & & {\ronaldo $145\pm 4^e$, 150-160$^f$, 160$^g$} \\
			$\Delta_{cond}$ & 3 & 3 & 4 & 3 & 3 & 4 & {\ronaldo $4.3\pm 0.1^h$}\\ 
			$m_e(\mathrm{K\Gamma})$ & 0.40&     0.41&     0.37&     0.42&     0.42&     0.39 & \multirow{2}{*}{\ronaldo $0.67 \pm 0.08^i$}\\
			$m_e(\mathrm{KM})$ & 0.42&     0.42&     0.38&     0.42&     0.42&     0.39\\
			$m_h(\mathrm{K\Gamma})$ & 0.49&    0.50&    0.44&    0.42&    0.42&    0.42 & \multirow{2}{*}{\ronaldo$0.43 \pm 0.02^j$}\\
			$m_h(\mathrm{KM})$ & 0.53&    0.53&    0.47&    0.44&    0.44&    0.44\\
			\hline
		\end{tabular}    
	\end{table}
	
	$\Delta_{val}$ exhibits a very narrow spread among all the methods employed here (DFT and \GW{}), \ie a range of 143-149 meV. These values are in excellent agreement with the experimental counterparts of $\Delta_{val}=$130-160 meV \cite{Dou_2016,Zhang_2015,Splendiani_2010,Miwa_2014,Klots_2014,Li_2014,Shen_2013}. The value for the conduction band, $\Delta_{cond}$, is 3~meV for LDA, PBE, \GW{@LDA}, and {\ronaldo\GW{@PBE}}; it is only slightly higher for HSE and \GW{@HSE}, namely 4~meV. Again, there is excellent agreement with the measured value of $\Delta_{cond}=4.3\pm 0.1$~meV \cite{Schmidt_2015}.
	
	The effective hole mass, $m_h$, exhibits minor variations, not larger than $0.04m_0$, when going from $\mathrm{K \Gamma }$ to $\mathrm{KM}$. This is in line with other calculations \cite{Peelaers_2012,Cheiwchanchamnangij_2012}. The measured value for freestanding MoS$_2$ is $m_h = 0.43 \pm 0.02$ \cite{Jin_2015} and, apart from LDA and PBE, all theoretical results show excellent agreement. The electron mass, $m_e$, is more isotropic than $m_h$. For LDA, it differs by at most $0.02m_0$ between $\mathrm{K \Gamma }$ and $\mathrm{KM}$. Our values are in line with other calculated results\cite{Peelaers_2012,Cheiwchanchamnangij_2012,Zibouche_2021,MolinaSanchez2015,Pulkin_2020,Shi_2013,Kormanyos_2015}. The measured counterpart of $0.67 \pm 0.08$ \cite{Eknapakul_2014}, is significantly larger than the calculated value reported here and in other theoretical works \cite{Peelaers_2012,Cheiwchanchamnangij_2012,Zibouche_2021,MolinaSanchez2015,Pulkin_2020,Shi_2013,Kormanyos_2015}. As discussed in Ref.\cite{Zibouche_2021}, the difference could originate from the heavy doping of the measured sample, which may introduce metallic screening.
	
	\subsection{Discussion of energy gaps}
	
	\subsubsection{Comparison with experiment}\label{subsubsec:experimental_data}
	
	The experimental band gap for free-standing MoS$_2$, determined by photocurrent spectroscopy, is 2.5 eV \cite{Klots_2014}. In order to compare with experiment, it is important to account for the zero-point renormalization energy of 75 meV \cite{MolinaSanchez_2016}. This means that the theoretical value computed without this correction should be $2.575 \cong 2.6$~eV to match its measured counterpart. For our discussion here, we consider {\ronaldo the HSE and PBE geometries, which are closer to experiment than that obtained by LDA. For the following assessment, we refer to the values in Table \ref{tab:bandgaps}. For the structure optimized with HSE, \GW{} performed on LDA, PBE, and HSE as starting points gives $E_g(\mathrm{KK})$ of {\ronaldo 2.52, 2.52, and 2.76} eV, respectively, \ie \GW{@LDA} and \GW{@PBE} underestimate the measured value by about 0.08 eV, whereas \GW{@HSE} overestimates it by 0.16 eV. However, even though \GW{@LDA} agrees better with experiment than \GW{@HSE}, it erroneously predicts an indirect band gap $E_g(\mathrm{KT})$ which is {\ronaldo 0.02~eV} smaller than $E_g(\mathrm{KK})$. \GW{@PBE} shows the best agreement with experiment and also predicts the gap to be direct. Interestingly, {\ronaldo considering the PBE geometry, as done in several works \cite{Ramasubramaniam_2012, Rasmussen_2015,Komsa_2012,Liang_2013, Soklaski_2014, Rasmussen_2016, Conley_2013, Shi_2013, Hueser_2013, Jiang_2021,Pulkin_2020,Haastrup_2018,Gjerding_2021,Zhuang_2013,Kim_2021}}, \GW{@HSE}, giving a direct band gap of {\ronaldo $E_g(\mathrm{KK})=2.68$~eV}, agrees best with experiment, the deviation being 0.08 eV only. {\ronaldo With LDA and PBE as starting points, the calculated \GW{} band gap is direct, but 0.16 and 0.15 eV, respectively, below the experimental value. }
		
		\begin{table}[htb]
			\centering
			\caption{{\ronaldo $\Delta_{\mathrm{KT}}$ and $\Delta_{\mathrm{\Gamma K}}$ from \GW{} calculations including SOC, compared to measured values (a \cite{Kormanyos_2015,Eknapakul_2014}; b \cite{Kormanyos_2015,Jin_2013}). All values are given in eV.}}
			\label{tab:gap_exp}
			\begin{tabular}{c|c|ccc|ccc|c}
				{\ronaldo Geometry} & & {\ronaldo LDA} & {\ronaldo PBE} & {\ronaldo HSE} & {\ronaldo \GW{@LDA}} &  {\ronaldo \GW{@PBE}} & {\ronaldo \GW{@HSE}} & {\ronaldo Experiment}\\ \hline
				\multirow{2}{*}{\ronaldo PBE}& {\ronaldo $\Delta_{\mathrm{KT}}$} & {\ronaldo 0.24} & {\ronaldo 0.25} & {\ronaldo 0.40} & {\ronaldo $0.00$} & {\ronaldo 0.03} & {\ronaldo 0.14} & {\ronaldo $\gtrsim 0.06^a$} \\
				& {\ronaldo $\Delta_{\mathrm{\Gamma K}}$} & {\ronaldo 0.05} & {\ronaldo 0.04} & {\ronaldo 0.09} & {\ronaldo 0.23} & {\ronaldo 0.21} & {\ronaldo 0.21} & {\ronaldo 0.14$^b$} \\ \hline
				\multirow{2}{*}{\ronaldo HSE}& {\ronaldo $\Delta_{\mathrm{KT}}$} & {\ronaldo 0.23} & {\ronaldo 0.24} & {\ronaldo 0.38} & {\ronaldo $-0.02$} & {\ronaldo 0.01} & {\ronaldo 0.12} & {\ronaldo $\gtrsim 0.06^a$} \\
				& {\ronaldo $\Delta_{\mathrm{\Gamma K}}$} & {\ronaldo 0.05} & {\ronaldo 0.05} & {\ronaldo 0.12} & {\ronaldo 0.25} & {\ronaldo 0.24} & {\ronaldo 0.23} & {\ronaldo 0.14$^b$} \\
				\hline
			\end{tabular}    
	\end{table}}
	
	Other relevant aspects of the band structure concern relative energy differences, in particular $\Delta_{\mathrm{KT}}$ (introduced above) as well as the maximum energy at the $\Gamma$ point wrt the VBM at the $\mathrm{K}$ point, $\Delta_{\mathrm{\Gamma K}} = E_g(\mathrm{KK})-E_g(\mathrm{\Gamma K})$ (see Fig. \ref{fig:bandstructure}). Experimentally, $\Delta_{\mathrm{KT}}$ is expected to be  $\gtrsim 60$~meV \cite{Kormanyos_2015,Eknapakul_2014} and $\Delta_{\mathrm{\Gamma K}} \approx 140$~meV \cite{Kormanyos_2015,Jin_2013}. Taking our calculations with SOC at the HSE geometry, {\ronaldo (see Table \ref{tab:gap_exp})}, the \GW{@HSE} value of {\ronaldo 0.12 eV} reproduces $\Delta_{\mathrm{KT}}$ best. On the other hand, HSE satisfies $\Delta_{\mathrm{\Gamma K}}$ best with a value of {\ronaldo 0.12~eV (0.02 eV} smaller than in experiment), while the values obtained with the other methods differ from experiment by {\ronaldo 0.09 eV} (\GW{@HSE}), {\ronaldo -0.09 eV} (LDA and PBE), {\ronaldo 0.10 eV} (\GW{@PBE}), and {\ronaldo 0.11 eV} (\GW{@LDA}), respectively. {\ronaldo At the PBE geometry, \GW{@HSE} and \GW{@PBE} give the same value for $\Delta_{\mathrm{\Gamma K}}$. With an overestimation of 0.07~eV, it is closer to experiment than the value at the HSE geometry. Again, HSE is the only starting point for which \GW{} predicts $\Delta_{\mathrm{KT}}$ in agreement with experiment.}
	
	In summary, considering the band gap as well as the energy differences $\Delta_{\mathrm{KT}}$ and $\Delta_{\mathrm{\Gamma K}}$, we conclude that {\ronaldo at the PBE geometry, \GW{@HSE} including SOC} shows the best overall agreement with experimental data. {\ronaldo Also for the HSE structure, HSE is the best starting point, with results that are overall only slightly worse. Overall, \GW{@}HSE at the HSE geometry can be considered more appropriate, since only one xc functional is needed for providing decent results for both, the geometry and the electronic properties and thus the most consistent picture. Also for other materials, HSE has been found to be a superior \GW{} starting point \cite{Fuchs_2007,Gant_2022,Yadav_2012,CamarasaGomez_2023} compared to LDA and PBE. For such materials with intermediate band gaps, \cite{Chen_2014,Leppert_2019,Rinke_2005,Atalla_2013,Bruneval_2013,Marom_2012,Koerzdoerfer_2012,Ren_2009}, this functional better justifies the perturbative self-energy correction \cite{Bechstedt,Fuchs_2007,Pela_2016}. Figure \ref{fig:delta-gaps} confirms this for MoS$_2$.
		}
		
		\subsubsection{Comparison with theoretical works}
		By employing coupled-cluster calculations including singles and doubles (CCSD) \cite{Carsky, Helgaker_2008}, Pulkin \textit{et al.} obtained energy gaps of $E_g(\mathrm{\Gamma K})=2.93$~eV and $E_g(\mathrm{KK})=3.00$~eV with an error bar of $\pm0.05$~eV \cite{Pulkin_2020} for the PBE geometry of Ref. \cite{Zhu_2011}; SOC was not included. For our PBE geometry and also omitting SOC, the \GW{@HSE} results are the ones closest to these values, with $E_g(\mathrm{\Gamma K})$ differing by 0.04~eV and $E_g(\mathrm{KK})$ by 0.24~eV.
		
		For a fair comparison with other published \GW{} values with LDA and PBE as starting points, we restrict ourselves here to results obtained by using a Coulomb truncation in combination with either a special treatment of the $\mathbf{q=0}$ singularity or a nonuniform \kgrid{} sampling, since these methods ensure well converged gaps. In Ref. \cite{Rasmussen_2016}, disregarding SOC and adopting the PBE geometry ($a=3.184$~\AA, $d_{\mathrm{S}\mathrm{S}}=3.127$~\AA), a direct band gap of 2.54~eV was reported for \GW{@PBE} which is very close to ours ({\ronaldo $E_g(\mathrm{KK})=2.52$~eV}), \ie differing by only $0.02$~eV. Including SOC and the thus slightly changed PBE geometry ($a=3.18$~\AA, $d_{\mathrm{S}\mathrm{S}}=3.13$~\AA), a \GW{@LDA} value of 2.48 eV was obtained in Ref. \cite{Rasmussen_2015}; at basically the same geometry (differences in the order of 10$^{-3}$ \AA), our results of {\ronaldo $E_g(\mathrm{KK})=2.44$}~eV is only 0.04~eV smaller.
		
		For a lattice parameter of 3.15 \AA, Rasmussen \textit{et al.} calculated a \GW{@PBE} band gap of 2.64 eV without SOC \cite{Rasmussen_2016}. For the same lattice constant, but including SOC, Qiu \textit{et al.} reported a \GW{@LDA} band gap of \cite{Qiu_2016} of $E_g(\mathrm{KK})=2.59$~eV with the plasmon-pole model and $E_g(\mathrm{KK})=2.45$~eV with the contour deformation method. In our case, the structure optimized with HSE ({\ronaldo $a=3.160$~\AA}) is closest to $a=3.15$~\AA. For this structure, without including SOC, we compute a \GW{@PBE} band gap of {\ronaldo $E_g(\mathrm{KK})=2.60$~eV}, which agrees quite well with the one by \cite{Rasmussen_2016}, differing by less than $0.04$~eV. When we include SOC, we obtain {\ronaldo $E_g(\mathrm{KK})=2.52$~eV} with \GW{@LDA}, although at this geometry, we obtain an indirect gap that is $18$~meV smaller than $E_g(\mathrm{KK})$. This is in good agreement with Ref. \cite{Qiu_2016}, with a difference of $0.07$~meV only.
		
		For the experimental geometry and neglecting SOC, H\"{u}ser \textit{et al.}  \cite{Hueser_2013} reported values of $E_g(\mathrm{KT})=2.58$~eV and $E_g(\mathrm{KK})=2.77$~eV for \GW{@LDA}. In our case, at the HSE geometry, we obtain {\ronaldo $E_g(\mathrm{KT})=2.61$~eV} and {\ronaldo $E_g(\mathrm{KK})=2.60$~eV}. As the HSE geometry is close to experiment, we may attribute the discrepancies mainly to the different underlying KS states. Indeed, at the LDA level, the energy gaps in Ref. \cite{Hueser_2013} are $E_g(\mathrm{KK})=1.77$~eV and $E_g(\mathrm{\Gamma K})=1.83$~eV \cite{Hueser_2013}, while ours are {\ronaldo $E_g(\mathrm{KK})=1.73$~eV and $E_g(\mathrm{\Gamma K})=1.71$~eV}. The values for $\Delta E_g$, however, compare fairly well ($\Delta E_g(\mathrm{KK})=1.00$~eV, $\Delta E_g(\mathrm{KT})=$0.6-0.7~eV in Ref. \cite{Hueser_2013}, compared to $\Delta E_g(\mathrm{KK})=0.87$~eV, $\Delta E_g(\mathrm{KT})=0.63$~eV in the present work).
		
		When it comes to \GW{@HSE}, there are only a few results for MoS$_2$ in the literature, neither obtained with Coulomb truncation nor by any special treatment of the $q=0$ singularity. For MoS$_2$, these two aspects lead to opposite effects, competing with each other when converging band gaps with respect to the vacuum size and the number of \kpts{} \cite{Rasmussen_2016}: Neglecting them, band gaps increase when the vacuum layer is enlarged, whereas denser \kgrid{s} make them decrease. Hence, due to fortunate error cancellation, an insufficient vacuum length combined with a coarse \kgrid{} may lead to \GW{} band gaps that agree well with those obtained in a highly converged situation \cite{Rasmussen_2016,Echeverry_2016}. In Ref. \cite{Jiang_2021}, using the PBE geometry and taking SOC into account, a $\mathrm{KK}$ gap of 2.66 eV was reported. With 15~\AA{} of vacuum, a $6\times 6 \times 1$ \kgrid{}, and adopting the PBE geometry, in Ref. \cite{Ramasubramaniam_2012}, band gaps of 2.05 and 2.82~eV at the HSE and \GW{} levels, respectively, have been obtained. The HSE band gap agrees quite well with ours ({\ronaldo 2.04} eV), whereas our \GW{@HSE06} gap is 0.14 eV smaller. The band gap of 2.72 eV reported in Ref. \cite{Echeverry_2016} is based on the experimental lattice parameter of 3.16 \AA, a $12\times 12\times 1$ \kpts{} grid, and a vacuum layer of 17 \AA, and includes SOC effects. The authors state to have chosen these settings to take advantage of error cancellation in the band gap \cite{Echeverry_2016}, and, surprisingly, our band gap of {\ronaldo 2.76} eV obtained with \GW{@HSE} at the HSE geometry ({\ronaldo $a=3.160$~\AA}) agrees quite well.
		
		\section{Conclusions}
		We have employed the LAPW+LO method to provide a set of benchmark \GW{} calculations of the electronic structure of two-dimensional MoS$_2$. We have addressed the impact of geometry, SOC, and DFT starting point on the energy gaps, spin-orbit splittings, and effective masses. {\ronaldo We find that the self-energy corrections to the band gaps hardly depend on the adopted geometry. As could be expected, employing LDA and PBE as starting points does not make a significant difference when the same structure is used. The best agreement with experimental results is achieved by  \GW{@HSE} at either the HSE or PBE geometry, considering SOC. The spin-splittings obtained with all methods agree well with experimental results. This also holds true for the effective hole mass, using either HSE or \GW{} on top of any of the considered starting points (LDA, PBE, and HSE).} In line with other theoretical {\ronaldo works}, we highlight the importance of a Coulomb truncation and an adequate treatment of the Coulomb singularity around $q=0$ as being fundamental for high-quality calculations. {\ronaldo Our findings are expected to be valid for other two-dimensional materials as well.}
		
		\begin{acknowledgments}
			This work received funding from the German Research Foundation, projects 182087777 (CRC HIOS) and 424709454 (SPP 2196, Perovskite Semiconductors). I.G.O. thanks the DAAD (Deutscher Akademischer Austauschdienst) for financial support. {\ronaldo Partial funding is appreciated from the European Union’s Horizon 2020 research and innovation program under the grant agreement Nº 951786 (NOMAD CoE).} Computing time on the supercomputers Lise and Emmy at NHR@ZIB and NHR@Göttingen as part of the NHR infrastructure is gratefully acknowledged.
			
			This version of the article has been accepted for publication, after peer review but is not the Version of Record and does not reflect post-acceptance improvements, or any
			corrections. The Version of Record is available online at: \url{http://dx.doi.org/10.1038/s41524-024-01253-2}.
		\end{acknowledgments}
		
		\section{Author contributions}
		R.R.P. carried out the \GW{}(@LDA and @HSE) calculations with SOC; collected, analyzed, and interpreted all the results; and wrote the first version of the manuscript. C.V. carried out the majority of calculations involving HSE with \exciting; tested the Coulomb truncation and singularity treatment for GW@HSE. S.L. carried out \GW{} (@LDA and @PBE) calculations with SOC. I.G.O. carried out the FHI-aims calculations. B.A. implemented the Coulomb truncation in \exciting. C.D. initiated and guided the overall work. All authors contributed to regular discussions, the next steps to be taken, and to the writing.
		
		{\ronaldo 
			\section{Competing interests}
			The authors declare no competing interests.
			
			\section{Data availability}
			All input and output files are available at NOMAD \cite{Draxl2019,nomad-doi} under \url{https://doi.org/10.17172/NOMAD/2023.09.16-1}. 
		}
		
		\appendix
		
		\section{Comparison between \exciting{} and FHI-aims}\label{sec:fhi_vs_exciting}
		
		Structure optimizations with the all-electron code FHI-aims \cite{Blum2009} are performed by minimizing the amplitude of the interatomic forces below a threshold value of $10^{-3}$~eV \AA$^{-1}$. TIER1 and TIER1+dg basis sets are used for Mo and S, respectively. The sampling of the Brillouin zone (BZ) is carried out with a 12$\times$12$\times$1 $\mathbf{k}$-grid. The HSE screening parameter $\omega$ is set to $0.11 ~\text{bohr}^{-1}$. 
		
		\begin{figure}[h]
			\centering
			\includegraphics[scale=1]{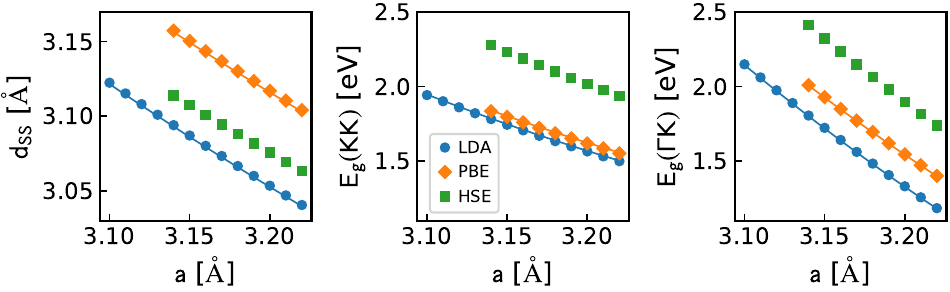}
			\caption{Dependence of $d_{\mathrm{SS}}$, $E_g(\mathrm{KK})$, and $E_g(\mathrm{\Gamma K})$ on the lattice constant $a$ of MoS$_2$ obtained  with \exciting{} (straight lines) and \FHIaims\ (dots) with different xc functionals. }
			\label{fig:FHI-vs-exciting}
		\end{figure}
		
		\begin{table}[htb]
			\centering
			\caption{Maximum absolute difference in $d_{\mathrm{SS}}$, $E_g(\mathrm{KK})$, and $E_g(\mathrm{\Gamma K})$ between results obtained with \exciting{} and \FHIaims. }
			\label{tab:max_diff_FHI_exciting}
			\setlength{\tabcolsep}{5pt}
			\begin{tabular}{cccc}
				\hline
				xc functional & $\Delta d_{\mathrm{SS}}^{max}$ [$10^{-3}$\AA] & $\Delta E_g^{max}(\mathrm{KK})$ [meV] & $\Delta E_g^{max}(\mathrm{\Gamma K})$ [meV] \\ \hline
				LDA & 0.9 & 3.8 & 2.7\\
				PBE & 1.0 & 5.2 & 2.6\\ \hline
			\end{tabular}
		\end{table}
		
		Figure \ref{fig:FHI-vs-exciting} depicts for LDA, PBE, and HSE the dependence of $d_{\mathrm{SS}}$, $E_g(\mathrm{KK})$, and $E_g(\mathrm{\Gamma K})$ on the lattice parameter $a$. For LDA and PBE, comparison of \exciting{} (straight lines) and \FHIaims{} (dots) reveals excellent agreement. Table \ref{tab:max_diff_FHI_exciting} presents the maximum absolute differences of these quantities which are smaller than $1.0\times 10^{-3}$ \AA~for $d_{SS}$ and smaller than 5.2 meV for the energy gaps. From this comparison we expect a similar level of agreement for HSE.
		
		\section{Impact of treatment of the singularity in \GW{}}
		
		\begin{table}[htb]
			\centering
			\caption{Convergence behavior of the energy gaps $E_g(\mathrm{KK})$ and $E_g(\mathrm{\Gamma K})$ with the  \kgrid{} ($n_k \times n_k \times 1$) with and without using the special treatment of the $q=0$ singularity (indicated by the Y and N in the first column).}
			\label{tab:two_schemes_bandgap}
			\setlength{\tabcolsep}{10pt}
			\begin{tabular}{cccccccccc}
				\hline
				& & \multicolumn{8}{c}{$n_k$} \\  \cline{3-10}
				$q=0$ & Gap &  6  & 9 & 12	& 15 & 18 & 21  & 24 & 30 \\ \hline
				N & $E_g (\mathrm{KK})$  & 3.674	& 3.288	& 3.070	& 2.936	& 2.847 & 2.785	& 2.740 & 2.679\\
				N & $E_g (\mathrm{\Gamma K})$ &	3.798	& 3.429	& 3.215	& 3.083	& 2.994 & 2.932	&2.886& 2.825 \\
				\hline 
				Y &$E_g (\mathrm{KK})$       & 2.790	& 2.621 & 2.562	& 2.537	& 2.525 \\
				Y &$E_g (\mathrm{\Gamma K})$ & 2.903	& 2.754	& 2.702	& 2.679	& 2.668 \\
				\hline
			\end{tabular}
		\end{table}
		
		\begin{table}[htb]
			\centering
			\caption{Extrapolated \GW{} energy gaps depending on the range $[n_{k,i},n_{k,f}]$ of the data used in the fit (\ref{eq:extrapolated_gap}) with and without using the special treatment of the $q=0$ singularity (indicated by the Y and N in the first column).}
			\label{tab:extrapolated_bandgap}
			\setlength{\tabcolsep}{5pt}
			\begin{tabular}{ccccccccccc}
				& & \multicolumn{9}{c}{$[n_{k,i}$,$n_{k,f}$]} \\ \cline{3-11}
				$q=0$ & 	& [12,21] & [15,21] & [12,24] & [15,24] & [18,24] & [12,30] & [15,30] & [18,30] & [21,30] \\ 
				\hline
				N & $E_g(\mathrm{KK})$       & 2.58 & 2.57	& 2.57	& 2.56	& 2.56 & 2.56 &	2.55 & 2.55 & 2.54 \\
				N & $E_g(\mathrm{\Gamma K})$ & 2.72	& 2.72	& 2.71	& 2.71	& 2.71 & 2.71 & 2.70 & 2.69 & 2.69 \\
				\hline \hline 
				&                 & [6,18]   & [9,18]  & [12,18] \\
				\hline 
				Y & $E_g(\mathrm{KK})$       & 2.49	& 2.50  & 2.50	\\
				Y & $E_g(\mathrm{\Gamma K})$ & 2.64	& 2.64	& 2.65	\\
				\hline
			\end{tabular}
		\end{table}	
		
		In Table \ref{tab:two_schemes_bandgap}, we compare \GW{@}PBE energy gaps at the PBE geometry obtained with and without the special treatment of the $q=0$ singularity, as described in Section \ref{sec:singularity}, for different numbers of \kpts{}. To extrapolate to an infinitely dense \kgrid{}, we consider an expression that has been adopted frequently in literature  \cite{Friedrich_2011,Friedrich_2012,Pela_2016,Nabok_2016,Rasmussen_2016},
		\begin{equation}\label{eq:extrapolated_gap}
			E_g(n_k)=\frac{A}{B+n_k^2}+E_g(\infty)  .
		\end{equation}
		The fitting coefficient $E_g(\infty)$, is shown in the table as a function of the used data range $[n_{k,i},n_{k,f}]$. The results obtained with the analytic treatment are overall less sensitive to this range, rapidly converging to 2.50 eV and 2.65 eV for $\mathrm{KK}$ and $\mathrm{\Gamma K}$, respectively. In contrast, not using the analytic scheme, shows slow convergence to 2.54 eV and 2.69 eV, respectively. For the analytic treatment, the energy gaps from the $18\times 18 \times 1$ mesh deviate from $E_g(\infty)$ by only 20-30 meV. Without the special treatment, even a \kgrid{} of $30\times 30 \times 1$ leads to errors of 130-140 meV. To obtain a precision of 20-30 meV in this case, we estimate that \kgrid{s} in the range $66\times 66 \times 1$ to $84\times 84 \times 1$ may be necessary.
		
		\section{$GW$ energy gaps from the literature}
		\label{sec:gw_gaps_literature}
		
		{\ronaldo \subsection{MoS$_2$}}
		
		\setlength{\LTcapwidth}{0.9\textwidth}
		\begin{longtable}[c]{c|lll|llc|lrc|lll}
			\caption{\GW{} energy gaps from the literature, indicating the used supercell geometry, the method / approximation, and computational parameters. If no value for $d_{\mathrm{S}\mathrm{S}}$ is provided in the respective reference, only the functional used for relaxation is shown. SP stands for the \GW\ starting point, KS refers to the specific pseudopotential method (NC means norm-conserving, PAW stands for the projector augmented wave method). FI indicates the method for frequency integration, where AC stands for imaginary frequencies and analytic continuation; CD for the contour-deformation approach; PP for plasmon-pole model; RA for full-frequency integration along the real axis. The symbols $^*$, $^\diamond$, and $^\bullet$ mean, respectively, analytical treatment of the $q=0$ singularity, nonuniform neck subsampling, and Sternheimer \GW. The symbol $^\dagger$ indicates the use of Coulomb truncation; $^\oplus$ refers to extrapolation of the gap assuming a $1/L$ behavior with the vacuum size. } \label{tab:gw_gaps_literature}\\
			\hline
			& \multicolumn{3}{c|}{Geometry} &  \multicolumn{3}{c|}{Method} & \multicolumn{3}{c|}{Comp. details} & \multicolumn{3}{c}{$E_g$ [eV]} \\
			Ref. & $a$ [\AA] & $d_{\mathrm{S}\mathrm{S}}$ [\AA] & $L$ [\AA]& SP & KS &   SOC & FI & \kgrid & Misc. & $\mathrm{KK}$ & $\mathrm{\Gamma K}$ & $\mathrm{KT}$\\
			\hline
			\cite{MolinaSanchez_2013} & 3.15 &  & 21 & LDA & NC & Y & PP & $18\times 18 \times 1 $ &  & 2.41 \\
			\cite{Qiu_2016} & 3.15 & LDA & 25  & LDA & NC & Y & CD & $24\times 24 \times 1$&  $\diamond$, $^\dagger$ & 2.45 \\
			\ronaldo{\cite{Gillen_2016}} & \ronaldo{3.15} & \ronaldo{PBE} & \ronaldo{25} & \ronaldo{PBE} & \ronaldo{NC} & \ronaldo{Y} & \ronaldo{PP} & \ronaldo{$24\times 24 \times 1 $} & \ronaldo{$\dagger$} & \ronaldo{2.69}\\
			\cite{Shi_2013} & 3.16& & 19  & PBE & PAW & N & RA & $12\times 12 \times 1 $ & & 2.45  \\
			\cite{Zibouche_2021} & 3.16 & PBE & 20  & PBE & NC & N & AC & $24\times 24 \times 1$ &$\bullet$, $\dagger$ &  2.73  \\
			\cite{Hueser_2013} & 3.16 & & 23  &LDA & PAW & N & PP & $45\times 45 \times 1 $& *, $\dagger$ &  2.77 & & 2.58 \\
			\cite{Echeverry_2016} & 3.16 & PBE&  17  & HSE & PAW & Y & RA & $12\times 12 \times 1$ &  & {\ronaldo{2.72}} \\
			\cite{Schmidt_2017} & 3.160 & 3.170 & 15 & LDA & PAW & Y & RA & $18\times 18 \times 1 $ & *, $\dagger$  & 2.47 \\
			\cite{MolinaSanchez2015} & 3.162 & LDA & 20  & LDA & PAW & Y& RA & $12\times 12 \times 1 $ &  & 2.45 & 2.61 & 2.59 \\
			\cite{Ramasubramaniam_2012} & 3.18 & 3.12 &15 & HSE & PAW & Y& RA & $6\times 6 \times 1$ &  & 2.82 \\
			\cite{Rasmussen_2015} & 3.18 & 3.13 &20 & LDA & PAW & Y & RA & $30\times 30 \times 1$ & *, $\dagger$ &  2.48 \\
			\ronaldo{\cite{Zhuang_2013}} & \ronaldo{3.18} & \ronaldo{3.13} & \ronaldo{18} & \ronaldo{PBE} & \ronaldo{PAW} & \ronaldo{N} & \ronaldo{RA} & \ronaldo{$18\times 18 \times 1 $} &  & \ronaldo{2.36}\\
			\cite{Xia_2020} & 3.18 & 3.16 & 25 & LDA & NC & N & PP & $6\times 6 \times 1$ & *, $\diamond$, $\dagger$ &  2.56 \\
			\cite{Komsa_2012} & 3.18  & PBE & 20 & PBE & PAW & N & RA & $12\times 12 \times 1$ & $\oplus$ &  2.97 \\
			\cite{Liang_2013} & 3.18 & PBE &23 & PBE & NC & N & PP & $12\times 12 \times 1$ & $\dagger$ & 2.75 & 2.91 & 2.98 \\
			\cite{Soklaski_2014} & 3.18 & PBE &  & PBE & NC & N & & &  & 2.63 \\
			\cite{Rasmussen_2016} & 3.184 & 3.127 & 10& PBE & PAW & N & & $18\times 18 \times$ 1 & *, $\dagger$  & 2.54 \\
			\ronaldo{\cite{Haastrup_2018,Gjerding_2021}} & \ronaldo{3.184} & \ronaldo{3.127} & \ronaldo{18} & \ronaldo{PBE} & \ronaldo{PAW} & \ronaldo{Y} & \ronaldo{RA} & \ronaldo{$12\times 12 \times 1 $} & \ronaldo{*, $\dagger$}  & \ronaldo{2.53}\\
			\cite{Shi_2013} & 3.19 &  &19 & PBE & PAW & N& RA & $12\times 12 \times 1 $ &  & 2.50  \\
			\cite{Hueser_2013} & 3.19 &  & 23 & LDA & PAW & N & PP & $45\times 45 \times 1 $&*, $\dagger$  &  2.65 & & 2.57  \\
			\ronaldo{\cite{Kim_2021}} & \ronaldo{3.191} & \ronaldo{3.116} & \ronaldo{25} & \ronaldo{PBE} & \ronaldo{NC} & \ronaldo{Y} & \ronaldo{PP} & \ronaldo{$12\times 12 \times 1 $} & \ronaldo{$\diamond,\dagger$} &  \ronaldo{2.40} & \ronaldo{2.48} & \ronaldo{2.60} \\
			\cite{Ataca_2011} & 3.20 & 3.13 &10 & HSE & PAW & N& RA & $12\times 12 \times 1 $ &  & 2.78 \\
			\cite{Soklaski_2014} & 3.20 & PBE & & PBE & NC & N & & & &  2.55 \\
			\ronaldo{\cite{Echeverry_2016}} & \ronaldo{3.22} & \ronaldo{PBE}&  \ronaldo{17}  & \ronaldo{PBE} & \ronaldo{PAW} & \ronaldo{Y} & \ronaldo{RA} & \ronaldo{$12\times 12 \times 1$} &  & \ronaldo{2.31} \\
			\ronaldo{\cite{Echeverry_2016}} & \ronaldo{3.22} & \ronaldo{PBE}&  \ronaldo{17}  & \ronaldo{HSE} & \ronaldo{PAW} & \ronaldo{Y} & \ronaldo{RA} & \ronaldo{$12\times 12 \times 1$} &  & \ronaldo{2.46} \\
			\ronaldo{\cite{Smart_2018}} & \ronaldo{PBE} & \ronaldo{PBE}&  \ronaldo{16}  & \ronaldo{PBE} & \ronaldo{NC} & \ronaldo{N} & \ronaldo{CD} & \ronaldo{extrapolation} & $\dagger$ & \ronaldo{2.82} \\
			\cite{Jiang_2021} & PBE & PBE & & HSE & PAW & Y & RA & $24\times 24 \times 1 $  & & 2.66 \\
			\cite{Gao_2016} & exp. & exp. & 25& LDA & NC &  & PP & $24\times 24 \times 1 $ & $\dagger$ &  2.64 \\
		\end{longtable}
		
		{\ronaldo
			\subsection{MoSe$_2$}
			
			\setlength{\LTcapwidth}{0.9\textwidth}
			\begin{longtable}[c]{c|lll|llc|lrc|lll}
				\caption{Same as Table \ref{tab:gw_gaps_literature}, but for MoSe$_2$. $d_{\mathrm{Se}\mathrm{Se}}$ is the distance between two Se atoms, analogous to $d_{\mathrm{S}\mathrm{S}}$.} \label{tab:gw_gaps_literature_mose2}\\
				\hline
				& \multicolumn{3}{c|}{Geometry} &  \multicolumn{3}{c|}{Method} & \multicolumn{3}{c|}{Comp. details} & \multicolumn{3}{c}{$E_g$ [eV]} \\
				Ref. & $a$ [\AA] & $d_{\mathrm{Se}\mathrm{Se}}$ [\AA] & $L$ [\AA]& SP & KS &   SOC & FI & \kgrid & Misc. & $\mathrm{KK}$ & $\mathrm{\Gamma K}$ & $\mathrm{KT}$\\
				\hline
				\cite{Gillen_2016} & 3.279 & PBE & 25 & PBE & NC & Y & PP & $24\times 24 \times 1 $ & $\dagger$ & 2.33\\
				\cite{Schmidt_2017} & 3.289 & 3.340 & 15 & LDA & PAW & Y & RA & $18\times 18 \times 1 $ & $*,\dagger$ & 2.08 \\
				\cite{Liang_2013} & 3.31 & PBE & 23 & PBE & NC & N & PP & $12\times 12 \times 1 $ & $\dagger$ & 2.33 & 2.67 & 2.66 \\
				\cite{Kim_2021} & 3.316 & 3.332 & 25 & PBE & NC & Y & PP & $12\times 12 \times 1 $ & $\diamond,\dagger$ &  2.08 & 2.47 & 2.17\\
				\cite{Rasmussen_2015} & 3.32 & 3.34 & 20 & LDA & PAW & Y & RA & $30\times 30 \times 1 $ & $*,\dagger$ & 2.18 \\
				\cite{Zhuang_2013} & 3.32 & 3.34 & 18 & PBE & PAW & N & RA & $18\times 18 \times 1 $ &  &  2.04\\
				\cite{Ramasubramaniam_2012} & 3.32 & 3.34 &15 & HSE & PAW & Y& RA & $6\times 6 \times 1$ &  & 2.41 \\
				\cite{Haastrup_2018,Gjerding_2021} & 3.32 & 3.34 & 18.3 & PBE & PAW & Y& RA& $9\times 9 \times$ 1 &*, $\dagger$ & 2.12\\
				\cite{Echeverry_2016} & 3.32 & PBE & 17 & PBE & PAW & Y & RA & $12\times 12 \times 1 $ & & 2.13 \\
				\cite{Echeverry_2016} & 3.32 & PBE & 17 & HSE & PAW & Y & RA & $12\times 12 \times 1 $ & & 2.24 \\
				\cite{Horzum_2013} & 3.321 & 3.293 & 30 & PBE & PAW & N & RA & $12\times 12 \times 1 $ & $\oplus$ & 2.33\\
				\cite{Gao_2016} & exp. & exp. & 25& LDA & NC &  & PP & $24\times 24 \times 1 $ & $\dagger$ &  2.28 \\
			\end{longtable}
			
			\subsection{MoTe$_2$}
			\setlength{\LTcapwidth}{0.9\textwidth}
			\begin{longtable}[c]{c|lll|llc|lrc|lll}
				\caption{Same as Table \ref{tab:gw_gaps_literature} but for MoTe$_2$. $d_{\mathrm{Te}\mathrm{Te}}$ is the distance between two Se atoms, analogous to $d_{\mathrm{S}\mathrm{S}}$.} \label{tab:gw_gaps_literature_mote2}\\
				\hline
				& \multicolumn{3}{c|}{Geometry} &  \multicolumn{3}{c|}{Method} & \multicolumn{3}{c|}{Comp. details} & \multicolumn{3}{c}{$E_g$ [eV]} \\
				Ref. & $a$ [\AA] & $d_{\mathrm{Te}\mathrm{Te}}$ [\AA] & $L$ [\AA]& SP & KS &   SOC & FI & \kgrid & Misc. & $\mathrm{KK}$ & $\mathrm{\Gamma K}$ & $\mathrm{KT}$\\
				\hline
				\cite{Liang_2013} & 3.51 & PBE & 23 & PBE & NC & N & PP & $12\times 12 \times 1 $ & $\dagger$ & 1.82 & 2.65 & 2.16 \\
				\cite{Gillen_2016} & 3.515 & PBE & 25 & PBE & NC & Y & PP & $24\times 24 \times 1 $ & $\dagger$ & 1.78\\
				\cite{Echeverry_2016} & 3.52 & PBE & 17 & HSE & PAW & Y & RA & $12\times 12 \times 1 $ & & 1.80 \\
				\cite{Kim_2021} & 3.53 & 3.636 & 25 & PBE & NC & Y & PP & $12\times 12 \times 1 $ & $\diamond,\dagger$ & 1.65 & 2.28 & 1.54 \\
				\cite{Haastrup_2018,Gjerding_2021} & 3.547 & 3.61 & 18.6 & PBE & PAW & Y& RA& $12\times 12 \times$ 1 &*, $\dagger$ & 1.56\\
				\cite{Rasmussen_2015} & 3.55 & 3.61 & 20 & LDA & PAW & Y & RA & $30\times 30 \times 1 $ & $*,\dagger$ & 1.72 \\
				\cite{Zhuang_2013} & 3.32 & 3.61 & 18 & PBE & PAW & N & RA & $18\times 18 \times 1 $ &  & 1.54  \\
				\cite{Ramasubramaniam_2012} & 3.55 & 3.62 &15 & HSE & PAW & Y& RA & $6\times 6 \times 1$ &  & 1.77 \\
				\cite{Echeverry_2016} & 3.55 & PBE & 17 & PBE & PAW & Y & RA & $12\times 12 \times 1 $ & & 1.61 \\
				\cite{Echeverry_2016} & 3.55 & PBE & 17 & HSE & PAW & Y & RA & $12\times 12 \times 1 $ & & 1.72 \\
			\end{longtable}
			
			\subsection{WS$_2$}
			\setlength{\LTcapwidth}{0.9\textwidth}
			\begin{longtable}[c]{c|lll|llc|lrc|lll}
				\caption{Same as Table \ref{tab:gw_gaps_literature} but for WS$_2$.} \label{tab:gw_gaps_literature_ws2}\\
				\hline
				& \multicolumn{3}{c|}{Geometry} &  \multicolumn{3}{c|}{Method} & \multicolumn{3}{c|}{Comp. details} & \multicolumn{3}{c}{$E_g$ [eV]} \\ 
				Ref. & $a$ [\AA] & $d_{\mathrm{S}\mathrm{S}}$ [\AA] & $L$ [\AA]& SP & KS &   SOC & FI & \kgrid & Misc. & $\mathrm{KK}$ & $\mathrm{\Gamma K}$ & $\mathrm{KT}$\\
				\hline
				\cite{Schmidt_2017} & 3.153 & 3.36 & 15 & LDA & PAW & Y & RA & $18\times 18 \times 1 $ & $*,\dagger$ & 2.75 \\
				\cite{Shi_2013} & 3.155& & 19  & PBE & PAW & N & RA & $12\times 12 \times 1 $ & & 2.91& 3.25 & 2.71  \\
				\cite{Zhuang_2013} & 3.18 & 3.14 & 18 & PBE & PAW & N & RA & $18\times 18 \times 1 $ & & 2.64 &  \\
				\cite{Echeverry_2016} & 3.18 & PBE & 17 & HSE & PAW & Y & RA & $12\times 12 \times 1 $ & & 2.73 \\
				\cite{Haastrup_2018,Gjerding_2021} & 3.186 & 3.146 & 18.1 & PBE & PAW & Y& RA& $12\times 12 \times$ 1 &*, $\dagger$ & 2.53\\
				\cite{Kim_2021} & 3.189 & 3.136 & 25 & PBE & NC & Y & PP & $12\times 12 \times 1 $ & $\diamond,\dagger$ & 2.46 & 2.65& 2.50\\
				\cite{Ramasubramaniam_2012} & 3.19 & 3.14 &15 & HSE & PAW & Y& RA & $6\times 6 \times 1$ &  & 2.88 \\
				\cite{Rasmussen_2015} & 3.19 & 3.15 & 20 & LDA & PAW & Y & RA & $30\times 30 \times 1 $ & $*,\dagger$ & 2.43 \\
				\cite{Shi_2013} & 3.190& & 19  & PBE & PAW & N & RA & $12\times 12 \times 1 $ & & 2.73& 2.92 & 2.76  \\
				\cite{Liang_2013} & 3.20 & PBE & 23 & PBE & NC & N & PP & $12\times 12 \times 1 $ & $\dagger$ & 2.88 & 2.94 & 3.13 \\
				\cite{Shi_2013} & 3.250& & 19  & PBE & PAW & N & RA & $12\times 12 \times 1 $ & & 2.46& 2.39 & 2.83  \\
				\cite{Lee_2017} & LDA & LDA & 20 & LDA & PAW & N & RA & $4\times 12 \times 1$ & & 3.19\\
			\end{longtable}
			
			\subsection{WSe$_2$}
			\setlength{\LTcapwidth}{0.9\textwidth}
			\begin{longtable}[c]{c|lll|llc|lrc|lll}
				\caption{Same as Table \ref{tab:gw_gaps_literature} but for WSe$_2$.} \label{tab:gw_gaps_literature_wse2}\\
				\hline
				& \multicolumn{3}{c|}{Geometry} &  \multicolumn{3}{c|}{Method} & \multicolumn{3}{c|}{Comp. details} & \multicolumn{3}{c}{$E_g$ [eV]} \\
				Ref. & $a$ [\AA] & $d_{\mathrm{Se}\mathrm{Se}}$ [\AA] & $L$ [\AA]& SP & KS &   SOC & FI & \kgrid & Misc. & $\mathrm{KK}$ & $\mathrm{\Gamma K}$ & $\mathrm{KT}$\\
				\hline
				\cite{Echeverry_2016} & 3.28 & PBE & 17 & PBE & PAW & Y & RA & $12\times 12 \times 1 $ & & 2.06 \\
				\cite{Echeverry_2016} & 3.28 & PBE & 17 & HSE & PAW & Y & RA & $12\times 12 \times 1 $ & & 2.18 \\
				\cite{Elliott_2020}& 3.29& 3.34 & 25.2 & LDA & NC & Y & & $24\times 24 \times 1 $ & $\dagger$ & 2.89 & 3.52 & 2.60 \\
				\cite{Haastrup_2018,Gjerding_2021} & 3.319 & 3.356 & 18.4 & PBE & PAW & Y& RA& $12\times 12 \times$ 1 &*, $\dagger$ & 2.13 & & 2.10\\
				\cite{Echeverry_2016} & 3.32 & PBE & 17 & HSE & PAW & Y & RA & $12\times 12 \times 1 $ & & 2.27 \\
				\cite{Rasmussen_2015} & 3.32 & 3.36 & 20 & LDA & PAW & Y & RA & $30\times 30 \times 1 $ & $*,\dagger$ & 2.08 \\
				\cite{Zhuang_2013} & 3.32 & 3.36 & 18 & PBE & PAW & N & RA & $18\times 18 \times 1 $ &  &  2.26 \\
				\cite{Ramasubramaniam_2012} & 3.32 & 3.36 &15 & HSE & PAW & Y& RA & $6\times 6 \times 1$ &  & 2.42 & & 2.34 \\
				\cite{Liang_2013} & 3.33 & PBE & 23 & PBE & NC & N & PP & $12\times 12 \times 1 $ & $\dagger$ & 2.38 & 2.72 & 2.74 \\
				\cite{Kim_2021} & 3.334 & 3.340 & 25 & PBE & NC & Y & PP & $12\times 12 \times 1 $ & $\diamond,\dagger$ & 2.01 & 2.47 & 2.06\\
				\cite{Lee_2017} & LDA & LDA & 20 & LDA & PAW & N & RA & $4\times 12 \times 1$ & & 1.7\\
			\end{longtable}

			\subsection{BN}
			\setlength{\LTcapwidth}{0.9\textwidth}
			\begin{longtable}[c]{c|ll|llc|lrc|ll}
				\caption{Same as Table \ref{tab:gw_gaps_literature} but for BN. G stands for atom-centered Gaussian orbitals and US for ultra-soft pseudopotentials.} \label{tab:gw_gaps_literature_bn}\\
				\hline
				& \multicolumn{2}{c|}{Geometry} &  \multicolumn{3}{c|}{Method} & \multicolumn{3}{c|}{Comp. details} & \multicolumn{2}{c}{$E_g$ [eV]} \\
				Ref. & $a$ [\AA] & $L$ [\AA]& SP & KS &   SOC & FI & \kgrid & Misc. & $\mathrm{KK}$ & $\mathrm{K\Gamma}$ \\
				\hline
				\cite{Kirchhoff_2022} & 2.479 & & LDA & NC+G & Y & & $12\times 12 \times$ 1 & & 7.60 & 7.27 \\
				\cite{Hueser_2013_b} & 2.5 & 30& PBE & PAW & N& PP & $45\times 45 \times$ 1 & $\dagger$  &  7.37 & 6.58 \\
				\cite{Ferreira_2019} & 2.5 & 15.8& PBE & NC & N& & $16\times 16 \times$ 1 & $\dagger$  &  7.77 & 7.32 \\
				\cite{Attaccalite_2011} & 2.50 & 10.6& LDA & NC & Y & & $10\times 10 \times$ 1 & & & 7.03 \\
				\cite{Galvani_2016} & 2.50 & & LDA & & N & PP & $24\times 24 \times$ 1 & $\dagger$ & 7.25& \\
				\cite{Wu_2017} & 2.50 & 16& PBE & NC & Y & CD & $12\times 12 \times$ 1 & & 7.17\\
				\cite{Rasmussen_2016} & 2.504 & 10& PBE & PAW & N & & $18\times 18 \times$ 1 & *, $\dagger$  &  7.80 & 7.06 \\
				\cite{Blase_1995} & 2.51 & 13.5 & LDA & NC & N & PP & $4\times 4 \times$ 2 & & 6.37 & 6.00 \\
				\cite{Haastrup_2018,Gjerding_2021} & 2.51 & 15 & PBE & PAW & Y& RA& $12\times 12 \times$ 1 &*, $\dagger$ & 7.74& 7.12\\
				\cite{Wang_2020} & 2.51 & 16 & PBE & US & N & & extrapolation& $\dagger$ & & 7.20\\
				\cite{Mengle_2019} & 2.51 & 20 & LDA & & N & PP & $24\times 24 \times$ 1 & $\dagger$& & 7.74\\
				\cite{Fu_2016} & 2.517 & 18.5& PBE & LAPW & N & & $5\times 9 \times$ 1 & $\dagger$  &  7.60 & 6.95 \\    
				\cite{Berseneva_2013} & PBE &  40 & PBE & PAW & Y & RA & $6\times 6 \times 1 $ & $\oplus$ & & 7.40 \\
				\cite{Smart_2018} & PBE & 16 & PBE & NC & N & CD & extrapolation & $\dagger$ & & 7.01 \\
				\cite{Cudazzo_2016}& exp & 20 & LDA & NC & N& &$48\times 48 \times 1 $ & $\dagger$ & & 7.36 \\
			\end{longtable}

			\subsection{Phosphorene}
			\setlength{\LTcapwidth}{0.9\textwidth}
			\begin{longtable}[c]{c|clll|llc|lrc|c}
				\caption{Same as Table \ref{tab:gw_gaps_literature} but for phosphorene. $a$ and $b$ are the two in-plane lattice parameters; $d$ refers to the layer thickness; and $\theta$ is the in-plane P-P-P angle. 
				} \label{tab:gw_gaps_literature_phosphorene}\\
				\hline
				& \multicolumn{4}{c|}{Geometry} &  \multicolumn{3}{c|}{Method} & \multicolumn{3}{c|}{Comp. details} & $E_g$ [eV] \\
				Ref. & $a\times b$ [\AA$^2$] & $d$ [\AA] & $\theta$ [°] & $L$ [\AA]& SP & KS &   SOC & FI & \kgrid & Misc. & $\mathrm{\Gamma \Gamma}$ \\
				\hline
				\cite{Ferreira_2017} & 3.30$\times$4.59 & & & 26.5 & PBE & NC & N & PP & $17\times 17 \times$ 1 & $\dagger$ & 2.06 \\
				\cite{Rasmussen_2016} & 3.306$\times$4.630 & 2.110 & 95.8 & 10& PBE & PAW & N & & $14\times 10 \times$ 1 & *, $\dagger$  & 2.03 \\
				\cite{Steinkasserer_2016} & 3.31$\times$4.52 &  &  & 10& PBE & PAW & N & & $12\times 8 \times$ 1 & *, $\dagger$  & 1.89 \\
				\cite{Steinkasserer_2016} & 3.32$\times$4.42 &  &  & 10& PBE & PAW & N & & $12\times 8 \times$ 1 & *, $\dagger$  & 1.72 \\
				\cite{Rudenko_2014} & exp.& & & 20& PBE & PAW & N & RA & $12\times 10 \times$ 1 &  & 1.60 \\
				\cite{Tran_2014} & PBE& & & & PBE & & N &  & $14\times 10 \times$ 1 &  & 2.00 \\
				\cite{Cakir_2014} & PBE& & & 15& HSE & PAW & N& RA  & $13\times 9 \times$ 1 &  & 2.31 \\
				\cite{Tran_2015} & & & & &PBE &NC & N & PP & $21\times 15 \times$ 1 & $\dagger$ & 2.0 \\ 
			\end{longtable}
			
		}
		
		\section{Spin-orbit splittings and effective masses for different geometries} 
		Table \ref{tab:SOC_and_meff_all_geometries} provides spin-orbit splittings in the valence and conduction band at the K point for the LDA, PBE, and HSE geometries together with effective electron and hole masses at the $\mathrm{K}$ point along the directions K$\Gamma$ and KM. $\Delta_{val}$ and $\Delta_{cond}$ are not very sensitive to the geometry in the sense that, for a given methodology, $\Delta_{val}$ ($\Delta_{cond}$) varies less than 5 meV (1 meV) between the different geometries. A similar trend is also observed for the effective masses.
		
		\setlength{\LTcapwidth}{0.9\textwidth}
		\begin{table}[htb]
			\caption{Spin-orbit splittings in the valence (conduction) band $\Delta_{val}$ ($\Delta_{cond}$), in meV, evaluated at the {\textrm K} point for the various geometries considered in this work as well as effective electron (hole) masses $m_h$ ($m_e$), in units of $m_0$, at the K point along different directions (in parentheses).}
			\label{tab:SOC_and_meff_all_geometries}
			\renewcommand{\arraystretch}{0.6}
			\vspace{0.4cm}
			\begin{tabular}{l|l|ccc|ccc}
				Geometry  &  & LDA & PBE & HSE & \GW{@LDA} & \GW{@PBE} & \GW{@HSE}\\ \hline
				\multirow{6}{*}{LDA}       & $\Delta_{val}$ & 146&      146&      141&      146&      146&      141\\
				& $\Delta_{cond}$ & 3&        3&        4&        2&        2&        3\\ 
				& $m_e(\mathrm{K\Gamma})$ &0.42&     0.42&     0.39&     0.45&     0.45&     0.41\\
				& $m_e(\mathrm{KM})$ & 0.43&     0.43&     0.40&     0.45&     0.44&     0.42\\
				& $m_h(\mathrm{K\Gamma})$ & 0.52&    0.52&    0.46&    0.43&    0.44&    0.43\\
				& $m_h(\mathrm{KM})$ & 0.56&    0.56&    0.49&    0.45&    0.46&    0.45\\
				\hline
				\multirow{6}{*}{PBE}       & $\Delta_{val}$ & 150&      149&      144&      150&      149&      144\\
				& $\Delta_{cond}$ & 3&        3&        4&        3&        3&        4 \\ 
				& $m_e(\mathrm{K\Gamma})$  & 0.41&     0.42& 0.38&    0.42&     0.42 & 0.40\\
				& $m_e(\mathrm{KM})$ & 0.42&     0.43&  0.39&   0.43&     0.43 & 0.40\\
				& $m_h(\mathrm{K\Gamma})$ & 0.52&    0.52& 0.45 &    0.43&    0.43 & 0.42\\
				& $m_h(\mathrm{KM})$ & 0.56&    0.56& 0.48 &  0.45&    0.45 & 0.45\\
				\hline
				\multirow{6}{*}{HSE}       & $\Delta_{val}$ & 149 & 148 & 144& 149 & 148 & 143\\
				& $\Delta_{cond}$ & 3 & 3 & 4 & 3 & 3 & 4\\ 
				& $m_e(\mathrm{K\Gamma})$ & 0.40&     0.41&     0.37&     0.42&     0.42&     0.39\\
				& $m_e(\mathrm{KM})$ & 0.42&     0.42&     0.38&     0.42&     0.42&     0.39\\
				& $m_h(\mathrm{K\Gamma})$ & 0.49&    0.50&    0.44&    0.42&    0.42&    0.42\\
				& $m_h(\mathrm{KM})$ & 0.53&    0.53&    0.47&    0.44&    0.44&    0.44\\
			\end{tabular}
		\end{table}


\begin{thebibliography}{153}%
			\makeatletter
			\providecommand \@ifxundefined [1]{%
				\@ifx{#1\undefined}
			}%
			\providecommand \@ifnum [1]{%
				\ifnum #1\expandafter \@firstoftwo
				\else \expandafter \@secondoftwo
				\fi
			}%
			\providecommand \@ifx [1]{%
				\ifx #1\expandafter \@firstoftwo
				\else \expandafter \@secondoftwo
				\fi
			}%
			\providecommand \natexlab [1]{#1}%
			\providecommand \enquote  [1]{``#1''}%
			\providecommand \bibnamefont  [1]{#1}%
			\providecommand \bibfnamefont [1]{#1}%
			\providecommand \citenamefont [1]{#1}%
			\providecommand \href@noop [0]{\@secondoftwo}%
			\providecommand \href [0]{\begingroup \@sanitize@url \@href}%
			\providecommand \@href[1]{\@@startlink{#1}\@@href}%
			\providecommand \@@href[1]{\endgroup#1\@@endlink}%
			\providecommand \@sanitize@url [0]{\catcode `\\12\catcode `\$12\catcode
				`\&12\catcode `\#12\catcode `\^12\catcode `\_12\catcode `\%12\relax}%
			\providecommand \@@startlink[1]{}%
			\providecommand \@@endlink[0]{}%
			\providecommand \url  [0]{\begingroup\@sanitize@url \@url }%
			\providecommand \@url [1]{\endgroup\@href {#1}{\urlprefix }}%
			\providecommand \urlprefix  [0]{URL }%
			\providecommand \Eprint [0]{\href }%
			\providecommand \doibase [0]{https://doi.org/}%
			\providecommand \selectlanguage [0]{\@gobble}%
			\providecommand \bibinfo  [0]{\@secondoftwo}%
			\providecommand \bibfield  [0]{\@secondoftwo}%
			\providecommand \translation [1]{[#1]}%
			\providecommand \BibitemOpen [0]{}%
			\providecommand \bibitemStop [0]{}%
			\providecommand \bibitemNoStop [0]{.\EOS\space}%
			\providecommand \EOS [0]{\spacefactor3000\relax}%
			\providecommand \BibitemShut  [1]{\csname bibitem#1\endcsname}%
			\let\auto@bib@innerbib\@empty
			\bibitem [{\citenamefont {Geim}\ and\ \citenamefont
				{Novoselov}(2007)}]{Geim_2007}%
			\BibitemOpen
			\bibfield  {author} {\bibinfo {author} {\bibfnamefont {A.~K.}\ \bibnamefont
					{Geim}}\ and\ \bibinfo {author} {\bibfnamefont {K.~S.}\ \bibnamefont
					{Novoselov}},\ }\bibfield  {title} {\bibinfo {title} {{The rise of
						graphene}},\ }\href {https://doi.org/10.1038/nmat1849} {\bibfield  {journal}
				{\bibinfo  {journal} {Nature Materials}\ }\textbf {\bibinfo {volume} {6}},\
				\bibinfo {pages} {183} (\bibinfo {year} {2007})}\BibitemShut {NoStop}%
			\bibitem [{\citenamefont {Shanmugam}\ \emph {et~al.}(2022)\citenamefont
				{Shanmugam}, \citenamefont {Mensah}, \citenamefont {Babu}, \citenamefont
				{Gawusu}, \citenamefont {Chanda}, \citenamefont {Tu}, \citenamefont
				{Neisiany}, \citenamefont {F{\"o}rsth}, \citenamefont {Sas},\ and\
				\citenamefont {Das}}]{Shanmugam_2022}%
			\BibitemOpen
			\bibfield  {author} {\bibinfo {author} {\bibfnamefont {V.}~\bibnamefont
					{Shanmugam}}, \bibinfo {author} {\bibfnamefont {R.~A.}\ \bibnamefont
					{Mensah}}, \bibinfo {author} {\bibfnamefont {K.}~\bibnamefont {Babu}},
				\bibinfo {author} {\bibfnamefont {S.}~\bibnamefont {Gawusu}}, \bibinfo
				{author} {\bibfnamefont {A.}~\bibnamefont {Chanda}}, \bibinfo {author}
				{\bibfnamefont {Y.}~\bibnamefont {Tu}}, \bibinfo {author} {\bibfnamefont
					{R.~E.}\ \bibnamefont {Neisiany}}, \bibinfo {author} {\bibfnamefont
					{M.}~\bibnamefont {F{\"o}rsth}}, \bibinfo {author} {\bibfnamefont
					{G.}~\bibnamefont {Sas}},\ and\ \bibinfo {author} {\bibfnamefont
					{O.}~\bibnamefont {Das}},\ }\bibfield  {title} {\bibinfo {title} {A review of
					the synthesis, properties, and applications of 2d materials},\ }\href@noop {}
			{\bibfield  {journal} {\bibinfo  {journal} {Particle \& Particle Systems
						Characterization}\ }\textbf {\bibinfo {volume} {39}},\ \bibinfo {pages}
				{2200031} (\bibinfo {year} {2022})}\BibitemShut {NoStop}%
			\bibitem [{\citenamefont {Kumbhakar}\ \emph {et~al.}(2023)\citenamefont
				{Kumbhakar}, \citenamefont {Jayan}, \citenamefont {Madhavikutty},
				\citenamefont {Sreeram}, \citenamefont {Appukuttan}, \citenamefont {Ito},\
				and\ \citenamefont {Tiwary}}]{Kumbhakar_2023}%
			\BibitemOpen
			\bibfield  {author} {\bibinfo {author} {\bibfnamefont {P.}~\bibnamefont
					{Kumbhakar}}, \bibinfo {author} {\bibfnamefont {J.~S.}\ \bibnamefont
					{Jayan}}, \bibinfo {author} {\bibfnamefont {A.~S.}\ \bibnamefont
					{Madhavikutty}}, \bibinfo {author} {\bibfnamefont {P.}~\bibnamefont
					{Sreeram}}, \bibinfo {author} {\bibfnamefont {S.}~\bibnamefont {Appukuttan}},
				\bibinfo {author} {\bibfnamefont {T.}~\bibnamefont {Ito}},\ and\ \bibinfo
				{author} {\bibfnamefont {C.~S.}\ \bibnamefont {Tiwary}},\ }\bibfield  {title}
			{\bibinfo {title} {Prospective applications of two-dimensional materials
					beyond laboratory frontiers: A review},\ }\href
			{https://doi.org/10.1016/j.isci.2023.106671} {\bibfield  {journal} {\bibinfo
					{journal} {Iscience}\ }\textbf {\bibinfo {volume} {26}},\ \bibinfo {pages}
				{106671} (\bibinfo {year} {2023})}\BibitemShut {NoStop}%
			\bibitem [{\citenamefont {Mir}\ \emph {et~al.}(2020)\citenamefont {Mir},
				\citenamefont {Yadav},\ and\ \citenamefont {Singh}}]{Mir_2020}%
			\BibitemOpen
			\bibfield  {author} {\bibinfo {author} {\bibfnamefont {S.~H.}\ \bibnamefont
					{Mir}}, \bibinfo {author} {\bibfnamefont {V.~K.}\ \bibnamefont {Yadav}},\
				and\ \bibinfo {author} {\bibfnamefont {J.~K.}\ \bibnamefont {Singh}},\
			}\bibfield  {title} {\bibinfo {title} {Recent advances in the carrier
					mobility of two-dimensional materials: a theoretical perspective},\
			}\href@noop {} {\bibfield  {journal} {\bibinfo  {journal} {ACS omega}\
				}\textbf {\bibinfo {volume} {5}},\ \bibinfo {pages} {14203} (\bibinfo {year}
				{2020})}\BibitemShut {NoStop}%
			\bibitem [{\citenamefont {Zeng}\ \emph {et~al.}(2018)\citenamefont {Zeng},
				\citenamefont {Xiao}, \citenamefont {Liu}, \citenamefont {Yang},\ and\
				\citenamefont {Fu}}]{Zeng_2018}%
			\BibitemOpen
			\bibfield  {author} {\bibinfo {author} {\bibfnamefont {M.}~\bibnamefont
					{Zeng}}, \bibinfo {author} {\bibfnamefont {Y.}~\bibnamefont {Xiao}}, \bibinfo
				{author} {\bibfnamefont {J.}~\bibnamefont {Liu}}, \bibinfo {author}
				{\bibfnamefont {K.}~\bibnamefont {Yang}},\ and\ \bibinfo {author}
				{\bibfnamefont {L.}~\bibnamefont {Fu}},\ }\bibfield  {title} {\bibinfo
				{title} {Exploring two-dimensional materials toward the next-generation
					circuits: from monomer design to assembly control},\ }\href@noop {}
			{\bibfield  {journal} {\bibinfo  {journal} {Chemical reviews}\ }\textbf
				{\bibinfo {volume} {118}},\ \bibinfo {pages} {6236} (\bibinfo {year}
				{2018})}\BibitemShut {NoStop}%
			\bibitem [{\citenamefont {Zhang}(2018)}]{Zhang_2018}%
			\BibitemOpen
			\bibfield  {author} {\bibinfo {author} {\bibfnamefont {H.}~\bibnamefont
					{Zhang}},\ }\bibfield  {title} {\bibinfo {title} {Introduction: 2d materials
					chemistry},\ }\href@noop {} {\bibfield  {journal} {\bibinfo  {journal}
					{Chemical reviews}\ }\textbf {\bibinfo {volume} {118}},\ \bibinfo {pages}
				{6089} (\bibinfo {year} {2018})}\BibitemShut {NoStop}%
			\bibitem [{\citenamefont {Chaves}\ \emph {et~al.}(2020)\citenamefont {Chaves},
				\citenamefont {Azadani}, \citenamefont {Alsalman}, \citenamefont {Da~Costa},
				\citenamefont {Frisenda}, \citenamefont {Chaves}, \citenamefont {Song},
				\citenamefont {Kim}, \citenamefont {He}, \citenamefont {Zhou} \emph
				{et~al.}}]{Chaves_2020}%
			\BibitemOpen
			\bibfield  {author} {\bibinfo {author} {\bibfnamefont {A.}~\bibnamefont
					{Chaves}}, \bibinfo {author} {\bibfnamefont {J.~G.}\ \bibnamefont {Azadani}},
				\bibinfo {author} {\bibfnamefont {H.}~\bibnamefont {Alsalman}}, \bibinfo
				{author} {\bibfnamefont {D.}~\bibnamefont {Da~Costa}}, \bibinfo {author}
				{\bibfnamefont {R.}~\bibnamefont {Frisenda}}, \bibinfo {author}
				{\bibfnamefont {A.}~\bibnamefont {Chaves}}, \bibinfo {author} {\bibfnamefont
					{S.~H.}\ \bibnamefont {Song}}, \bibinfo {author} {\bibfnamefont {Y.~D.}\
					\bibnamefont {Kim}}, \bibinfo {author} {\bibfnamefont {D.}~\bibnamefont
					{He}}, \bibinfo {author} {\bibfnamefont {J.}~\bibnamefont {Zhou}}, \emph
				{et~al.},\ }\bibfield  {title} {\bibinfo {title} {Bandgap engineering of
					two-dimensional semiconductor materials},\ }\href@noop {} {\bibfield
				{journal} {\bibinfo  {journal} {npj 2D Materials and Applications}\ }\textbf
				{\bibinfo {volume} {4}},\ \bibinfo {pages} {29} (\bibinfo {year}
				{2020})}\BibitemShut {NoStop}%
			\bibitem [{\citenamefont {Guinea}\ \emph {et~al.}(2014)\citenamefont {Guinea},
				\citenamefont {Katsnelson},\ and\ \citenamefont {Wehling}}]{Guinea_2014}%
			\BibitemOpen
			\bibfield  {author} {\bibinfo {author} {\bibfnamefont {F.}~\bibnamefont
					{Guinea}}, \bibinfo {author} {\bibfnamefont {M.~I.}\ \bibnamefont
					{Katsnelson}},\ and\ \bibinfo {author} {\bibfnamefont {T.~O.}\ \bibnamefont
					{Wehling}},\ }\bibfield  {title} {\bibinfo {title} {Two-dimensional
					materials: Electronic structure and many-body effects},\ }\href
			{https://doi.org/10.1002/andp.201470096} {\bibfield  {journal} {\bibinfo
					{journal} {Annalen der Physik}\ }\textbf {\bibinfo {volume} {526}},\ \bibinfo
				{pages} {A81} (\bibinfo {year} {2014})}\BibitemShut {NoStop}%
			\bibitem [{\citenamefont {Xiao}\ \emph {et~al.}(2017)\citenamefont {Xiao},
				\citenamefont {Zhao}, \citenamefont {Wang},\ and\ \citenamefont
				{Zhang}}]{Xiao_2017}%
			\BibitemOpen
			\bibfield  {author} {\bibinfo {author} {\bibfnamefont {J.}~\bibnamefont
					{Xiao}}, \bibinfo {author} {\bibfnamefont {M.}~\bibnamefont {Zhao}}, \bibinfo
				{author} {\bibfnamefont {Y.}~\bibnamefont {Wang}},\ and\ \bibinfo {author}
				{\bibfnamefont {X.}~\bibnamefont {Zhang}},\ }\bibfield  {title} {\bibinfo
				{title} {{Excitons in atomically thin 2D semiconductors and their
						applications}},\ }\href@noop {} {\bibfield  {journal} {\bibinfo  {journal}
					{Nanophotonics}\ }\textbf {\bibinfo {volume} {6}},\ \bibinfo {pages} {1309}
				(\bibinfo {year} {2017})}\BibitemShut {NoStop}%
			\bibitem [{\citenamefont {Mueller}\ and\ \citenamefont
				{Malic}(2018)}]{Mueller_2018}%
			\BibitemOpen
			\bibfield  {author} {\bibinfo {author} {\bibfnamefont {T.}~\bibnamefont
					{Mueller}}\ and\ \bibinfo {author} {\bibfnamefont {E.}~\bibnamefont
					{Malic}},\ }\bibfield  {title} {\bibinfo {title} {{Exciton physics and device
						application of two-dimensional transition metal dichalcogenide
						semiconductors}},\ }\href {https://doi.org/10.1038/s41699-018-0074-2}
			{\bibfield  {journal} {\bibinfo  {journal} {npj 2D Materials and
						Applications}\ }\textbf {\bibinfo {volume} {2}},\ \bibinfo {pages} {29}
				(\bibinfo {year} {2018})},\ \Eprint {https://arxiv.org/abs/1903.02962}
			{1903.02962} \BibitemShut {NoStop}%
			\bibitem [{\citenamefont {Thygesen}(2017)}]{Thygesen_2017}%
			\BibitemOpen
			\bibfield  {author} {\bibinfo {author} {\bibfnamefont {K.~S.}\ \bibnamefont
					{Thygesen}},\ }\bibfield  {title} {\bibinfo {title} {{Calculating excitons,
						plasmons, and quasiparticles in 2D materials and van der Waals
						heterostructures}},\ }\href {https://doi.org/10.1088/2053-1583/aa6432}
			{\bibfield  {journal} {\bibinfo  {journal} {2D Materials}\ }\textbf {\bibinfo
					{volume} {4}},\ \bibinfo {pages} {022004} (\bibinfo {year}
				{2017})}\BibitemShut {NoStop}%
			\bibitem [{\citenamefont {Mir}(2019)}]{Mir_2019}%
			\BibitemOpen
			\bibfield  {author} {\bibinfo {author} {\bibfnamefont {S.~H.}\ \bibnamefont
					{Mir}},\ }\bibfield  {title} {\bibinfo {title} {Exploring the electronic,
					charge transport and lattice dynamic properties of two-dimensional
					phosphorene},\ }\href@noop {} {\bibfield  {journal} {\bibinfo  {journal}
					{Physica B: Condensed Matter}\ }\textbf {\bibinfo {volume} {572}},\ \bibinfo
				{pages} {88} (\bibinfo {year} {2019})}\BibitemShut {NoStop}%
			\bibitem [{\citenamefont {Bolotin}\ \emph {et~al.}(2008)\citenamefont
				{Bolotin}, \citenamefont {Sikes}, \citenamefont {Jiang}, \citenamefont
				{Klima}, \citenamefont {Fudenberg}, \citenamefont {Hone}, \citenamefont
				{Kim},\ and\ \citenamefont {Stormer}}]{Bolotin_2008}%
			\BibitemOpen
			\bibfield  {author} {\bibinfo {author} {\bibfnamefont {K.~I.}\ \bibnamefont
					{Bolotin}}, \bibinfo {author} {\bibfnamefont {K.}~\bibnamefont {Sikes}},
				\bibinfo {author} {\bibfnamefont {Z.}~\bibnamefont {Jiang}}, \bibinfo
				{author} {\bibfnamefont {M.}~\bibnamefont {Klima}}, \bibinfo {author}
				{\bibfnamefont {G.}~\bibnamefont {Fudenberg}}, \bibinfo {author}
				{\bibfnamefont {J.}~\bibnamefont {Hone}}, \bibinfo {author} {\bibfnamefont
					{P.}~\bibnamefont {Kim}},\ and\ \bibinfo {author} {\bibfnamefont {H.~L.}\
					\bibnamefont {Stormer}},\ }\bibfield  {title} {\bibinfo {title} {Ultrahigh
					electron mobility in suspended graphene},\ }\href@noop {} {\bibfield
				{journal} {\bibinfo  {journal} {Solid state communications}\ }\textbf
				{\bibinfo {volume} {146}},\ \bibinfo {pages} {351} (\bibinfo {year}
				{2008})}\BibitemShut {NoStop}%
			\bibitem [{\citenamefont {Radisavljevic}\ \emph {et~al.}(2011)\citenamefont
				{Radisavljevic}, \citenamefont {Radenovic}, \citenamefont {Brivio},
				\citenamefont {Giacometti},\ and\ \citenamefont {Kis}}]{Radisavljevic_2011}%
			\BibitemOpen
			\bibfield  {author} {\bibinfo {author} {\bibfnamefont {B.}~\bibnamefont
					{Radisavljevic}}, \bibinfo {author} {\bibfnamefont {A.}~\bibnamefont
					{Radenovic}}, \bibinfo {author} {\bibfnamefont {J.}~\bibnamefont {Brivio}},
				\bibinfo {author} {\bibfnamefont {V.}~\bibnamefont {Giacometti}},\ and\
				\bibinfo {author} {\bibfnamefont {A.}~\bibnamefont {Kis}},\ }\bibfield
			{title} {\bibinfo {title} {{Single-layer MoS$_2$ transistors}},\ }\href
			{https://doi.org/10.1038/nnano.2010.279} {\bibfield  {journal} {\bibinfo
					{journal} {Nature~Nanotechnol.}\ }\textbf {\bibinfo {volume} {6}},\ \bibinfo
				{pages} {147} (\bibinfo {year} {2011})}\BibitemShut {NoStop}%
			\bibitem [{\citenamefont {Liu}\ \emph {et~al.}(2016)\citenamefont {Liu},
				\citenamefont {Weiss}, \citenamefont {Duan}, \citenamefont {Cheng},
				\citenamefont {Huang},\ and\ \citenamefont {Duan}}]{Liu_2016}%
			\BibitemOpen
			\bibfield  {author} {\bibinfo {author} {\bibfnamefont {Y.}~\bibnamefont
					{Liu}}, \bibinfo {author} {\bibfnamefont {N.~O.}\ \bibnamefont {Weiss}},
				\bibinfo {author} {\bibfnamefont {X.}~\bibnamefont {Duan}}, \bibinfo {author}
				{\bibfnamefont {H.-C.}\ \bibnamefont {Cheng}}, \bibinfo {author}
				{\bibfnamefont {Y.}~\bibnamefont {Huang}},\ and\ \bibinfo {author}
				{\bibfnamefont {X.}~\bibnamefont {Duan}},\ }\bibfield  {title} {\bibinfo
				{title} {Van der waals heterostructures and devices},\ }\href@noop {}
			{\bibfield  {journal} {\bibinfo  {journal} {Nature Reviews Materials}\
				}\textbf {\bibinfo {volume} {1}},\ \bibinfo {pages} {1} (\bibinfo {year}
				{2016})}\BibitemShut {NoStop}%
			\bibitem [{\citenamefont {Jariwala}\ \emph {et~al.}(2014)\citenamefont
				{Jariwala}, \citenamefont {Sangwan}, \citenamefont {Lauhon}, \citenamefont
				{Marks},\ and\ \citenamefont {Hersam}}]{Jariwala_2014}%
			\BibitemOpen
			\bibfield  {author} {\bibinfo {author} {\bibfnamefont {D.}~\bibnamefont
					{Jariwala}}, \bibinfo {author} {\bibfnamefont {V.~K.}\ \bibnamefont
					{Sangwan}}, \bibinfo {author} {\bibfnamefont {L.~J.}\ \bibnamefont {Lauhon}},
				\bibinfo {author} {\bibfnamefont {T.~J.}\ \bibnamefont {Marks}},\ and\
				\bibinfo {author} {\bibfnamefont {M.~C.}\ \bibnamefont {Hersam}},\ }\bibfield
			{title} {\bibinfo {title} {Emerging device applications for semiconducting
					two-dimensional transition metal dichalcogenides},\ }\href@noop {} {\bibfield
				{journal} {\bibinfo  {journal} {ACS nano}\ }\textbf {\bibinfo {volume}
					{8}},\ \bibinfo {pages} {1102} (\bibinfo {year} {2014})}\BibitemShut
			{NoStop}%
			\bibitem [{\citenamefont {Wang}\ \emph {et~al.}(2012)\citenamefont {Wang},
				\citenamefont {Kalantar-Zadeh}, \citenamefont {Kis}, \citenamefont
				{Coleman},\ and\ \citenamefont {Strano}}]{Wang_2012}%
			\BibitemOpen
			\bibfield  {author} {\bibinfo {author} {\bibfnamefont {Q.~H.}\ \bibnamefont
					{Wang}}, \bibinfo {author} {\bibfnamefont {K.}~\bibnamefont
					{Kalantar-Zadeh}}, \bibinfo {author} {\bibfnamefont {A.}~\bibnamefont {Kis}},
				\bibinfo {author} {\bibfnamefont {J.~N.}\ \bibnamefont {Coleman}},\ and\
				\bibinfo {author} {\bibfnamefont {M.~S.}\ \bibnamefont {Strano}},\ }\bibfield
			{title} {\bibinfo {title} {Electronics and optoelectronics of
					two-dimensional transition metal dichalcogenides},\ }\href@noop {} {\bibfield
				{journal} {\bibinfo  {journal} {Nature nanotechnology}\ }\textbf {\bibinfo
					{volume} {7}},\ \bibinfo {pages} {699} (\bibinfo {year} {2012})}\BibitemShut
			{NoStop}%
			\bibitem [{\citenamefont {Molina-Sánchez}\ \emph {et~al.}(2015)\citenamefont
				{Molina-Sánchez}, \citenamefont {Hummer},\ and\ \citenamefont
				{Wirtz}}]{MolinaSanchez2015}%
			\BibitemOpen
			\bibfield  {author} {\bibinfo {author} {\bibfnamefont {A.}~\bibnamefont
					{Molina-Sánchez}}, \bibinfo {author} {\bibfnamefont {K.}~\bibnamefont
					{Hummer}},\ and\ \bibinfo {author} {\bibfnamefont {L.}~\bibnamefont
					{Wirtz}},\ }\bibfield  {title} {\bibinfo {title} {Vibrational and optical
					properties of {MoS}$_2$: From monolayer to bulk},\ }\href
			{https://doi.org/10.1016/j.surfrep.2015.10.001} {\bibfield  {journal}
				{\bibinfo  {journal} {Surf.~Sci.~Rep.~}\ }\textbf {\bibinfo {volume} {70}},\
				\bibinfo {pages} {554} (\bibinfo {year} {2015})}\BibitemShut {NoStop}%
			\bibitem [{\citenamefont {Conley}\ \emph {et~al.}(2013)\citenamefont {Conley},
				\citenamefont {Wang}, \citenamefont {Ziegler}, \citenamefont {Haglund},
				\citenamefont {Pantelides},\ and\ \citenamefont {Bolotin}}]{Conley_2013}%
			\BibitemOpen
			\bibfield  {author} {\bibinfo {author} {\bibfnamefont {H.~J.}\ \bibnamefont
					{Conley}}, \bibinfo {author} {\bibfnamefont {B.}~\bibnamefont {Wang}},
				\bibinfo {author} {\bibfnamefont {J.~I.}\ \bibnamefont {Ziegler}}, \bibinfo
				{author} {\bibfnamefont {R.~F.}\ \bibnamefont {Haglund}}, \bibinfo {author}
				{\bibfnamefont {S.~T.}\ \bibnamefont {Pantelides}},\ and\ \bibinfo {author}
				{\bibfnamefont {K.~I.}\ \bibnamefont {Bolotin}},\ }\bibfield  {title}
			{\bibinfo {title} {{Bandgap Engineering of Strained Monolayer and Bilayer
						MoS$_2$}},\ }\href {https://doi.org/10.1021/nl4014748} {\bibfield  {journal}
				{\bibinfo  {journal} {Nano Letters}\ }\textbf {\bibinfo {volume} {13}},\
				\bibinfo {pages} {3626} (\bibinfo {year} {2013})},\ \Eprint
			{https://arxiv.org/abs/1305.3880} {1305.3880} \BibitemShut {NoStop}%
			\bibitem [{\citenamefont {Ataca}\ and\ \citenamefont
				{Ciraci}(2011)}]{Ataca_2011}%
			\BibitemOpen
			\bibfield  {author} {\bibinfo {author} {\bibfnamefont {C.}~\bibnamefont
					{Ataca}}\ and\ \bibinfo {author} {\bibfnamefont {S.}~\bibnamefont {Ciraci}},\
			}\bibfield  {title} {\bibinfo {title} {{Functionalization of Single-Layer
						MoS$_2$ Honeycomb Structures}},\ }\href {https://doi.org/10.1021/jp2000442}
			{\bibfield  {journal} {\bibinfo  {journal} {J.~Phys.~Chem.~C}\ }\textbf
				{\bibinfo {volume} {115}},\ \bibinfo {pages} {13303} (\bibinfo {year}
				{2011})}\BibitemShut {NoStop}%
			\bibitem [{\citenamefont {Shi}\ \emph {et~al.}(2013)\citenamefont {Shi},
				\citenamefont {Pan}, \citenamefont {Zhang},\ and\ \citenamefont
				{Yakobson}}]{Shi_2013}%
			\BibitemOpen
			\bibfield  {author} {\bibinfo {author} {\bibfnamefont {H.}~\bibnamefont
					{Shi}}, \bibinfo {author} {\bibfnamefont {H.}~\bibnamefont {Pan}}, \bibinfo
				{author} {\bibfnamefont {Y.-W.}\ \bibnamefont {Zhang}},\ and\ \bibinfo
				{author} {\bibfnamefont {B.~I.}\ \bibnamefont {Yakobson}},\ }\bibfield
			{title} {\bibinfo {title} {{Quasiparticle band structures and optical
						properties of strained monolayer MoS$_2$ and WS$_2$}},\ }\bibfield  {journal}
			{\bibinfo  {journal} {Phys.~Rev.~B}\ }\textbf {\bibinfo {volume} {87}},\
			\href {https://doi.org/10.1103/physrevb.87.155304}
			{10.1103/physrevb.87.155304} (\bibinfo {year} {2013})\BibitemShut {NoStop}%
			\bibitem [{\citenamefont {Ramasubramaniam}(2012)}]{Ramasubramaniam_2012}%
			\BibitemOpen
			\bibfield  {author} {\bibinfo {author} {\bibfnamefont {A.}~\bibnamefont
					{Ramasubramaniam}},\ }\bibfield  {title} {\bibinfo {title} {{Large excitonic
						effects in monolayers of molybdenum and tungsten dichalcogenides}},\
			}\bibfield  {journal} {\bibinfo  {journal} {Phys.~Rev.~B}\ }\textbf {\bibinfo
				{volume} {86}},\ \href {https://doi.org/10.1103/physrevb.86.115409}
			{10.1103/physrevb.86.115409} (\bibinfo {year} {2012})\BibitemShut {NoStop}%
			\bibitem [{\citenamefont {Rasmussen}\ and\ \citenamefont
				{Thygesen}(2015)}]{Rasmussen_2015}%
			\BibitemOpen
			\bibfield  {author} {\bibinfo {author} {\bibfnamefont {F.~A.}\ \bibnamefont
					{Rasmussen}}\ and\ \bibinfo {author} {\bibfnamefont {K.~S.}\ \bibnamefont
					{Thygesen}},\ }\bibfield  {title} {\bibinfo {title} {{Computational 2D
						Materials Database: Electronic Structure of Transition-Metal Dichalcogenides
						and Oxides}},\ }\href {https://doi.org/10.1021/acs.jpcc.5b02950} {\bibfield
				{journal} {\bibinfo  {journal} {The Journal of Physical Chemistry C}\
				}\textbf {\bibinfo {volume} {119}},\ \bibinfo {pages} {13169} (\bibinfo
				{year} {2015})}\BibitemShut {NoStop}%
			\bibitem [{\citenamefont {Rasmussen}\ \emph {et~al.}(2016)\citenamefont
				{Rasmussen}, \citenamefont {Schmidt}, \citenamefont {Winther},\ and\
				\citenamefont {Thygesen}}]{Rasmussen_2016}%
			\BibitemOpen
			\bibfield  {author} {\bibinfo {author} {\bibfnamefont {F.~A.}\ \bibnamefont
					{Rasmussen}}, \bibinfo {author} {\bibfnamefont {P.~S.}\ \bibnamefont
					{Schmidt}}, \bibinfo {author} {\bibfnamefont {K.~T.}\ \bibnamefont
					{Winther}},\ and\ \bibinfo {author} {\bibfnamefont {K.~S.}\ \bibnamefont
					{Thygesen}},\ }\bibfield  {title} {\bibinfo {title} {{Efficient many-body
						calculations for two-dimensional materials using exact limits for the
						screened potential: Band gaps of MoS$_2$, h-BN, and phosphorene}},\
			}\bibfield  {journal} {\bibinfo  {journal} {Phys.~Rev.~B}\ }\textbf {\bibinfo
				{volume} {94}},\ \href {https://doi.org/10.1103/physrevb.94.155406}
			{10.1103/physrevb.94.155406} (\bibinfo {year} {2016})\BibitemShut {NoStop}%
			\bibitem [{\citenamefont {Molina-Sánchez}\ \emph {et~al.}(2013)\citenamefont
				{Molina-Sánchez}, \citenamefont {Sangalli}, \citenamefont {Hummer},
				\citenamefont {Marini},\ and\ \citenamefont {Wirtz}}]{MolinaSanchez_2013}%
			\BibitemOpen
			\bibfield  {author} {\bibinfo {author} {\bibfnamefont {A.}~\bibnamefont
					{Molina-Sánchez}}, \bibinfo {author} {\bibfnamefont {D.}~\bibnamefont
					{Sangalli}}, \bibinfo {author} {\bibfnamefont {K.}~\bibnamefont {Hummer}},
				\bibinfo {author} {\bibfnamefont {A.}~\bibnamefont {Marini}},\ and\ \bibinfo
				{author} {\bibfnamefont {L.}~\bibnamefont {Wirtz}},\ }\bibfield  {title}
			{\bibinfo {title} {{Effect of spin-orbit interaction on the optical spectra
						of single-layer, double-layer, and bulk MoS$_2$}},\ }\href
			{https://doi.org/10.1103/physrevb.88.045412} {\bibfield  {journal} {\bibinfo
					{journal} {Phys.~Rev.~B}\ }\textbf {\bibinfo {volume} {88}},\ \bibinfo
				{pages} {045412} (\bibinfo {year} {2013})},\ \Eprint
			{https://arxiv.org/abs/1306.4257} {1306.4257} \BibitemShut {NoStop}%
			\bibitem [{\citenamefont {Qiu}\ \emph {et~al.}(2013)\citenamefont {Qiu},
				\citenamefont {Jornada},\ and\ \citenamefont {Louie}}]{Qiu_2013}%
			\BibitemOpen
			\bibfield  {author} {\bibinfo {author} {\bibfnamefont {D.~Y.}\ \bibnamefont
					{Qiu}}, \bibinfo {author} {\bibfnamefont {F.~H.~d.}\ \bibnamefont
					{Jornada}},\ and\ \bibinfo {author} {\bibfnamefont {S.~G.}\ \bibnamefont
					{Louie}},\ }\bibfield  {title} {\bibinfo {title} {{Optical Spectrum of
						MoS$_2$: Many-Body Effects and Diversity of Exciton States}},\ }\href
			{https://doi.org/10.1103/physrevlett.111.216805} {\bibfield  {journal}
				{\bibinfo  {journal} {Phys.~Rev.~Lett.~}\ }\textbf {\bibinfo {volume}
					{111}},\ \bibinfo {pages} {216805} (\bibinfo {year} {2013})}\BibitemShut
			{NoStop}%
			\bibitem [{\citenamefont {Cheiwchanchamnangij}\ and\ \citenamefont
				{Lambrecht}(2012)}]{Cheiwchanchamnangij_2012}%
			\BibitemOpen
			\bibfield  {author} {\bibinfo {author} {\bibfnamefont {T.}~\bibnamefont
					{Cheiwchanchamnangij}}\ and\ \bibinfo {author} {\bibfnamefont {W.~R.~L.}\
					\bibnamefont {Lambrecht}},\ }\bibfield  {title} {\bibinfo {title}
				{{Quasiparticle band structure calculation of monolayer, bilayer, and bulk
						MoS$_2$}},\ }\href {https://doi.org/10.1103/physrevb.85.205302} {\bibfield
				{journal} {\bibinfo  {journal} {Phys.~Rev.~B}\ }\textbf {\bibinfo {volume}
					{85}},\ \bibinfo {pages} {205302} (\bibinfo {year} {2012})}\BibitemShut
			{NoStop}%
			\bibitem [{\citenamefont {Hüser}\ \emph
				{et~al.}(2013{\natexlab{a}})\citenamefont {Hüser}, \citenamefont {Olsen},\
				and\ \citenamefont {Thygesen}}]{Hueser_2013}%
			\BibitemOpen
			\bibfield  {author} {\bibinfo {author} {\bibfnamefont {F.}~\bibnamefont
					{Hüser}}, \bibinfo {author} {\bibfnamefont {T.}~\bibnamefont {Olsen}},\ and\
				\bibinfo {author} {\bibfnamefont {K.~S.}\ \bibnamefont {Thygesen}},\
			}\bibfield  {title} {\bibinfo {title} {{How dielectric screening in
						two-dimensional crystals affects the convergence of excited-state
						calculations: Monolayer MoS$_2$}},\ }\href
			{https://doi.org/10.1103/physrevb.88.245309} {\bibfield  {journal} {\bibinfo
					{journal} {Phys.~Rev.~B}\ }\textbf {\bibinfo {volume} {88}},\ \bibinfo
				{pages} {245309} (\bibinfo {year} {2013}{\natexlab{a}})}\BibitemShut
			{NoStop}%
			\bibitem [{\citenamefont {Echeverry}\ \emph {et~al.}(2016)\citenamefont
				{Echeverry}, \citenamefont {Urbaszek}, \citenamefont {Amand}, \citenamefont
				{Marie},\ and\ \citenamefont {Gerber}}]{Echeverry_2016}%
			\BibitemOpen
			\bibfield  {author} {\bibinfo {author} {\bibfnamefont {J.~P.}\ \bibnamefont
					{Echeverry}}, \bibinfo {author} {\bibfnamefont {B.}~\bibnamefont {Urbaszek}},
				\bibinfo {author} {\bibfnamefont {T.}~\bibnamefont {Amand}}, \bibinfo
				{author} {\bibfnamefont {X.}~\bibnamefont {Marie}},\ and\ \bibinfo {author}
				{\bibfnamefont {I.~C.}\ \bibnamefont {Gerber}},\ }\bibfield  {title}
			{\bibinfo {title} {{Splitting between bright and dark excitons in transition
						metal dichalcogenide monolayers}},\ }\href
			{https://doi.org/10.1103/physrevb.93.121107} {\bibfield  {journal} {\bibinfo
					{journal} {Phys.~Rev.~B}\ }\textbf {\bibinfo {volume} {93}},\ \bibinfo
				{pages} {121107} (\bibinfo {year} {2016})},\ \Eprint
			{https://arxiv.org/abs/1601.07351} {1601.07351} \BibitemShut {NoStop}%
			\bibitem [{\citenamefont {Schmidt}\ \emph {et~al.}(2017)\citenamefont
				{Schmidt}, \citenamefont {Patrick},\ and\ \citenamefont
				{Thygesen}}]{Schmidt_2017}%
			\BibitemOpen
			\bibfield  {author} {\bibinfo {author} {\bibfnamefont {P.~S.}\ \bibnamefont
					{Schmidt}}, \bibinfo {author} {\bibfnamefont {C.~E.}\ \bibnamefont
					{Patrick}},\ and\ \bibinfo {author} {\bibfnamefont {K.~S.}\ \bibnamefont
					{Thygesen}},\ }\bibfield  {title} {\bibinfo {title} {{Simple vertex
						correction improves GW band energies of bulk and two-dimensional crystals}},\
			}\href {https://doi.org/10.1103/physrevb.96.205206} {\bibfield  {journal}
				{\bibinfo  {journal} {Physical Review B}\ }\textbf {\bibinfo {volume} {96}},\
				\bibinfo {pages} {205206} (\bibinfo {year} {2017})}\BibitemShut {NoStop}%
			\bibitem [{\citenamefont {Komsa}\ and\ \citenamefont
				{Krasheninnikov}(2012)}]{Komsa_2012}%
			\BibitemOpen
			\bibfield  {author} {\bibinfo {author} {\bibfnamefont {H.-P.}\ \bibnamefont
					{Komsa}}\ and\ \bibinfo {author} {\bibfnamefont {A.~V.}\ \bibnamefont
					{Krasheninnikov}},\ }\bibfield  {title} {\bibinfo {title} {{Effects of
						confinement and environment on the electronic structure and exciton binding
						energy of MoS$_2$ from first principles}},\ }\href
			{https://doi.org/10.1103/physrevb.86.241201} {\bibfield  {journal} {\bibinfo
					{journal} {Physical Review B}\ }\textbf {\bibinfo {volume} {86}},\ \bibinfo
				{pages} {241201} (\bibinfo {year} {2012})}\BibitemShut {NoStop}%
			\bibitem [{\citenamefont {Liang}\ \emph {et~al.}(2013)\citenamefont {Liang},
				\citenamefont {Huang}, \citenamefont {Soklaski},\ and\ \citenamefont
				{Yang}}]{Liang_2013}%
			\BibitemOpen
			\bibfield  {author} {\bibinfo {author} {\bibfnamefont {Y.}~\bibnamefont
					{Liang}}, \bibinfo {author} {\bibfnamefont {S.}~\bibnamefont {Huang}},
				\bibinfo {author} {\bibfnamefont {R.}~\bibnamefont {Soklaski}},\ and\
				\bibinfo {author} {\bibfnamefont {L.}~\bibnamefont {Yang}},\ }\bibfield
			{title} {\bibinfo {title} {{Quasiparticle band-edge energy and band offsets
						of monolayer of molybdenum and tungsten chalcogenides}},\ }\href
			{https://doi.org/10.1063/1.4816517} {\bibfield  {journal} {\bibinfo
					{journal} {Appl.~Phys.~Lett.~}\ }\textbf {\bibinfo {volume} {103}},\ \bibinfo
				{pages} {042106} (\bibinfo {year} {2013})}\BibitemShut {NoStop}%
			\bibitem [{\citenamefont {Jiang}\ \emph {et~al.}(2021)\citenamefont {Jiang},
				\citenamefont {Zheng}, \citenamefont {Lan}, \citenamefont {Saidi},
				\citenamefont {Ren},\ and\ \citenamefont {Zhao}}]{Jiang_2021}%
			\BibitemOpen
			\bibfield  {author} {\bibinfo {author} {\bibfnamefont {X.}~\bibnamefont
					{Jiang}}, \bibinfo {author} {\bibfnamefont {Q.}~\bibnamefont {Zheng}},
				\bibinfo {author} {\bibfnamefont {Z.}~\bibnamefont {Lan}}, \bibinfo {author}
				{\bibfnamefont {W.~A.}\ \bibnamefont {Saidi}}, \bibinfo {author}
				{\bibfnamefont {X.}~\bibnamefont {Ren}},\ and\ \bibinfo {author}
				{\bibfnamefont {J.}~\bibnamefont {Zhao}},\ }\bibfield  {title} {\bibinfo
				{title} {Real-time \textit{GW}-{BSE} investigations on spin-valley exciton
					dynamics in monolayer transition metal dichalcogenide},\ }\href
			{https://doi.org/10.1126/sciadv.abf3759} {\bibfield  {journal} {\bibinfo
					{journal} {Science Advances}\ }\textbf {\bibinfo {volume} {7}},\ \bibinfo
				{pages} {eabf3759} (\bibinfo {year} {2021})}\BibitemShut {NoStop}%
			\bibitem [{\citenamefont {Zibouche}\ \emph {et~al.}(2021)\citenamefont
				{Zibouche}, \citenamefont {Schlipf},\ and\ \citenamefont
				{Giustino}}]{Zibouche_2021}%
			\BibitemOpen
			\bibfield  {author} {\bibinfo {author} {\bibfnamefont {N.}~\bibnamefont
					{Zibouche}}, \bibinfo {author} {\bibfnamefont {M.}~\bibnamefont {Schlipf}},\
				and\ \bibinfo {author} {\bibfnamefont {F.}~\bibnamefont {Giustino}},\
			}\bibfield  {title} {\bibinfo {title} {{GW band structure of monolayer
						MoS$_2$ using the SternheimerGW method and effect of dielectric
						environment}},\ }\href {https://doi.org/10.1103/physrevb.103.125401}
			{\bibfield  {journal} {\bibinfo  {journal} {Phys.~Rev.~B}\ }\textbf {\bibinfo
					{volume} {103}},\ \bibinfo {pages} {125401} (\bibinfo {year} {2021})},\
			\Eprint {https://arxiv.org/abs/2008.03980} {2008.03980} \BibitemShut
			{NoStop}%
			\bibitem [{\citenamefont {Soklaski}\ \emph {et~al.}(2014)\citenamefont
				{Soklaski}, \citenamefont {Liang}, \citenamefont {Zhang}, \citenamefont
				{Wang}, \citenamefont {Rana},\ and\ \citenamefont {Yang}}]{Soklaski_2014}%
			\BibitemOpen
			\bibfield  {author} {\bibinfo {author} {\bibfnamefont {R.}~\bibnamefont
					{Soklaski}}, \bibinfo {author} {\bibfnamefont {Y.}~\bibnamefont {Liang}},
				\bibinfo {author} {\bibfnamefont {C.}~\bibnamefont {Zhang}}, \bibinfo
				{author} {\bibfnamefont {H.}~\bibnamefont {Wang}}, \bibinfo {author}
				{\bibfnamefont {F.}~\bibnamefont {Rana}},\ and\ \bibinfo {author}
				{\bibfnamefont {L.}~\bibnamefont {Yang}},\ }\bibfield  {title} {\bibinfo
				{title} {{Temperature effect on optical spectra of monolayer molybdenum
						disulfide}},\ }\href {https://doi.org/10.1063/1.4878098} {\bibfield
				{journal} {\bibinfo  {journal} {Appl.~Phys.~Lett.~}\ }\textbf {\bibinfo
					{volume} {104}},\ \bibinfo {pages} {193110} (\bibinfo {year} {2014})},\
			\Eprint {https://arxiv.org/abs/1401.5732} {1401.5732} \BibitemShut {NoStop}%
			\bibitem [{\citenamefont {Xia}\ \emph {et~al.}(2020)\citenamefont {Xia},
				\citenamefont {Gao}, \citenamefont {Lopez-Candales}, \citenamefont {Wu},
				\citenamefont {Ren}, \citenamefont {Zhang},\ and\ \citenamefont
				{Zhang}}]{Xia_2020}%
			\BibitemOpen
			\bibfield  {author} {\bibinfo {author} {\bibfnamefont {W.}~\bibnamefont
					{Xia}}, \bibinfo {author} {\bibfnamefont {W.}~\bibnamefont {Gao}}, \bibinfo
				{author} {\bibfnamefont {G.}~\bibnamefont {Lopez-Candales}}, \bibinfo
				{author} {\bibfnamefont {Y.}~\bibnamefont {Wu}}, \bibinfo {author}
				{\bibfnamefont {W.}~\bibnamefont {Ren}}, \bibinfo {author} {\bibfnamefont
					{W.}~\bibnamefont {Zhang}},\ and\ \bibinfo {author} {\bibfnamefont
					{P.}~\bibnamefont {Zhang}},\ }\bibfield  {title} {\bibinfo {title} {{Combined
						subsampling and analytical integration for efficient large-scale GW
						calculations for 2D systems}},\ }\href
			{https://doi.org/10.1038/s41524-020-00385-5} {\bibfield  {journal} {\bibinfo
					{journal} {npj Computational Materials}\ }\textbf {\bibinfo {volume} {6}},\
				\bibinfo {pages} {118} (\bibinfo {year} {2020})}\BibitemShut {NoStop}%
			\bibitem [{\citenamefont {Gao}\ \emph {et~al.}(2016)\citenamefont {Gao},
				\citenamefont {Xia}, \citenamefont {Gao},\ and\ \citenamefont
				{Zhang}}]{Gao_2016}%
			\BibitemOpen
			\bibfield  {author} {\bibinfo {author} {\bibfnamefont {W.}~\bibnamefont
					{Gao}}, \bibinfo {author} {\bibfnamefont {W.}~\bibnamefont {Xia}}, \bibinfo
				{author} {\bibfnamefont {X.}~\bibnamefont {Gao}},\ and\ \bibinfo {author}
				{\bibfnamefont {P.}~\bibnamefont {Zhang}},\ }\bibfield  {title} {\bibinfo
				{title} {{Speeding up GW Calculations to Meet the Challenge of Large Scale
						Quasiparticle Predictions}},\ }\href {https://doi.org/10.1038/srep36849}
			{\bibfield  {journal} {\bibinfo  {journal} {Scientific Reports}\ }\textbf
				{\bibinfo {volume} {6}},\ \bibinfo {pages} {36849} (\bibinfo {year}
				{2016})}\BibitemShut {NoStop}%
			\bibitem [{\citenamefont {Qiu}\ \emph {et~al.}(2016)\citenamefont {Qiu},
				\citenamefont {Jornada},\ and\ \citenamefont {Louie}}]{Qiu_2016}%
			\BibitemOpen
			\bibfield  {author} {\bibinfo {author} {\bibfnamefont {D.~Y.}\ \bibnamefont
					{Qiu}}, \bibinfo {author} {\bibfnamefont {F.~H.~d.}\ \bibnamefont
					{Jornada}},\ and\ \bibinfo {author} {\bibfnamefont {S.~G.}\ \bibnamefont
					{Louie}},\ }\bibfield  {title} {\bibinfo {title} {{Screening and many-body
						effects in two-dimensional crystals: Monolayer MoS$_2$}},\ }\href
			{https://doi.org/10.1103/physrevb.93.235435} {\bibfield  {journal} {\bibinfo
					{journal} {Phys.~Rev.~B}\ }\textbf {\bibinfo {volume} {93}},\ \bibinfo
				{pages} {235435} (\bibinfo {year} {2016})},\ \Eprint
			{https://arxiv.org/abs/1605.08733} {1605.08733} \BibitemShut {NoStop}%
			\bibitem [{\citenamefont {Gillen}\ and\ \citenamefont
				{Maultzsch}(2016)}]{Gillen_2016}%
			\BibitemOpen
			\bibfield  {author} {\bibinfo {author} {\bibfnamefont {R.}~\bibnamefont
					{Gillen}}\ and\ \bibinfo {author} {\bibfnamefont {J.}~\bibnamefont
					{Maultzsch}},\ }\bibfield  {title} {\bibinfo {title} {{Light-Matter
						Interactions in Two-Dimensional Transition Metal Dichalcogenides: Dominant
						Excitonic Transitions in Mono-and Few-Layer MoX$_2 $ and Band Nesting}},\
			}\href@noop {} {\bibfield  {journal} {\bibinfo  {journal} {IEEE Journal of
						Selected Topics in Quantum Electronics}\ }\textbf {\bibinfo {volume} {23}},\
				\bibinfo {pages} {219} (\bibinfo {year} {2016})}\BibitemShut {NoStop}%
			\bibitem [{\citenamefont {Zhuang}\ and\ \citenamefont
				{Hennig}(2013)}]{Zhuang_2013}%
			\BibitemOpen
			\bibfield  {author} {\bibinfo {author} {\bibfnamefont {H.~L.}\ \bibnamefont
					{Zhuang}}\ and\ \bibinfo {author} {\bibfnamefont {R.~G.}\ \bibnamefont
					{Hennig}},\ }\bibfield  {title} {\bibinfo {title} {Computational search for
					single-layer transition-metal dichalcogenide photocatalysts},\ }\href@noop {}
			{\bibfield  {journal} {\bibinfo  {journal} {The Journal of Physical Chemistry
						C}\ }\textbf {\bibinfo {volume} {117}},\ \bibinfo {pages} {20440} (\bibinfo
				{year} {2013})}\BibitemShut {NoStop}%
			\bibitem [{\citenamefont {Haastrup}\ \emph {et~al.}(2018)\citenamefont
				{Haastrup}, \citenamefont {Strange}, \citenamefont {Pandey}, \citenamefont
				{Deilmann}, \citenamefont {Schmidt}, \citenamefont {Hinsche}, \citenamefont
				{Gjerding}, \citenamefont {Torelli}, \citenamefont {Larsen}, \citenamefont
				{Riis-Jensen}, \citenamefont {Gath}, \citenamefont {Jacobsen}, \citenamefont
				{Mortensen}, \citenamefont {Olsen},\ and\ \citenamefont
				{Thygesen}}]{Haastrup_2018}%
			\BibitemOpen
			\bibfield  {author} {\bibinfo {author} {\bibfnamefont {S.}~\bibnamefont
					{Haastrup}}, \bibinfo {author} {\bibfnamefont {M.}~\bibnamefont {Strange}},
				\bibinfo {author} {\bibfnamefont {M.}~\bibnamefont {Pandey}}, \bibinfo
				{author} {\bibfnamefont {T.}~\bibnamefont {Deilmann}}, \bibinfo {author}
				{\bibfnamefont {P.~S.}\ \bibnamefont {Schmidt}}, \bibinfo {author}
				{\bibfnamefont {N.~F.}\ \bibnamefont {Hinsche}}, \bibinfo {author}
				{\bibfnamefont {M.~N.}\ \bibnamefont {Gjerding}}, \bibinfo {author}
				{\bibfnamefont {D.}~\bibnamefont {Torelli}}, \bibinfo {author} {\bibfnamefont
					{P.~M.}\ \bibnamefont {Larsen}}, \bibinfo {author} {\bibfnamefont {A.~C.}\
					\bibnamefont {Riis-Jensen}}, \bibinfo {author} {\bibfnamefont
					{J.}~\bibnamefont {Gath}}, \bibinfo {author} {\bibfnamefont {K.~W.}\
					\bibnamefont {Jacobsen}}, \bibinfo {author} {\bibfnamefont {J.~J.}\
					\bibnamefont {Mortensen}}, \bibinfo {author} {\bibfnamefont {T.}~\bibnamefont
					{Olsen}},\ and\ \bibinfo {author} {\bibfnamefont {K.~S.}\ \bibnamefont
					{Thygesen}},\ }\bibfield  {title} {\bibinfo {title} {{The Computational 2D
						Materials Database: high-throughput modeling and discovery of atomically thin
						crystals}},\ }\href {https://doi.org/10.1088/2053-1583/aacfc1} {\bibfield
				{journal} {\bibinfo  {journal} {2D Materials}\ }\textbf {\bibinfo {volume}
					{5}},\ \bibinfo {pages} {042002} (\bibinfo {year} {2018})}\BibitemShut
			{NoStop}%
			\bibitem [{\citenamefont {Gjerding}\ \emph {et~al.}(2021)\citenamefont
				{Gjerding}, \citenamefont {Taghizadeh}, \citenamefont {Rasmussen},
				\citenamefont {Ali}, \citenamefont {Bertoldo}, \citenamefont {Deilmann},
				\citenamefont {Kn{\o}sgaard}, \citenamefont {Kruse}, \citenamefont {Larsen},
				\citenamefont {Manti} \emph {et~al.}}]{Gjerding_2021}%
			\BibitemOpen
			\bibfield  {author} {\bibinfo {author} {\bibfnamefont {M.~N.}\ \bibnamefont
					{Gjerding}}, \bibinfo {author} {\bibfnamefont {A.}~\bibnamefont
					{Taghizadeh}}, \bibinfo {author} {\bibfnamefont {A.}~\bibnamefont
					{Rasmussen}}, \bibinfo {author} {\bibfnamefont {S.}~\bibnamefont {Ali}},
				\bibinfo {author} {\bibfnamefont {F.}~\bibnamefont {Bertoldo}}, \bibinfo
				{author} {\bibfnamefont {T.}~\bibnamefont {Deilmann}}, \bibinfo {author}
				{\bibfnamefont {N.~R.}\ \bibnamefont {Kn{\o}sgaard}}, \bibinfo {author}
				{\bibfnamefont {M.}~\bibnamefont {Kruse}}, \bibinfo {author} {\bibfnamefont
					{A.~H.}\ \bibnamefont {Larsen}}, \bibinfo {author} {\bibfnamefont
					{S.}~\bibnamefont {Manti}}, \emph {et~al.},\ }\bibfield  {title} {\bibinfo
				{title} {Recent progress of the computational 2d materials database (c2db)},\
			}\href@noop {} {\bibfield  {journal} {\bibinfo  {journal} {2D Materials}\
				}\textbf {\bibinfo {volume} {8}},\ \bibinfo {pages} {044002} (\bibinfo {year}
				{2021})}\BibitemShut {NoStop}%
			\bibitem [{\citenamefont {Kim}\ and\ \citenamefont {Choi}(2021)}]{Kim_2021}%
			\BibitemOpen
			\bibfield  {author} {\bibinfo {author} {\bibfnamefont {H.-g.}\ \bibnamefont
					{Kim}}\ and\ \bibinfo {author} {\bibfnamefont {H.~J.}\ \bibnamefont {Choi}},\
			}\bibfield  {title} {\bibinfo {title} {{Thickness dependence of work
						function, ionization energy, and electron affinity of Mo and W
						dichalcogenides from DFT and GW calculations}},\ }\href
			{https://doi.org/10.1103/PhysRevB.103.085404} {\bibfield  {journal} {\bibinfo
					{journal} {Phys. Rev. B}\ }\textbf {\bibinfo {volume} {103}},\ \bibinfo
				{pages} {085404} (\bibinfo {year} {2021})}\BibitemShut {NoStop}%
			\bibitem [{\citenamefont {Smart}\ \emph {et~al.}(2018)\citenamefont {Smart},
				\citenamefont {Wu}, \citenamefont {Govoni},\ and\ \citenamefont
				{Ping}}]{Smart_2018}%
			\BibitemOpen
			\bibfield  {author} {\bibinfo {author} {\bibfnamefont {T.~J.}\ \bibnamefont
					{Smart}}, \bibinfo {author} {\bibfnamefont {F.}~\bibnamefont {Wu}}, \bibinfo
				{author} {\bibfnamefont {M.}~\bibnamefont {Govoni}},\ and\ \bibinfo {author}
				{\bibfnamefont {Y.}~\bibnamefont {Ping}},\ }\bibfield  {title} {\bibinfo
				{title} {{Fundamental principles for calculating charged defect ionization
						energies in ultrathin two-dimensional materials}},\ }\href
			{https://doi.org/10.1103/physrevmaterials.2.124002} {\bibfield  {journal}
				{\bibinfo  {journal} {Physical Review Materials}\ }\textbf {\bibinfo {volume}
					{2}},\ \bibinfo {pages} {124002} (\bibinfo {year} {2018})},\ \Eprint
			{https://arxiv.org/abs/1808.03221} {1808.03221} \BibitemShut {NoStop}%
			\bibitem [{\citenamefont {Horzum}\ \emph {et~al.}(2013)\citenamefont {Horzum},
				\citenamefont {Sahin}, \citenamefont {Cahangirov}, \citenamefont {Cudazzo},
				\citenamefont {Rubio}, \citenamefont {Serin},\ and\ \citenamefont
				{Peeters}}]{Horzum_2013}%
			\BibitemOpen
			\bibfield  {author} {\bibinfo {author} {\bibfnamefont {S.}~\bibnamefont
					{Horzum}}, \bibinfo {author} {\bibfnamefont {H.}~\bibnamefont {Sahin}},
				\bibinfo {author} {\bibfnamefont {S.}~\bibnamefont {Cahangirov}}, \bibinfo
				{author} {\bibfnamefont {P.}~\bibnamefont {Cudazzo}}, \bibinfo {author}
				{\bibfnamefont {A.}~\bibnamefont {Rubio}}, \bibinfo {author} {\bibfnamefont
					{T.}~\bibnamefont {Serin}},\ and\ \bibinfo {author} {\bibfnamefont {F.~M.}\
					\bibnamefont {Peeters}},\ }\bibfield  {title} {\bibinfo {title} {{Phonon
						softening and direct to indirect band gap crossover in strained single-layer
						MoSe${}_{2}$}},\ }\href {https://doi.org/10.1103/PhysRevB.87.125415}
			{\bibfield  {journal} {\bibinfo  {journal} {Phys. Rev. B}\ }\textbf {\bibinfo
					{volume} {87}},\ \bibinfo {pages} {125415} (\bibinfo {year}
				{2013})}\BibitemShut {NoStop}%
			\bibitem [{\citenamefont {Lee}\ \emph {et~al.}(2017)\citenamefont {Lee},
				\citenamefont {Huang}, \citenamefont {Sumpter},\ and\ \citenamefont
				{Yoon}}]{Lee_2017}%
			\BibitemOpen
			\bibfield  {author} {\bibinfo {author} {\bibfnamefont {J.}~\bibnamefont
					{Lee}}, \bibinfo {author} {\bibfnamefont {J.}~\bibnamefont {Huang}}, \bibinfo
				{author} {\bibfnamefont {B.~G.}\ \bibnamefont {Sumpter}},\ and\ \bibinfo
				{author} {\bibfnamefont {M.}~\bibnamefont {Yoon}},\ }\bibfield  {title}
			{\bibinfo {title} {Strain-engineered optoelectronic properties of 2d
					transition metal dichalcogenide lateral heterostructures},\ }\href@noop {}
			{\bibfield  {journal} {\bibinfo  {journal} {2D Materials}\ }\textbf {\bibinfo
					{volume} {4}},\ \bibinfo {pages} {021016} (\bibinfo {year}
				{2017})}\BibitemShut {NoStop}%
			\bibitem [{\citenamefont {Elliott}\ \emph {et~al.}(2020)\citenamefont
				{Elliott}, \citenamefont {Xu}, \citenamefont {Umari}, \citenamefont
				{Jayaswal}, \citenamefont {Chen}, \citenamefont {Zhang}, \citenamefont
				{Martucci}, \citenamefont {Marsili},\ and\ \citenamefont
				{Merano}}]{Elliott_2020}%
			\BibitemOpen
			\bibfield  {author} {\bibinfo {author} {\bibfnamefont {J.~D.}\ \bibnamefont
					{Elliott}}, \bibinfo {author} {\bibfnamefont {Z.}~\bibnamefont {Xu}},
				\bibinfo {author} {\bibfnamefont {P.}~\bibnamefont {Umari}}, \bibinfo
				{author} {\bibfnamefont {G.}~\bibnamefont {Jayaswal}}, \bibinfo {author}
				{\bibfnamefont {M.}~\bibnamefont {Chen}}, \bibinfo {author} {\bibfnamefont
					{X.}~\bibnamefont {Zhang}}, \bibinfo {author} {\bibfnamefont
					{A.}~\bibnamefont {Martucci}}, \bibinfo {author} {\bibfnamefont
					{M.}~\bibnamefont {Marsili}},\ and\ \bibinfo {author} {\bibfnamefont
					{M.}~\bibnamefont {Merano}},\ }\bibfield  {title} {\bibinfo {title} {{Surface
						susceptibility and conductivity of ${\mathrm{MoS}}_{2}$ and
						${\mathrm{WSe}}_{2}$ monolayers: A first-principles and ellipsometry
						characterization}},\ }\href {https://doi.org/10.1103/PhysRevB.101.045414}
			{\bibfield  {journal} {\bibinfo  {journal} {Phys. Rev. B}\ }\textbf {\bibinfo
					{volume} {101}},\ \bibinfo {pages} {045414} (\bibinfo {year}
				{2020})}\BibitemShut {NoStop}%
			\bibitem [{\citenamefont {Kirchhoff}\ \emph {et~al.}(2022)\citenamefont
				{Kirchhoff}, \citenamefont {Deilmann}, \citenamefont {Kr\"uger},\ and\
				\citenamefont {Rohlfing}}]{Kirchhoff_2022}%
			\BibitemOpen
			\bibfield  {author} {\bibinfo {author} {\bibfnamefont {A.}~\bibnamefont
					{Kirchhoff}}, \bibinfo {author} {\bibfnamefont {T.}~\bibnamefont {Deilmann}},
				\bibinfo {author} {\bibfnamefont {P.}~\bibnamefont {Kr\"uger}},\ and\
				\bibinfo {author} {\bibfnamefont {M.}~\bibnamefont {Rohlfing}},\ }\bibfield
			{title} {\bibinfo {title} {Electronic and optical properties of a hexagonal
					boron nitride monolayer in its pristine form and with point defects from
					first principles},\ }\href {https://doi.org/10.1103/PhysRevB.106.045118}
			{\bibfield  {journal} {\bibinfo  {journal} {Phys. Rev. B}\ }\textbf {\bibinfo
					{volume} {106}},\ \bibinfo {pages} {045118} (\bibinfo {year}
				{2022})}\BibitemShut {NoStop}%
			\bibitem [{\citenamefont {Hüser}\ \emph
				{et~al.}(2013{\natexlab{b}})\citenamefont {Hüser}, \citenamefont {Olsen},\
				and\ \citenamefont {Thygesen}}]{Hueser_2013_b}%
			\BibitemOpen
			\bibfield  {author} {\bibinfo {author} {\bibfnamefont {F.}~\bibnamefont
					{Hüser}}, \bibinfo {author} {\bibfnamefont {T.}~\bibnamefont {Olsen}},\ and\
				\bibinfo {author} {\bibfnamefont {K.~S.}\ \bibnamefont {Thygesen}},\
			}\bibfield  {title} {\bibinfo {title} {{Quasiparticle GW calculations for
						solids, molecules, and two-dimensional materials}},\ }\href
			{https://doi.org/10.1103/physrevb.87.235132} {\bibfield  {journal} {\bibinfo
					{journal} {Phys.~Rev.~B}\ }\textbf {\bibinfo {volume} {87}},\ \bibinfo
				{pages} {235132} (\bibinfo {year} {2013}{\natexlab{b}})}\BibitemShut
			{NoStop}%
			\bibitem [{\citenamefont {Ferreira}\ \emph {et~al.}(2019)\citenamefont
				{Ferreira}, \citenamefont {Chaves}, \citenamefont {Peres},\ and\
				\citenamefont {Ribeiro}}]{Ferreira_2019}%
			\BibitemOpen
			\bibfield  {author} {\bibinfo {author} {\bibfnamefont {F.}~\bibnamefont
					{Ferreira}}, \bibinfo {author} {\bibfnamefont {A.}~\bibnamefont {Chaves}},
				\bibinfo {author} {\bibfnamefont {N.}~\bibnamefont {Peres}},\ and\ \bibinfo
				{author} {\bibfnamefont {R.}~\bibnamefont {Ribeiro}},\ }\bibfield  {title}
			{\bibinfo {title} {Excitons in hexagonal boron nitride single-layer: a new
					platform for polaritonics in the ultraviolet},\ }\href@noop {} {\bibfield
				{journal} {\bibinfo  {journal} {Journal of the Optical Society of America B}\
				}\textbf {\bibinfo {volume} {36}},\ \bibinfo {pages} {674} (\bibinfo {year}
				{2019})}\BibitemShut {NoStop}%
			\bibitem [{\citenamefont {Attaccalite}\ \emph {et~al.}(2011)\citenamefont
				{Attaccalite}, \citenamefont {Bockstedte}, \citenamefont {Marini},
				\citenamefont {Rubio},\ and\ \citenamefont {Wirtz}}]{Attaccalite_2011}%
			\BibitemOpen
			\bibfield  {author} {\bibinfo {author} {\bibfnamefont {C.}~\bibnamefont
					{Attaccalite}}, \bibinfo {author} {\bibfnamefont {M.}~\bibnamefont
					{Bockstedte}}, \bibinfo {author} {\bibfnamefont {A.}~\bibnamefont {Marini}},
				\bibinfo {author} {\bibfnamefont {A.}~\bibnamefont {Rubio}},\ and\ \bibinfo
				{author} {\bibfnamefont {L.}~\bibnamefont {Wirtz}},\ }\bibfield  {title}
			{\bibinfo {title} {Coupling of excitons and defect states in boron-nitride
					nanostructures},\ }\href {https://doi.org/10.1103/PhysRevB.83.144115}
			{\bibfield  {journal} {\bibinfo  {journal} {Phys. Rev. B}\ }\textbf {\bibinfo
					{volume} {83}},\ \bibinfo {pages} {144115} (\bibinfo {year}
				{2011})}\BibitemShut {NoStop}%
			\bibitem [{\citenamefont {Galvani}\ \emph {et~al.}(2016)\citenamefont
				{Galvani}, \citenamefont {Paleari}, \citenamefont {Miranda}, \citenamefont
				{Molina-S\'anchez}, \citenamefont {Wirtz}, \citenamefont {Latil},
				\citenamefont {Amara},\ and\ \citenamefont {Ducastelle}}]{Galvani_2016}%
			\BibitemOpen
			\bibfield  {author} {\bibinfo {author} {\bibfnamefont {T.}~\bibnamefont
					{Galvani}}, \bibinfo {author} {\bibfnamefont {F.}~\bibnamefont {Paleari}},
				\bibinfo {author} {\bibfnamefont {H.~P.~C.}\ \bibnamefont {Miranda}},
				\bibinfo {author} {\bibfnamefont {A.}~\bibnamefont {Molina-S\'anchez}},
				\bibinfo {author} {\bibfnamefont {L.}~\bibnamefont {Wirtz}}, \bibinfo
				{author} {\bibfnamefont {S.}~\bibnamefont {Latil}}, \bibinfo {author}
				{\bibfnamefont {H.}~\bibnamefont {Amara}},\ and\ \bibinfo {author}
				{\bibfnamefont {F.~m.~c.}\ \bibnamefont {Ducastelle}},\ }\bibfield  {title}
			{\bibinfo {title} {Excitons in boron nitride single layer},\ }\href
			{https://doi.org/10.1103/PhysRevB.94.125303} {\bibfield  {journal} {\bibinfo
					{journal} {Phys. Rev. B}\ }\textbf {\bibinfo {volume} {94}},\ \bibinfo
				{pages} {125303} (\bibinfo {year} {2016})}\BibitemShut {NoStop}%
			\bibitem [{\citenamefont {Wu}\ \emph {et~al.}(2017)\citenamefont {Wu},
				\citenamefont {Galatas}, \citenamefont {Sundararaman}, \citenamefont
				{Rocca},\ and\ \citenamefont {Ping}}]{Wu_2017}%
			\BibitemOpen
			\bibfield  {author} {\bibinfo {author} {\bibfnamefont {F.}~\bibnamefont
					{Wu}}, \bibinfo {author} {\bibfnamefont {A.}~\bibnamefont {Galatas}},
				\bibinfo {author} {\bibfnamefont {R.}~\bibnamefont {Sundararaman}}, \bibinfo
				{author} {\bibfnamefont {D.}~\bibnamefont {Rocca}},\ and\ \bibinfo {author}
				{\bibfnamefont {Y.}~\bibnamefont {Ping}},\ }\bibfield  {title} {\bibinfo
				{title} {First-principles engineering of charged defects for two-dimensional
					quantum technologies},\ }\href
			{https://doi.org/10.1103/PhysRevMaterials.1.071001} {\bibfield  {journal}
				{\bibinfo  {journal} {Phys. Rev. Mater.}\ }\textbf {\bibinfo {volume} {1}},\
				\bibinfo {pages} {071001} (\bibinfo {year} {2017})}\BibitemShut {NoStop}%
			\bibitem [{\citenamefont {Blase}\ \emph {et~al.}(1995)\citenamefont {Blase},
				\citenamefont {Rubio}, \citenamefont {Louie},\ and\ \citenamefont
				{Cohen}}]{Blase_1995}%
			\BibitemOpen
			\bibfield  {author} {\bibinfo {author} {\bibfnamefont {X.}~\bibnamefont
					{Blase}}, \bibinfo {author} {\bibfnamefont {A.}~\bibnamefont {Rubio}},
				\bibinfo {author} {\bibfnamefont {S.~G.}\ \bibnamefont {Louie}},\ and\
				\bibinfo {author} {\bibfnamefont {M.~L.}\ \bibnamefont {Cohen}},\ }\bibfield
			{title} {\bibinfo {title} {Quasiparticle band structure of bulk hexagonal
					boron nitride and related systems},\ }\href
			{https://doi.org/10.1103/PhysRevB.51.6868} {\bibfield  {journal} {\bibinfo
					{journal} {Phys. Rev. B}\ }\textbf {\bibinfo {volume} {51}},\ \bibinfo
				{pages} {6868} (\bibinfo {year} {1995})}\BibitemShut {NoStop}%
			\bibitem [{\citenamefont {Wang}\ and\ \citenamefont
				{Sundararaman}(2020)}]{Wang_2020}%
			\BibitemOpen
			\bibfield  {author} {\bibinfo {author} {\bibfnamefont {D.}~\bibnamefont
					{Wang}}\ and\ \bibinfo {author} {\bibfnamefont {R.}~\bibnamefont
					{Sundararaman}},\ }\bibfield  {title} {\bibinfo {title} {Layer dependence of
					defect charge transition levels in two-dimensional materials},\ }\href
			{https://doi.org/10.1103/PhysRevB.101.054103} {\bibfield  {journal} {\bibinfo
					{journal} {Phys. Rev. B}\ }\textbf {\bibinfo {volume} {101}},\ \bibinfo
				{pages} {054103} (\bibinfo {year} {2020})}\BibitemShut {NoStop}%
			\bibitem [{\citenamefont {Mengle}\ and\ \citenamefont
				{Kioupakis}(2019)}]{Mengle_2019}%
			\BibitemOpen
			\bibfield  {author} {\bibinfo {author} {\bibfnamefont {K.}~\bibnamefont
					{Mengle}}\ and\ \bibinfo {author} {\bibfnamefont {E.}~\bibnamefont
					{Kioupakis}},\ }\bibfield  {title} {\bibinfo {title} {Impact of the stacking
					sequence on the bandgap and luminescence properties of bulk, bilayer, and
					monolayer hexagonal boron nitride},\ }\href@noop {} {\bibfield  {journal}
				{\bibinfo  {journal} {APL Materials}\ }\textbf {\bibinfo {volume} {7}}
				(\bibinfo {year} {2019})}\BibitemShut {NoStop}%
			\bibitem [{\citenamefont {Fu}\ \emph {et~al.}(2016)\citenamefont {Fu},
				\citenamefont {Nabok},\ and\ \citenamefont {Draxl}}]{Fu_2016}%
			\BibitemOpen
			\bibfield  {author} {\bibinfo {author} {\bibfnamefont {Q.}~\bibnamefont
					{Fu}}, \bibinfo {author} {\bibfnamefont {D.}~\bibnamefont {Nabok}},\ and\
				\bibinfo {author} {\bibfnamefont {C.}~\bibnamefont {Draxl}},\ }\bibfield
			{title} {\bibinfo {title} {Energy-level alignment at the interface of
					graphene fluoride and boron nitride monolayers: an investigation by many-body
					perturbation theory},\ }\href@noop {} {\bibfield  {journal} {\bibinfo
					{journal} {The Journal of Physical Chemistry C}\ }\textbf {\bibinfo {volume}
					{120}},\ \bibinfo {pages} {11671} (\bibinfo {year} {2016})}\BibitemShut
			{NoStop}%
			\bibitem [{\citenamefont {Berseneva}\ \emph {et~al.}(2013)\citenamefont
				{Berseneva}, \citenamefont {Gulans}, \citenamefont {Krasheninnikov},\ and\
				\citenamefont {Nieminen}}]{Berseneva_2013}%
			\BibitemOpen
			\bibfield  {author} {\bibinfo {author} {\bibfnamefont {N.}~\bibnamefont
					{Berseneva}}, \bibinfo {author} {\bibfnamefont {A.}~\bibnamefont {Gulans}},
				\bibinfo {author} {\bibfnamefont {A.~V.}\ \bibnamefont {Krasheninnikov}},\
				and\ \bibinfo {author} {\bibfnamefont {R.~M.}\ \bibnamefont {Nieminen}},\
			}\bibfield  {title} {\bibinfo {title} {Electronic structure of boron nitride
					sheets doped with carbon from first-principles calculations},\ }\href
			{https://doi.org/10.1103/PhysRevB.87.035404} {\bibfield  {journal} {\bibinfo
					{journal} {Phys. Rev. B}\ }\textbf {\bibinfo {volume} {87}},\ \bibinfo
				{pages} {035404} (\bibinfo {year} {2013})}\BibitemShut {NoStop}%
			\bibitem [{\citenamefont {Cudazzo}\ \emph {et~al.}(2016)\citenamefont
				{Cudazzo}, \citenamefont {Sponza}, \citenamefont {Giorgetti}, \citenamefont
				{Reining}, \citenamefont {Sottile},\ and\ \citenamefont
				{Gatti}}]{Cudazzo_2016}%
			\BibitemOpen
			\bibfield  {author} {\bibinfo {author} {\bibfnamefont {P.}~\bibnamefont
					{Cudazzo}}, \bibinfo {author} {\bibfnamefont {L.}~\bibnamefont {Sponza}},
				\bibinfo {author} {\bibfnamefont {C.}~\bibnamefont {Giorgetti}}, \bibinfo
				{author} {\bibfnamefont {L.}~\bibnamefont {Reining}}, \bibinfo {author}
				{\bibfnamefont {F.}~\bibnamefont {Sottile}},\ and\ \bibinfo {author}
				{\bibfnamefont {M.}~\bibnamefont {Gatti}},\ }\bibfield  {title} {\bibinfo
				{title} {Exciton band structure in two-dimensional materials},\ }\href
			{https://doi.org/10.1103/PhysRevLett.116.066803} {\bibfield  {journal}
				{\bibinfo  {journal} {Phys. Rev. Lett.}\ }\textbf {\bibinfo {volume} {116}},\
				\bibinfo {pages} {066803} (\bibinfo {year} {2016})}\BibitemShut {NoStop}%
			\bibitem [{\citenamefont {Ferreira}\ and\ \citenamefont
				{Ribeiro}(2017)}]{Ferreira_2017}%
			\BibitemOpen
			\bibfield  {author} {\bibinfo {author} {\bibfnamefont {F.}~\bibnamefont
					{Ferreira}}\ and\ \bibinfo {author} {\bibfnamefont {R.~M.}\ \bibnamefont
					{Ribeiro}},\ }\bibfield  {title} {\bibinfo {title} {Improvements in the $gw$
					and bethe-salpeter-equation calculations on phosphorene},\ }\href
			{https://doi.org/10.1103/PhysRevB.96.115431} {\bibfield  {journal} {\bibinfo
					{journal} {Phys. Rev. B}\ }\textbf {\bibinfo {volume} {96}},\ \bibinfo
				{pages} {115431} (\bibinfo {year} {2017})}\BibitemShut {NoStop}%
			\bibitem [{\citenamefont {Marsoner~Steinkasserer}\ \emph
				{et~al.}(2016)\citenamefont {Marsoner~Steinkasserer}, \citenamefont {Suhr},\
				and\ \citenamefont {Paulus}}]{Steinkasserer_2016}%
			\BibitemOpen
			\bibfield  {author} {\bibinfo {author} {\bibfnamefont {L.~E.}\ \bibnamefont
					{Marsoner~Steinkasserer}}, \bibinfo {author} {\bibfnamefont {S.}~\bibnamefont
					{Suhr}},\ and\ \bibinfo {author} {\bibfnamefont {B.}~\bibnamefont {Paulus}},\
			}\bibfield  {title} {\bibinfo {title} {{Band-gap control in phosphorene/BN
						structures from first-principles calculations}},\ }\href
			{https://doi.org/10.1103/PhysRevB.94.125444} {\bibfield  {journal} {\bibinfo
					{journal} {Phys. Rev. B}\ }\textbf {\bibinfo {volume} {94}},\ \bibinfo
				{pages} {125444} (\bibinfo {year} {2016})}\BibitemShut {NoStop}%
			\bibitem [{\citenamefont {Rudenko}\ and\ \citenamefont
				{Katsnelson}(2014)}]{Rudenko_2014}%
			\BibitemOpen
			\bibfield  {author} {\bibinfo {author} {\bibfnamefont {A.~N.}\ \bibnamefont
					{Rudenko}}\ and\ \bibinfo {author} {\bibfnamefont {M.~I.}\ \bibnamefont
					{Katsnelson}},\ }\bibfield  {title} {\bibinfo {title} {Quasiparticle band
					structure and tight-binding model for single- and bilayer black phosphorus},\
			}\href {https://doi.org/10.1103/PhysRevB.89.201408} {\bibfield  {journal}
				{\bibinfo  {journal} {Phys. Rev. B}\ }\textbf {\bibinfo {volume} {89}},\
				\bibinfo {pages} {201408} (\bibinfo {year} {2014})}\BibitemShut {NoStop}%
			\bibitem [{\citenamefont {Tran}\ \emph {et~al.}(2014)\citenamefont {Tran},
				\citenamefont {Soklaski}, \citenamefont {Liang},\ and\ \citenamefont
				{Yang}}]{Tran_2014}%
			\BibitemOpen
			\bibfield  {author} {\bibinfo {author} {\bibfnamefont {V.}~\bibnamefont
					{Tran}}, \bibinfo {author} {\bibfnamefont {R.}~\bibnamefont {Soklaski}},
				\bibinfo {author} {\bibfnamefont {Y.}~\bibnamefont {Liang}},\ and\ \bibinfo
				{author} {\bibfnamefont {L.}~\bibnamefont {Yang}},\ }\bibfield  {title}
			{\bibinfo {title} {Layer-controlled band gap and anisotropic excitons in
					few-layer black phosphorus},\ }\href
			{https://doi.org/10.1103/PhysRevB.89.235319} {\bibfield  {journal} {\bibinfo
					{journal} {Phys. Rev. B}\ }\textbf {\bibinfo {volume} {89}},\ \bibinfo
				{pages} {235319} (\bibinfo {year} {2014})}\BibitemShut {NoStop}%
			\bibitem [{\citenamefont {\ifmmode \mbox{\c{C}}\else \c{C}\fi{}ak\ifmmode
					\imath \else~\i \fi{}r}\ \emph {et~al.}(2014)\citenamefont {\ifmmode
					\mbox{\c{C}}\else \c{C}\fi{}ak\ifmmode \imath \else~\i \fi{}r}, \citenamefont
				{Sahin},\ and\ \citenamefont {Peeters}}]{Cakir_2014}%
			\BibitemOpen
			\bibfield  {author} {\bibinfo {author} {\bibfnamefont {D.}~\bibnamefont
					{\ifmmode \mbox{\c{C}}\else \c{C}\fi{}ak\ifmmode \imath \else~\i \fi{}r}},
				\bibinfo {author} {\bibfnamefont {H.}~\bibnamefont {Sahin}},\ and\ \bibinfo
				{author} {\bibfnamefont {F.~m. c.~M.}\ \bibnamefont {Peeters}},\ }\bibfield
			{title} {\bibinfo {title} {Tuning of the electronic and optical properties of
					single-layer black phosphorus by strain},\ }\href
			{https://doi.org/10.1103/PhysRevB.90.205421} {\bibfield  {journal} {\bibinfo
					{journal} {Phys. Rev. B}\ }\textbf {\bibinfo {volume} {90}},\ \bibinfo
				{pages} {205421} (\bibinfo {year} {2014})}\BibitemShut {NoStop}%
			\bibitem [{\citenamefont {Tran}\ \emph {et~al.}(2015)\citenamefont {Tran},
				\citenamefont {Fei},\ and\ \citenamefont {Yang}}]{Tran_2015}%
			\BibitemOpen
			\bibfield  {author} {\bibinfo {author} {\bibfnamefont {V.}~\bibnamefont
					{Tran}}, \bibinfo {author} {\bibfnamefont {R.}~\bibnamefont {Fei}},\ and\
				\bibinfo {author} {\bibfnamefont {L.}~\bibnamefont {Yang}},\ }\bibfield
			{title} {\bibinfo {title} {Quasiparticle energies, excitons, and optical
					spectra of few-layer black phosphorus},\ }\href@noop {} {\bibfield  {journal}
				{\bibinfo  {journal} {2D Materials}\ }\textbf {\bibinfo {volume} {2}},\
				\bibinfo {pages} {044014} (\bibinfo {year} {2015})}\BibitemShut {NoStop}%
			\bibitem [{\citenamefont {Pisarra}\ \emph {et~al.}(2021)\citenamefont
				{Pisarra}, \citenamefont {Díaz},\ and\ \citenamefont
				{Martín}}]{Pisarra_2021}%
			\BibitemOpen
			\bibfield  {author} {\bibinfo {author} {\bibfnamefont {M.}~\bibnamefont
					{Pisarra}}, \bibinfo {author} {\bibfnamefont {C.}~\bibnamefont {Díaz}},\
				and\ \bibinfo {author} {\bibfnamefont {F.}~\bibnamefont {Martín}},\
			}\bibfield  {title} {\bibinfo {title} {{Theoretical study of structural and
						electronic properties of 2H-phase transition metal dichalcogenides}},\ }\href
			{https://doi.org/10.1103/physrevb.103.195416} {\bibfield  {journal} {\bibinfo
					{journal} {Phys.~Rev.~B}\ }\textbf {\bibinfo {volume} {103}},\ \bibinfo
				{pages} {195416} (\bibinfo {year} {2021})}\BibitemShut {NoStop}%
			\bibitem [{\citenamefont {Ortenzi}\ \emph {et~al.}(2018)\citenamefont
				{Ortenzi}, \citenamefont {Pietronero},\ and\ \citenamefont
				{Cappelluti}}]{Ortenzi_2018}%
			\BibitemOpen
			\bibfield  {author} {\bibinfo {author} {\bibfnamefont {L.}~\bibnamefont
					{Ortenzi}}, \bibinfo {author} {\bibfnamefont {L.}~\bibnamefont
					{Pietronero}},\ and\ \bibinfo {author} {\bibfnamefont {E.}~\bibnamefont
					{Cappelluti}},\ }\bibfield  {title} {\bibinfo {title} {{Zero-point motion and
						direct-indirect band-gap crossover in layered transition-metal
						dichalcogenides}},\ }\href {https://doi.org/10.1103/physrevb.98.195313}
			{\bibfield  {journal} {\bibinfo  {journal} {Phys.~Rev.~B}\ }\textbf {\bibinfo
					{volume} {98}},\ \bibinfo {pages} {195313} (\bibinfo {year} {2018})},\
			\Eprint {https://arxiv.org/abs/1901.08301} {1901.08301} \BibitemShut
			{NoStop}%
			\bibitem [{\citenamefont {Bechstedt}(2016)}]{Bechstedt}%
			\BibitemOpen
			\bibfield  {author} {\bibinfo {author} {\bibfnamefont {F.}~\bibnamefont
					{Bechstedt}},\ }\href@noop {} {\emph {\bibinfo {title} {Many-body approach to
						electronic excitations}}}\ (\bibinfo  {publisher} {Springer},\ \bibinfo
			{year} {2016})\BibitemShut {NoStop}%
			\bibitem [{\citenamefont {Onida}\ \emph {et~al.}(2002)\citenamefont {Onida},
				\citenamefont {Reining},\ and\ \citenamefont {Rubio}}]{Onida_2002}%
			\BibitemOpen
			\bibfield  {author} {\bibinfo {author} {\bibfnamefont {G.}~\bibnamefont
					{Onida}}, \bibinfo {author} {\bibfnamefont {L.}~\bibnamefont {Reining}},\
				and\ \bibinfo {author} {\bibfnamefont {A.}~\bibnamefont {Rubio}},\ }\bibfield
			{title} {\bibinfo {title} {{Electronic excitations: density-functional
						versus many-body Green’s-function approaches}},\ }\href
			{https://doi.org/10.1103/revmodphys.74.601} {\bibfield  {journal} {\bibinfo
					{journal} {Reviews of Modern Physics}\ }\textbf {\bibinfo {volume} {74}},\
				\bibinfo {pages} {601} (\bibinfo {year} {2002})}\BibitemShut {NoStop}%
			\bibitem [{\citenamefont {Li}\ \emph {et~al.}(2012)\citenamefont {Li},
				\citenamefont {Gómez-Abal}, \citenamefont {Jiang}, \citenamefont
				{Ambrosch-Draxl},\ and\ \citenamefont {Scheffler}}]{Li_2012}%
			\BibitemOpen
			\bibfield  {author} {\bibinfo {author} {\bibfnamefont {X.-Z.}\ \bibnamefont
					{Li}}, \bibinfo {author} {\bibfnamefont {R.}~\bibnamefont {Gómez-Abal}},
				\bibinfo {author} {\bibfnamefont {H.}~\bibnamefont {Jiang}}, \bibinfo
				{author} {\bibfnamefont {C.}~\bibnamefont {Ambrosch-Draxl}},\ and\ \bibinfo
				{author} {\bibfnamefont {M.}~\bibnamefont {Scheffler}},\ }\bibfield  {title}
			{\bibinfo {title} {{Impact of widely used approximations to the $G_0W_0$
						method: an all-electron perspective}},\ }\href
			{https://doi.org/10.1088/1367-2630/14/2/023006} {\bibfield  {journal}
				{\bibinfo  {journal} {New Journal of Physics}\ }\textbf {\bibinfo {volume}
					{14}},\ \bibinfo {pages} {023006} (\bibinfo {year} {2012})}\BibitemShut
			{NoStop}%
			\bibitem [{\citenamefont {K\"orzd\"orfer}\ and\ \citenamefont
				{Marom}(2012)}]{Koerzdoerfer_2012}%
			\BibitemOpen
			\bibfield  {author} {\bibinfo {author} {\bibfnamefont {T.}~\bibnamefont
					{K\"orzd\"orfer}}\ and\ \bibinfo {author} {\bibfnamefont {N.}~\bibnamefont
					{Marom}},\ }\bibfield  {title} {\bibinfo {title} {Strategy for finding a
					reliable starting point for ${G}_{0}{W}_{0}$ demonstrated for molecules},\
			}\href {https://doi.org/10.1103/PhysRevB.86.041110} {\bibfield  {journal}
				{\bibinfo  {journal} {Phys.~Rev.~B}\ }\textbf {\bibinfo {volume} {86}},\
				\bibinfo {pages} {041110} (\bibinfo {year} {2012})}\BibitemShut {NoStop}%
			\bibitem [{\citenamefont {Chen}\ and\ \citenamefont
				{Pasquarello}(2014)}]{Chen_2014}%
			\BibitemOpen
			\bibfield  {author} {\bibinfo {author} {\bibfnamefont {W.}~\bibnamefont
					{Chen}}\ and\ \bibinfo {author} {\bibfnamefont {A.}~\bibnamefont
					{Pasquarello}},\ }\bibfield  {title} {\bibinfo {title} {{Band-edge positions
						in GW: Effects of starting point and self-consistency}},\ }\href
			{https://doi.org/10.1103/physrevb.90.165133} {\bibfield  {journal} {\bibinfo
					{journal} {Phys.~Rev.~B}\ }\textbf {\bibinfo {volume} {90}},\ \bibinfo
				{pages} {165133} (\bibinfo {year} {2014})}\BibitemShut {NoStop}%
			\bibitem [{\citenamefont {Pela}\ \emph {et~al.}(2016)\citenamefont {Pela},
				\citenamefont {Werner}, \citenamefont {Nabok},\ and\ \citenamefont
				{Draxl}}]{Pela_2016}%
			\BibitemOpen
			\bibfield  {author} {\bibinfo {author} {\bibfnamefont {R.~R.}\ \bibnamefont
					{Pela}}, \bibinfo {author} {\bibfnamefont {U.}~\bibnamefont {Werner}},
				\bibinfo {author} {\bibfnamefont {D.}~\bibnamefont {Nabok}},\ and\ \bibinfo
				{author} {\bibfnamefont {C.}~\bibnamefont {Draxl}},\ }\bibfield  {title}
			{\bibinfo {title} {{Probing the LDA-1/2 method as a starting point for
						$G_0W_0$ calculations}},\ }\href {https://doi.org/10.1103/physrevb.94.235141}
			{\bibfield  {journal} {\bibinfo  {journal} {Phys.~Rev.~B}\ }\textbf {\bibinfo
					{volume} {94}},\ \bibinfo {pages} {235141} (\bibinfo {year}
				{2016})}\BibitemShut {NoStop}%
			\bibitem [{\citenamefont {{McKeon}}\ \emph {et~al.}(2022)\citenamefont
				{{McKeon}}, \citenamefont {Hamed}, \citenamefont {Bruneval},\ and\
				\citenamefont {Neaton}}]{McKeon_2022}%
			\BibitemOpen
			\bibfield  {author} {\bibinfo {author} {\bibfnamefont {C.~A.}\ \bibnamefont
					{{McKeon}}}, \bibinfo {author} {\bibfnamefont {S.~M.}\ \bibnamefont {Hamed}},
				\bibinfo {author} {\bibfnamefont {F.}~\bibnamefont {Bruneval}},\ and\
				\bibinfo {author} {\bibfnamefont {J.~B.}\ \bibnamefont {Neaton}},\ }\bibfield
			{title} {\bibinfo {title} {{An optimally tuned range-separated hybrid
						starting point for ab initio GW plus Bethe–Salpeter equation calculations
						of molecules}},\ }\href {https://doi.org/10.1063/5.0097582} {\bibfield
				{journal} {\bibinfo  {journal} {The Journal of Chemical Physics}\ }\textbf
				{\bibinfo {volume} {157}},\ \bibinfo {pages} {074103} (\bibinfo {year}
				{2022})}\BibitemShut {NoStop}%
			\bibitem [{\citenamefont {Knight}\ \emph {et~al.}(2016)\citenamefont {Knight},
				\citenamefont {Wang}, \citenamefont {Gallandi}, \citenamefont
				{Dolgounitcheva}, \citenamefont {Ren}, \citenamefont {Ortiz}, \citenamefont
				{Rinke}, \citenamefont {Körzdörfer},\ and\ \citenamefont
				{Marom}}]{Knight_2016}%
			\BibitemOpen
			\bibfield  {author} {\bibinfo {author} {\bibfnamefont {J.~W.}\ \bibnamefont
					{Knight}}, \bibinfo {author} {\bibfnamefont {X.}~\bibnamefont {Wang}},
				\bibinfo {author} {\bibfnamefont {L.}~\bibnamefont {Gallandi}}, \bibinfo
				{author} {\bibfnamefont {O.}~\bibnamefont {Dolgounitcheva}}, \bibinfo
				{author} {\bibfnamefont {X.}~\bibnamefont {Ren}}, \bibinfo {author}
				{\bibfnamefont {J.~V.}\ \bibnamefont {Ortiz}}, \bibinfo {author}
				{\bibfnamefont {P.}~\bibnamefont {Rinke}}, \bibinfo {author} {\bibfnamefont
					{T.}~\bibnamefont {Körzdörfer}},\ and\ \bibinfo {author} {\bibfnamefont
					{N.}~\bibnamefont {Marom}},\ }\bibfield  {title} {\bibinfo {title} {{Accurate
						Ionization Potentials and Electron Affinities of Acceptor Molecules III: A
						Benchmark of {GW} Methods}},\ }\href
			{https://doi.org/10.1021/acs.jctc.5b00871} {\bibfield  {journal} {\bibinfo
					{journal} {Journal of Chemical Theory and Computation}\ }\textbf {\bibinfo
					{volume} {12}},\ \bibinfo {pages} {615} (\bibinfo {year} {2016})}\BibitemShut
			{NoStop}%
			\bibitem [{\citenamefont {Gant}\ \emph {et~al.}(2022)\citenamefont {Gant},
				\citenamefont {Haber}, \citenamefont {Filip}, \citenamefont {Sagredo},
				\citenamefont {Wing}, \citenamefont {Ohad}, \citenamefont {Kronik},\ and\
				\citenamefont {Neaton}}]{Gant_2022}%
			\BibitemOpen
			\bibfield  {author} {\bibinfo {author} {\bibfnamefont {S.~E.}\ \bibnamefont
					{Gant}}, \bibinfo {author} {\bibfnamefont {J.~B.}\ \bibnamefont {Haber}},
				\bibinfo {author} {\bibfnamefont {M.~R.}\ \bibnamefont {Filip}}, \bibinfo
				{author} {\bibfnamefont {F.}~\bibnamefont {Sagredo}}, \bibinfo {author}
				{\bibfnamefont {D.}~\bibnamefont {Wing}}, \bibinfo {author} {\bibfnamefont
					{G.}~\bibnamefont {Ohad}}, \bibinfo {author} {\bibfnamefont {L.}~\bibnamefont
					{Kronik}},\ and\ \bibinfo {author} {\bibfnamefont {J.~B.}\ \bibnamefont
					{Neaton}},\ }\bibfield  {title} {\bibinfo {title} {{Optimally tuned starting
						point for single-shot GW calculations of solids}},\ }\href
			{https://doi.org/10.1103/physrevmaterials.6.053802} {\bibfield  {journal}
				{\bibinfo  {journal} {Physical Review Materials}\ }\textbf {\bibinfo {volume}
					{6}},\ \bibinfo {pages} {053802} (\bibinfo {year} {2022})}\BibitemShut
			{NoStop}%
			\bibitem [{\citenamefont {Wang}\ \emph {et~al.}(2014)\citenamefont {Wang},
				\citenamefont {Kutana},\ and\ \citenamefont {Yakobson}}]{Wang_2014}%
			\BibitemOpen
			\bibfield  {author} {\bibinfo {author} {\bibfnamefont {L.}~\bibnamefont
					{Wang}}, \bibinfo {author} {\bibfnamefont {A.}~\bibnamefont {Kutana}},\ and\
				\bibinfo {author} {\bibfnamefont {B.~I.}\ \bibnamefont {Yakobson}},\
			}\bibfield  {title} {\bibinfo {title} {{Many‐body and spin‐orbit effects
						on direct‐indirect band gap transition of strained monolayer MoS$_2$ and
						WS$_2$}},\ }\href {https://doi.org/10.1002/andp.201400098} {\bibfield
				{journal} {\bibinfo  {journal} {Annalen der Physik}\ }\textbf {\bibinfo
					{volume} {526}},\ \bibinfo {pages} {L7} (\bibinfo {year} {2014})}\BibitemShut
			{NoStop}%
			\bibitem [{\citenamefont {Lembke}\ and\ \citenamefont
				{Kis}(2012)}]{Lembke_2012}%
			\BibitemOpen
			\bibfield  {author} {\bibinfo {author} {\bibfnamefont {D.}~\bibnamefont
					{Lembke}}\ and\ \bibinfo {author} {\bibfnamefont {A.}~\bibnamefont {Kis}},\
			}\bibfield  {title} {\bibinfo {title} {{Breakdown of High-Performance
						Monolayer MoS$_2$ Transistors}},\ }\href {https://doi.org/10.1021/nn303772b}
			{\bibfield  {journal} {\bibinfo  {journal} {ACS Nano}\ }\textbf {\bibinfo
					{volume} {6}},\ \bibinfo {pages} {10070–10075} (\bibinfo {year}
				{2012})}\BibitemShut {NoStop}%
			\bibitem [{\citenamefont {Xiao}\ \emph {et~al.}(2012)\citenamefont {Xiao},
				\citenamefont {Liu}, \citenamefont {Feng}, \citenamefont {Xu},\ and\
				\citenamefont {Yao}}]{Xiao_2012}%
			\BibitemOpen
			\bibfield  {author} {\bibinfo {author} {\bibfnamefont {D.}~\bibnamefont
					{Xiao}}, \bibinfo {author} {\bibfnamefont {G.-B.}\ \bibnamefont {Liu}},
				\bibinfo {author} {\bibfnamefont {W.}~\bibnamefont {Feng}}, \bibinfo {author}
				{\bibfnamefont {X.}~\bibnamefont {Xu}},\ and\ \bibinfo {author}
				{\bibfnamefont {W.}~\bibnamefont {Yao}},\ }\bibfield  {title} {\bibinfo
				{title} {{Coupled Spin and Valley Physics in Monolayers of {MoS}$_{2}$ and
						Other Group-VI Dichalcogenides}},\ }\href
			{https://doi.org/10.1103/PhysRevLett.108.196802} {\bibfield  {journal}
				{\bibinfo  {journal} {Phys.~Rev.~Lett.~}\ }\textbf {\bibinfo {volume}
					{108}},\ \bibinfo {pages} {196802} (\bibinfo {year} {2012})}\BibitemShut
			{NoStop}%
			\bibitem [{\citenamefont {Cao}\ \emph {et~al.}(2012)\citenamefont {Cao},
				\citenamefont {Wang}, \citenamefont {Han}, \citenamefont {Ye}, \citenamefont
				{Zhu}, \citenamefont {Shi}, \citenamefont {Niu}, \citenamefont {Tan},
				\citenamefont {Wang}, \citenamefont {Liu},\ and\ \citenamefont
				{Feng}}]{Cao_2012}%
			\BibitemOpen
			\bibfield  {author} {\bibinfo {author} {\bibfnamefont {T.}~\bibnamefont
					{Cao}}, \bibinfo {author} {\bibfnamefont {G.}~\bibnamefont {Wang}}, \bibinfo
				{author} {\bibfnamefont {W.}~\bibnamefont {Han}}, \bibinfo {author}
				{\bibfnamefont {H.}~\bibnamefont {Ye}}, \bibinfo {author} {\bibfnamefont
					{C.}~\bibnamefont {Zhu}}, \bibinfo {author} {\bibfnamefont {J.}~\bibnamefont
					{Shi}}, \bibinfo {author} {\bibfnamefont {Q.}~\bibnamefont {Niu}}, \bibinfo
				{author} {\bibfnamefont {P.}~\bibnamefont {Tan}}, \bibinfo {author}
				{\bibfnamefont {E.}~\bibnamefont {Wang}}, \bibinfo {author} {\bibfnamefont
					{B.}~\bibnamefont {Liu}},\ and\ \bibinfo {author} {\bibfnamefont
					{J.}~\bibnamefont {Feng}},\ }\bibfield  {title} {\bibinfo {title}
				{{Valley-selective circular dichroism of monolayer molybdenum disulphide}},\
			}\href {https://doi.org/10.1038/ncomms1882} {\bibfield  {journal} {\bibinfo
					{journal} {Nature Communications}\ }\textbf {\bibinfo {volume} {3}},\
				\bibinfo {pages} {887} (\bibinfo {year} {2012})},\ \Eprint
			{https://arxiv.org/abs/1112.4013} {1112.4013} \BibitemShut {NoStop}%
			\bibitem [{\citenamefont {Zeng}\ \emph {et~al.}(2012)\citenamefont {Zeng},
				\citenamefont {Dai}, \citenamefont {Yao}, \citenamefont {Xiao},\ and\
				\citenamefont {Cui}}]{Zeng_2012}%
			\BibitemOpen
			\bibfield  {author} {\bibinfo {author} {\bibfnamefont {H.}~\bibnamefont
					{Zeng}}, \bibinfo {author} {\bibfnamefont {J.}~\bibnamefont {Dai}}, \bibinfo
				{author} {\bibfnamefont {W.}~\bibnamefont {Yao}}, \bibinfo {author}
				{\bibfnamefont {D.}~\bibnamefont {Xiao}},\ and\ \bibinfo {author}
				{\bibfnamefont {X.}~\bibnamefont {Cui}},\ }\bibfield  {title} {\bibinfo
				{title} {{Valley polarization in MoS$_2$ monolayers by optical pumping}},\
			}\href {https://doi.org/10.1038/nnano.2012.95} {\bibfield  {journal}
				{\bibinfo  {journal} {Nature Nanotechnology}\ }\textbf {\bibinfo {volume}
					{7}},\ \bibinfo {pages} {490} (\bibinfo {year} {2012})},\ \Eprint
			{https://arxiv.org/abs/1202.1592} {1202.1592} \BibitemShut {NoStop}%
			\bibitem [{\citenamefont {Yang}\ \emph {et~al.}(2015)\citenamefont {Yang},
				\citenamefont {Sinitsyn}, \citenamefont {Chen}, \citenamefont {Yuan},
				\citenamefont {Zhang}, \citenamefont {Lou},\ and\ \citenamefont
				{Crooker}}]{Yang_2015}%
			\BibitemOpen
			\bibfield  {author} {\bibinfo {author} {\bibfnamefont {L.}~\bibnamefont
					{Yang}}, \bibinfo {author} {\bibfnamefont {N.~A.}\ \bibnamefont {Sinitsyn}},
				\bibinfo {author} {\bibfnamefont {W.}~\bibnamefont {Chen}}, \bibinfo {author}
				{\bibfnamefont {J.}~\bibnamefont {Yuan}}, \bibinfo {author} {\bibfnamefont
					{J.}~\bibnamefont {Zhang}}, \bibinfo {author} {\bibfnamefont
					{J.}~\bibnamefont {Lou}},\ and\ \bibinfo {author} {\bibfnamefont {S.~A.}\
					\bibnamefont {Crooker}},\ }\bibfield  {title} {\bibinfo {title} {{Long-lived
						nanosecond spin relaxation and spin coherence of electrons in monolayer
						MoS$_2$ and WS$_2$}},\ }\href@noop {} {\bibfield  {journal} {\bibinfo
					{journal} {Nature Physics}\ }\textbf {\bibinfo {volume} {11}},\ \bibinfo
				{pages} {830} (\bibinfo {year} {2015})}\BibitemShut {NoStop}%
			\bibitem [{\citenamefont {Bergh{\"a}user}\ \emph {et~al.}(2018)\citenamefont
				{Bergh{\"a}user}, \citenamefont {Bernal-Villamil}, \citenamefont {Schmidt},
				\citenamefont {Schneider}, \citenamefont {Niehues}, \citenamefont {Erhart},
				\citenamefont {Michaelis~de Vasconcellos}, \citenamefont {Bratschitsch},
				\citenamefont {Knorr},\ and\ \citenamefont {Malic}}]{Berghauser_2018}%
			\BibitemOpen
			\bibfield  {author} {\bibinfo {author} {\bibfnamefont {G.}~\bibnamefont
					{Bergh{\"a}user}}, \bibinfo {author} {\bibfnamefont {I.}~\bibnamefont
					{Bernal-Villamil}}, \bibinfo {author} {\bibfnamefont {R.}~\bibnamefont
					{Schmidt}}, \bibinfo {author} {\bibfnamefont {R.}~\bibnamefont {Schneider}},
				\bibinfo {author} {\bibfnamefont {I.}~\bibnamefont {Niehues}}, \bibinfo
				{author} {\bibfnamefont {P.}~\bibnamefont {Erhart}}, \bibinfo {author}
				{\bibfnamefont {S.}~\bibnamefont {Michaelis~de Vasconcellos}}, \bibinfo
				{author} {\bibfnamefont {R.}~\bibnamefont {Bratschitsch}}, \bibinfo {author}
				{\bibfnamefont {A.}~\bibnamefont {Knorr}},\ and\ \bibinfo {author}
				{\bibfnamefont {E.}~\bibnamefont {Malic}},\ }\bibfield  {title} {\bibinfo
				{title} {Inverted valley polarization in optically excited transition metal
					dichalcogenides},\ }\href@noop {} {\bibfield  {journal} {\bibinfo  {journal}
					{Nature communications}\ }\textbf {\bibinfo {volume} {9}},\ \bibinfo {pages}
				{971} (\bibinfo {year} {2018})}\BibitemShut {NoStop}%
			\bibitem [{\citenamefont {Caruso}\ \emph {et~al.}(2022)\citenamefont {Caruso},
				\citenamefont {Schebek}, \citenamefont {Pan}, \citenamefont {Vona},\ and\
				\citenamefont {Draxl}}]{Caruso_2022}%
			\BibitemOpen
			\bibfield  {author} {\bibinfo {author} {\bibfnamefont {F.}~\bibnamefont
					{Caruso}}, \bibinfo {author} {\bibfnamefont {M.}~\bibnamefont {Schebek}},
				\bibinfo {author} {\bibfnamefont {Y.}~\bibnamefont {Pan}}, \bibinfo {author}
				{\bibfnamefont {C.}~\bibnamefont {Vona}},\ and\ \bibinfo {author}
				{\bibfnamefont {C.}~\bibnamefont {Draxl}},\ }\bibfield  {title} {\bibinfo
				{title} {{Chirality of Valley Excitons in Monolayer Transition-Metal
						Dichalcogenides}},\ }\href {https://doi.org/10.1021/acs.jpclett.2c01034}
			{\bibfield  {journal} {\bibinfo  {journal} {The Journal of Physical Chemistry
						Letters}\ }\textbf {\bibinfo {volume} {13}},\ \bibinfo {pages} {5894}
				(\bibinfo {year} {2022})},\ \Eprint {https://arxiv.org/abs/2112.04781}
			{2112.04781} \BibitemShut {NoStop}%
			\bibitem [{\citenamefont {Ji}\ \emph {et~al.}(2023)\citenamefont {Ji},
				\citenamefont {Yao}, \citenamefont {Quan}, \citenamefont {Yang},
				\citenamefont {Caruso},\ and\ \citenamefont {Li}}]{Ji_2023}%
			\BibitemOpen
			\bibfield  {author} {\bibinfo {author} {\bibfnamefont {S.}~\bibnamefont
					{Ji}}, \bibinfo {author} {\bibfnamefont {R.}~\bibnamefont {Yao}}, \bibinfo
				{author} {\bibfnamefont {C.}~\bibnamefont {Quan}}, \bibinfo {author}
				{\bibfnamefont {J.}~\bibnamefont {Yang}}, \bibinfo {author} {\bibfnamefont
					{F.}~\bibnamefont {Caruso}},\ and\ \bibinfo {author} {\bibfnamefont
					{X.}~\bibnamefont {Li}},\ }\bibfield  {title} {\bibinfo {title} {{Anomalous
						valley Hall effect induced by mirror symmetry breaking in transition metal
						dichalcogenides}},\ }\href {https://doi.org/10.1103/physrevb.107.174434}
			{\bibfield  {journal} {\bibinfo  {journal} {Phys.~Rev.~B}\ }\textbf {\bibinfo
					{volume} {107}},\ \bibinfo {pages} {174434} (\bibinfo {year}
				{2023})}\BibitemShut {NoStop}%
			\bibitem [{\citenamefont {Klots}\ \emph {et~al.}(2014)\citenamefont {Klots},
				\citenamefont {Newaz}, \citenamefont {Wang}, \citenamefont {Prasai},
				\citenamefont {Krzyzanowska}, \citenamefont {Lin}, \citenamefont {Caudel},
				\citenamefont {Ghimire}, \citenamefont {Yan}, \citenamefont {Ivanov},
				\citenamefont {Velizhanin}, \citenamefont {Burger}, \citenamefont {Mandrus},
				\citenamefont {Tolk}, \citenamefont {Pantelides},\ and\ \citenamefont
				{Bolotin}}]{Klots_2014}%
			\BibitemOpen
			\bibfield  {author} {\bibinfo {author} {\bibfnamefont {A.~R.}\ \bibnamefont
					{Klots}}, \bibinfo {author} {\bibfnamefont {A.~K.~M.}\ \bibnamefont {Newaz}},
				\bibinfo {author} {\bibfnamefont {B.}~\bibnamefont {Wang}}, \bibinfo {author}
				{\bibfnamefont {D.}~\bibnamefont {Prasai}}, \bibinfo {author} {\bibfnamefont
					{H.}~\bibnamefont {Krzyzanowska}}, \bibinfo {author} {\bibfnamefont
					{J.}~\bibnamefont {Lin}}, \bibinfo {author} {\bibfnamefont {D.}~\bibnamefont
					{Caudel}}, \bibinfo {author} {\bibfnamefont {N.~J.}\ \bibnamefont {Ghimire}},
				\bibinfo {author} {\bibfnamefont {J.}~\bibnamefont {Yan}}, \bibinfo {author}
				{\bibfnamefont {B.~L.}\ \bibnamefont {Ivanov}}, \bibinfo {author}
				{\bibfnamefont {K.~A.}\ \bibnamefont {Velizhanin}}, \bibinfo {author}
				{\bibfnamefont {A.}~\bibnamefont {Burger}}, \bibinfo {author} {\bibfnamefont
					{D.~G.}\ \bibnamefont {Mandrus}}, \bibinfo {author} {\bibfnamefont {N.~H.}\
					\bibnamefont {Tolk}}, \bibinfo {author} {\bibfnamefont {S.~T.}\ \bibnamefont
					{Pantelides}},\ and\ \bibinfo {author} {\bibfnamefont {K.~I.}\ \bibnamefont
					{Bolotin}},\ }\bibfield  {title} {\bibinfo {title} {{Probing excitonic states
						in suspended two-dimensional semiconductors by photocurrent spectroscopy}},\
			}\href {https://doi.org/10.1038/srep06608} {\bibfield  {journal} {\bibinfo
					{journal} {Scientific Reports}\ }\textbf {\bibinfo {volume} {4}},\ \bibinfo
				{pages} {6608} (\bibinfo {year} {2014})}\BibitemShut {NoStop}%
			\bibitem [{\citenamefont {Gulans}\ \emph {et~al.}(2018)\citenamefont {Gulans},
				\citenamefont {Kozhevnikov},\ and\ \citenamefont {Draxl}}]{Gulans_2018}%
			\BibitemOpen
			\bibfield  {author} {\bibinfo {author} {\bibfnamefont {A.}~\bibnamefont
					{Gulans}}, \bibinfo {author} {\bibfnamefont {A.}~\bibnamefont
					{Kozhevnikov}},\ and\ \bibinfo {author} {\bibfnamefont {C.}~\bibnamefont
					{Draxl}},\ }\bibfield  {title} {\bibinfo {title} {{Microhartree precision in
						density functional theory calculations}},\ }\bibfield  {journal} {\bibinfo
				{journal} {Phys.~Rev.~B}\ }\textbf {\bibinfo {volume} {97}},\ \href
			{https://doi.org/10.1103/physrevb.97.161105} {10.1103/physrevb.97.161105}
			(\bibinfo {year} {2018})\BibitemShut {NoStop}%
			\bibitem [{\citenamefont {Nabok}\ \emph {et~al.}(2016)\citenamefont {Nabok},
				\citenamefont {Gulans},\ and\ \citenamefont {Draxl}}]{Nabok_2016}%
			\BibitemOpen
			\bibfield  {author} {\bibinfo {author} {\bibfnamefont {D.}~\bibnamefont
					{Nabok}}, \bibinfo {author} {\bibfnamefont {A.}~\bibnamefont {Gulans}},\ and\
				\bibinfo {author} {\bibfnamefont {C.}~\bibnamefont {Draxl}},\ }\bibfield
			{title} {\bibinfo {title} {{Accurate all-electron $G_0W_0$ quasiparticle
						energies employing the full-potential augmented planewave method}},\ }\href
			{https://doi.org/10.1103/physrevb.94.035118} {\bibfield  {journal} {\bibinfo
					{journal} {Phys.~Rev.~B}\ }\textbf {\bibinfo {volume} {94}},\ \bibinfo
				{pages} {035118} (\bibinfo {year} {2016})},\ \Eprint
			{https://arxiv.org/abs/1605.07351} {1605.07351} \BibitemShut {NoStop}%
			\bibitem [{\citenamefont {Perdew}\ and\ \citenamefont
				{Wang}(1992)}]{Perdew_1992}%
			\BibitemOpen
			\bibfield  {author} {\bibinfo {author} {\bibfnamefont {J.~P.}\ \bibnamefont
					{Perdew}}\ and\ \bibinfo {author} {\bibfnamefont {Y.}~\bibnamefont {Wang}},\
			}\bibfield  {title} {\bibinfo {title} {Accurate and simple analytic
					representation of the electron-gas correlation energy},\ }\href
			{https://doi.org/10.1103/PhysRevB.45.13244} {\bibfield  {journal} {\bibinfo
					{journal} {Phys. Rev. B}\ }\textbf {\bibinfo {volume} {45}},\ \bibinfo
				{pages} {13244} (\bibinfo {year} {1992})}\BibitemShut {NoStop}%
			\bibitem [{\citenamefont {Perdew}\ \emph {et~al.}(1996)\citenamefont {Perdew},
				\citenamefont {Burke},\ and\ \citenamefont {Ernzerhof}}]{Perdew_1996}%
			\BibitemOpen
			\bibfield  {author} {\bibinfo {author} {\bibfnamefont {J.~P.}\ \bibnamefont
					{Perdew}}, \bibinfo {author} {\bibfnamefont {K.}~\bibnamefont {Burke}},\ and\
				\bibinfo {author} {\bibfnamefont {M.}~\bibnamefont {Ernzerhof}},\ }\bibfield
			{title} {\bibinfo {title} {{Generalized Gradient Approximation Made
						Simple}},\ }\href {https://doi.org/10.1103/PhysRevLett.77.3865} {\bibfield
				{journal} {\bibinfo  {journal} {Phys. Rev. Lett.}\ }\textbf {\bibinfo
					{volume} {77}},\ \bibinfo {pages} {3865} (\bibinfo {year}
				{1996})}\BibitemShut {NoStop}%
			\bibitem [{\citenamefont {Perdew}\ \emph {et~al.}(1997)\citenamefont {Perdew},
				\citenamefont {Burke},\ and\ \citenamefont {Ernzerhof}}]{Perdew_1997_E}%
			\BibitemOpen
			\bibfield  {author} {\bibinfo {author} {\bibfnamefont {J.~P.}\ \bibnamefont
					{Perdew}}, \bibinfo {author} {\bibfnamefont {K.}~\bibnamefont {Burke}},\ and\
				\bibinfo {author} {\bibfnamefont {M.}~\bibnamefont {Ernzerhof}},\ }\bibfield
			{title} {\bibinfo {title} {Generalized gradient approximation made simple
					[phys. rev. lett. 77, 3865 (1996)]},\ }\href
			{https://doi.org/10.1103/PhysRevLett.78.1396} {\bibfield  {journal} {\bibinfo
					{journal} {Phys. Rev. Lett.}\ }\textbf {\bibinfo {volume} {78}},\ \bibinfo
				{pages} {1396} (\bibinfo {year} {1997})}\BibitemShut {NoStop}%
			\bibitem [{\citenamefont {Heyd}\ \emph {et~al.}(2003)\citenamefont {Heyd},
				\citenamefont {Scuseria},\ and\ \citenamefont {Ernzerhof}}]{Heyd_2003}%
			\BibitemOpen
			\bibfield  {author} {\bibinfo {author} {\bibfnamefont {J.}~\bibnamefont
					{Heyd}}, \bibinfo {author} {\bibfnamefont {G.~E.}\ \bibnamefont {Scuseria}},\
				and\ \bibinfo {author} {\bibfnamefont {M.}~\bibnamefont {Ernzerhof}},\
			}\bibfield  {title} {\bibinfo {title} {{Hybrid functionals based on a
						screened Coulomb potential}},\ }\href {https://doi.org/10.1063/1.1564060}
			{\bibfield  {journal} {\bibinfo  {journal} {The Journal of Chemical Physics}\
				}\textbf {\bibinfo {volume} {118}},\ \bibinfo {pages} {8207} (\bibinfo {year}
				{2003})}\BibitemShut {NoStop}%
			\bibitem [{\citenamefont {Heyd}\ \emph {et~al.}(2006)\citenamefont {Heyd},
				\citenamefont {Scuseria},\ and\ \citenamefont {Ernzerhof}}]{Heyd_2006_E}%
			\BibitemOpen
			\bibfield  {author} {\bibinfo {author} {\bibfnamefont {J.}~\bibnamefont
					{Heyd}}, \bibinfo {author} {\bibfnamefont {G.~E.}\ \bibnamefont {Scuseria}},\
				and\ \bibinfo {author} {\bibfnamefont {M.}~\bibnamefont {Ernzerhof}},\
			}\bibfield  {title} {\bibinfo {title} {{Erratum: “Hybrid functionals based
						on a screened Coulomb potential” [J. Chem. Phys. 118, 8207 (2003)]}},\
			}\href {https://doi.org/10.1063/1.2204597} {\bibfield  {journal} {\bibinfo
					{journal} {The Journal of Chemical Physics}\ }\textbf {\bibinfo {volume}
					{124}},\ \bibinfo {pages} {219906} (\bibinfo {year} {2006})}\BibitemShut
			{NoStop}%
			\bibitem [{\citenamefont {Krukau}\ \emph {et~al.}(2006)\citenamefont {Krukau},
				\citenamefont {Vydrov}, \citenamefont {Izmaylov},\ and\ \citenamefont
				{Scuseria}}]{Krukau_2006}%
			\BibitemOpen
			\bibfield  {author} {\bibinfo {author} {\bibfnamefont {A.~V.}\ \bibnamefont
					{Krukau}}, \bibinfo {author} {\bibfnamefont {O.~A.}\ \bibnamefont {Vydrov}},
				\bibinfo {author} {\bibfnamefont {A.~F.}\ \bibnamefont {Izmaylov}},\ and\
				\bibinfo {author} {\bibfnamefont {G.~E.}\ \bibnamefont {Scuseria}},\
			}\bibfield  {title} {\bibinfo {title} {{Influence of the exchange screening
						parameter on the performance of screened hybrid functionals}},\ }\href
			{https://doi.org/10.1063/1.2404663} {\bibfield  {journal} {\bibinfo
					{journal} {The Journal of Chemical Physics}\ }\textbf {\bibinfo {volume}
					{125}},\ \bibinfo {pages} {224106} (\bibinfo {year} {2006})}\BibitemShut
			{NoStop}%
			\bibitem [{\citenamefont {Ismail-Beigi}(2006)}]{Ismail-Beigi_2006}%
			\BibitemOpen
			\bibfield  {author} {\bibinfo {author} {\bibfnamefont {S.}~\bibnamefont
					{Ismail-Beigi}},\ }\bibfield  {title} {\bibinfo {title} {{Truncation of
						periodic image interactions for confined systems}},\ }\href
			{https://doi.org/10.1103/physrevb.73.233103} {\bibfield  {journal} {\bibinfo
					{journal} {Phys.~Rev.~B}\ }\textbf {\bibinfo {volume} {73}},\ \bibinfo
				{pages} {233103} (\bibinfo {year} {2006})}\BibitemShut {NoStop}%
			\bibitem [{\citenamefont {Gulans}\ \emph {et~al.}(2014)\citenamefont {Gulans},
				\citenamefont {Kontur}, \citenamefont {Meisenbichler}, \citenamefont {Nabok},
				\citenamefont {Pavone}, \citenamefont {Rigamonti}, \citenamefont
				{Sagmeister}, \citenamefont {Werner},\ and\ \citenamefont
				{Draxl}}]{Gulans_2014}%
			\BibitemOpen
			\bibfield  {author} {\bibinfo {author} {\bibfnamefont {A.}~\bibnamefont
					{Gulans}}, \bibinfo {author} {\bibfnamefont {S.}~\bibnamefont {Kontur}},
				\bibinfo {author} {\bibfnamefont {C.}~\bibnamefont {Meisenbichler}}, \bibinfo
				{author} {\bibfnamefont {D.}~\bibnamefont {Nabok}}, \bibinfo {author}
				{\bibfnamefont {P.}~\bibnamefont {Pavone}}, \bibinfo {author} {\bibfnamefont
					{S.}~\bibnamefont {Rigamonti}}, \bibinfo {author} {\bibfnamefont
					{S.}~\bibnamefont {Sagmeister}}, \bibinfo {author} {\bibfnamefont
					{U.}~\bibnamefont {Werner}},\ and\ \bibinfo {author} {\bibfnamefont
					{C.}~\bibnamefont {Draxl}},\ }\bibfield  {title} {\bibinfo {title}
				{{exciting: a full-potential all-electron package implementing
						density-functional theory and many-body perturbation theory}},\ }\href
			{https://doi.org/10.1088/0953-8984/26/36/363202} {\bibfield  {journal}
				{\bibinfo  {journal} {J.~Phys.~Condens.~Matter.~}\ }\textbf {\bibinfo
					{volume} {26}},\ \bibinfo {pages} {363202} (\bibinfo {year}
				{2014})}\BibitemShut {NoStop}%
			\bibitem [{\citenamefont {Sj{\"o}stedt}\ \emph {et~al.}(2000)\citenamefont
				{Sj{\"o}stedt}, \citenamefont {Nordstr{\"o}m},\ and\ \citenamefont
				{Singh}}]{sjostedt2000alternative}%
			\BibitemOpen
			\bibfield  {author} {\bibinfo {author} {\bibfnamefont {E.}~\bibnamefont
					{Sj{\"o}stedt}}, \bibinfo {author} {\bibfnamefont {L.}~\bibnamefont
					{Nordstr{\"o}m}},\ and\ \bibinfo {author} {\bibfnamefont {D.}~\bibnamefont
					{Singh}},\ }\bibfield  {title} {\bibinfo {title} {An alternative way of
					linearizing the augmented plane-wave method},\ }\href@noop {} {\bibfield
				{journal} {\bibinfo  {journal} {Solid state communications}\ }\textbf
				{\bibinfo {volume} {114}},\ \bibinfo {pages} {15} (\bibinfo {year}
				{2000})}\BibitemShut {NoStop}%
			\bibitem [{\citenamefont {Lejaeghere}\ \emph {et~al.}(2016)\citenamefont
				{Lejaeghere}, \citenamefont {Bihlmayer}, \citenamefont {Björkman},
				\citenamefont {Blaha}, \citenamefont {Blügel}, \citenamefont {Blum},
				\citenamefont {Caliste}, \citenamefont {Castelli}, \citenamefont {Clark},
				\citenamefont {Corso}, \citenamefont {Gironcoli}, \citenamefont {Deutsch},
				\citenamefont {Dewhurst}, \citenamefont {Marco}, \citenamefont {Draxl},
				\citenamefont {Dułak}, \citenamefont {Eriksson}, \citenamefont
				{Flores-Livas}, \citenamefont {Garrity}, \citenamefont {Genovese},
				\citenamefont {Giannozzi}, \citenamefont {Giantomassi}, \citenamefont
				{Goedecker}, \citenamefont {Gonze}, \citenamefont {Grånäs}, \citenamefont
				{Gross}, \citenamefont {Gulans}, \citenamefont {Gygi}, \citenamefont
				{Hamann}, \citenamefont {Hasnip}, \citenamefont {Holzwarth}, \citenamefont
				{Iuşan}, \citenamefont {Jochym}, \citenamefont {Jollet}, \citenamefont
				{Jones}, \citenamefont {Kresse}, \citenamefont {Koepernik}, \citenamefont
				{Küçükbenli}, \citenamefont {Kvashnin}, \citenamefont {Locht},
				\citenamefont {Lubeck}, \citenamefont {Marsman}, \citenamefont {Marzari},
				\citenamefont {Nitzsche}, \citenamefont {Nordström}, \citenamefont {Ozaki},
				\citenamefont {Paulatto}, \citenamefont {Pickard}, \citenamefont {Poelmans},
				\citenamefont {Probert}, \citenamefont {Refson}, \citenamefont {Richter},
				\citenamefont {Rignanese}, \citenamefont {Saha}, \citenamefont {Scheffler},
				\citenamefont {Schlipf}, \citenamefont {Schwarz}, \citenamefont {Sharma},
				\citenamefont {Tavazza}, \citenamefont {Thunström}, \citenamefont
				{Tkatchenko}, \citenamefont {Torrent}, \citenamefont {Vanderbilt},
				\citenamefont {Setten}, \citenamefont {Speybroeck}, \citenamefont {Wills},
				\citenamefont {Yates}, \citenamefont {Zhang},\ and\ \citenamefont
				{Cottenier}}]{Lejaeghere_2016}%
			\BibitemOpen
			\bibfield  {author} {\bibinfo {author} {\bibfnamefont {K.}~\bibnamefont
					{Lejaeghere}}, \bibinfo {author} {\bibfnamefont {G.}~\bibnamefont
					{Bihlmayer}}, \bibinfo {author} {\bibfnamefont {T.}~\bibnamefont
					{Björkman}}, \bibinfo {author} {\bibfnamefont {P.}~\bibnamefont {Blaha}},
				\bibinfo {author} {\bibfnamefont {S.}~\bibnamefont {Blügel}}, \bibinfo
				{author} {\bibfnamefont {V.}~\bibnamefont {Blum}}, \bibinfo {author}
				{\bibfnamefont {D.}~\bibnamefont {Caliste}}, \bibinfo {author} {\bibfnamefont
					{I.~E.}\ \bibnamefont {Castelli}}, \bibinfo {author} {\bibfnamefont {S.~J.}\
					\bibnamefont {Clark}}, \bibinfo {author} {\bibfnamefont {A.~D.}\ \bibnamefont
					{Corso}}, \bibinfo {author} {\bibfnamefont {S.~d.}\ \bibnamefont
					{Gironcoli}}, \bibinfo {author} {\bibfnamefont {T.}~\bibnamefont {Deutsch}},
				\bibinfo {author} {\bibfnamefont {J.~K.}\ \bibnamefont {Dewhurst}}, \bibinfo
				{author} {\bibfnamefont {I.~D.}\ \bibnamefont {Marco}}, \bibinfo {author}
				{\bibfnamefont {C.}~\bibnamefont {Draxl}}, \bibinfo {author} {\bibfnamefont
					{M.}~\bibnamefont {Dułak}}, \bibinfo {author} {\bibfnamefont
					{O.}~\bibnamefont {Eriksson}}, \bibinfo {author} {\bibfnamefont {J.~A.}\
					\bibnamefont {Flores-Livas}}, \bibinfo {author} {\bibfnamefont {K.~F.}\
					\bibnamefont {Garrity}}, \bibinfo {author} {\bibfnamefont {L.}~\bibnamefont
					{Genovese}}, \bibinfo {author} {\bibfnamefont {P.}~\bibnamefont {Giannozzi}},
				\bibinfo {author} {\bibfnamefont {M.}~\bibnamefont {Giantomassi}}, \bibinfo
				{author} {\bibfnamefont {S.}~\bibnamefont {Goedecker}}, \bibinfo {author}
				{\bibfnamefont {X.}~\bibnamefont {Gonze}}, \bibinfo {author} {\bibfnamefont
					{O.}~\bibnamefont {Grånäs}}, \bibinfo {author} {\bibfnamefont {E.~K.~U.}\
					\bibnamefont {Gross}}, \bibinfo {author} {\bibfnamefont {A.}~\bibnamefont
					{Gulans}}, \bibinfo {author} {\bibfnamefont {F.}~\bibnamefont {Gygi}},
				\bibinfo {author} {\bibfnamefont {D.~R.}\ \bibnamefont {Hamann}}, \bibinfo
				{author} {\bibfnamefont {P.~J.}\ \bibnamefont {Hasnip}}, \bibinfo {author}
				{\bibfnamefont {N.~A.~W.}\ \bibnamefont {Holzwarth}}, \bibinfo {author}
				{\bibfnamefont {D.}~\bibnamefont {Iuşan}}, \bibinfo {author} {\bibfnamefont
					{D.~B.}\ \bibnamefont {Jochym}}, \bibinfo {author} {\bibfnamefont
					{F.}~\bibnamefont {Jollet}}, \bibinfo {author} {\bibfnamefont
					{D.}~\bibnamefont {Jones}}, \bibinfo {author} {\bibfnamefont
					{G.}~\bibnamefont {Kresse}}, \bibinfo {author} {\bibfnamefont
					{K.}~\bibnamefont {Koepernik}}, \bibinfo {author} {\bibfnamefont
					{E.}~\bibnamefont {Küçükbenli}}, \bibinfo {author} {\bibfnamefont {Y.~O.}\
					\bibnamefont {Kvashnin}}, \bibinfo {author} {\bibfnamefont {I.~L.~M.}\
					\bibnamefont {Locht}}, \bibinfo {author} {\bibfnamefont {S.}~\bibnamefont
					{Lubeck}}, \bibinfo {author} {\bibfnamefont {M.}~\bibnamefont {Marsman}},
				\bibinfo {author} {\bibfnamefont {N.}~\bibnamefont {Marzari}}, \bibinfo
				{author} {\bibfnamefont {U.}~\bibnamefont {Nitzsche}}, \bibinfo {author}
				{\bibfnamefont {L.}~\bibnamefont {Nordström}}, \bibinfo {author}
				{\bibfnamefont {T.}~\bibnamefont {Ozaki}}, \bibinfo {author} {\bibfnamefont
					{L.}~\bibnamefont {Paulatto}}, \bibinfo {author} {\bibfnamefont {C.~J.}\
					\bibnamefont {Pickard}}, \bibinfo {author} {\bibfnamefont {W.}~\bibnamefont
					{Poelmans}}, \bibinfo {author} {\bibfnamefont {M.~I.~J.}\ \bibnamefont
					{Probert}}, \bibinfo {author} {\bibfnamefont {K.}~\bibnamefont {Refson}},
				\bibinfo {author} {\bibfnamefont {M.}~\bibnamefont {Richter}}, \bibinfo
				{author} {\bibfnamefont {G.-M.}\ \bibnamefont {Rignanese}}, \bibinfo {author}
				{\bibfnamefont {S.}~\bibnamefont {Saha}}, \bibinfo {author} {\bibfnamefont
					{M.}~\bibnamefont {Scheffler}}, \bibinfo {author} {\bibfnamefont
					{M.}~\bibnamefont {Schlipf}}, \bibinfo {author} {\bibfnamefont
					{K.}~\bibnamefont {Schwarz}}, \bibinfo {author} {\bibfnamefont
					{S.}~\bibnamefont {Sharma}}, \bibinfo {author} {\bibfnamefont
					{F.}~\bibnamefont {Tavazza}}, \bibinfo {author} {\bibfnamefont
					{P.}~\bibnamefont {Thunström}}, \bibinfo {author} {\bibfnamefont
					{A.}~\bibnamefont {Tkatchenko}}, \bibinfo {author} {\bibfnamefont
					{M.}~\bibnamefont {Torrent}}, \bibinfo {author} {\bibfnamefont
					{D.}~\bibnamefont {Vanderbilt}}, \bibinfo {author} {\bibfnamefont {M.~J.~v.}\
					\bibnamefont {Setten}}, \bibinfo {author} {\bibfnamefont {V.~V.}\
					\bibnamefont {Speybroeck}}, \bibinfo {author} {\bibfnamefont {J.~M.}\
					\bibnamefont {Wills}}, \bibinfo {author} {\bibfnamefont {J.~R.}\ \bibnamefont
					{Yates}}, \bibinfo {author} {\bibfnamefont {G.-X.}\ \bibnamefont {Zhang}},\
				and\ \bibinfo {author} {\bibfnamefont {S.}~\bibnamefont {Cottenier}},\
			}\bibfield  {title} {\bibinfo {title} {{Reproducibility in density functional
						theory calculations of solids}},\ }\href
			{https://doi.org/10.1126/science.aad3000} {\bibfield  {journal} {\bibinfo
					{journal} {Science}\ }\textbf {\bibinfo {volume} {351}},\ \bibinfo {pages}
				{aad3000} (\bibinfo {year} {2016})}\BibitemShut {NoStop}%
			\bibitem [{\citenamefont {Aryasetiawan}\ and\ \citenamefont
				{Gunnarsson}(1998)}]{Aryasetiawan_1998}%
			\BibitemOpen
			\bibfield  {author} {\bibinfo {author} {\bibfnamefont {F.}~\bibnamefont
					{Aryasetiawan}}\ and\ \bibinfo {author} {\bibfnamefont {O.}~\bibnamefont
					{Gunnarsson}},\ }\bibfield  {title} {\bibinfo {title} {{The GW method}},\
			}\href {https://doi.org/10.1088/0034-4885/61/3/002} {\bibfield  {journal}
				{\bibinfo  {journal} {Reports on Progress in Physics}\ }\textbf {\bibinfo
					{volume} {61}},\ \bibinfo {pages} {237} (\bibinfo {year} {1998})}\BibitemShut
			{NoStop}%
			\bibitem [{\citenamefont {Kotani}\ \emph {et~al.}(2007)\citenamefont {Kotani},
				\citenamefont {van Schilfgaarde},\ and\ \citenamefont
				{Faleev}}]{Kotani_2007}%
			\BibitemOpen
			\bibfield  {author} {\bibinfo {author} {\bibfnamefont {T.}~\bibnamefont
					{Kotani}}, \bibinfo {author} {\bibfnamefont {M.}~\bibnamefont {van
						Schilfgaarde}},\ and\ \bibinfo {author} {\bibfnamefont {S.~V.}\ \bibnamefont
					{Faleev}},\ }\bibfield  {title} {\bibinfo {title} {{Quasiparticle
						self-consistent $GW$ method: A basis for the independent-particle
						approximation}},\ }\href {https://doi.org/10.1103/PhysRevB.76.165106}
			{\bibfield  {journal} {\bibinfo  {journal} {Phys.~Rev.~B}\ }\textbf {\bibinfo
					{volume} {76}},\ \bibinfo {pages} {165106} (\bibinfo {year}
				{2007})}\BibitemShut {NoStop}%
			\bibitem [{\citenamefont {Leppert}\ \emph {et~al.}(2019)\citenamefont
				{Leppert}, \citenamefont {Rangel},\ and\ \citenamefont
				{Neaton}}]{Leppert_2019}%
			\BibitemOpen
			\bibfield  {author} {\bibinfo {author} {\bibfnamefont {L.}~\bibnamefont
					{Leppert}}, \bibinfo {author} {\bibfnamefont {T.}~\bibnamefont {Rangel}},\
				and\ \bibinfo {author} {\bibfnamefont {J.~B.}\ \bibnamefont {Neaton}},\
			}\bibfield  {title} {\bibinfo {title} {{Towards predictive band gaps for
						halide perovskites: Lessons from one-shot and eigenvalue self-consistent
						$GW$}},\ }\href {https://doi.org/10.1103/PhysRevMaterials.3.103803}
			{\bibfield  {journal} {\bibinfo  {journal} {Phys. Rev. Mater.}\ }\textbf
				{\bibinfo {volume} {3}},\ \bibinfo {pages} {103803} (\bibinfo {year}
				{2019})}\BibitemShut {NoStop}%
			\bibitem [{\citenamefont {Rinke}\ \emph {et~al.}(2005)\citenamefont {Rinke},
				\citenamefont {Qteish}, \citenamefont {Neugebauer}, \citenamefont
				{Freysoldt},\ and\ \citenamefont {Scheffler}}]{Rinke_2005}%
			\BibitemOpen
			\bibfield  {author} {\bibinfo {author} {\bibfnamefont {P.}~\bibnamefont
					{Rinke}}, \bibinfo {author} {\bibfnamefont {A.}~\bibnamefont {Qteish}},
				\bibinfo {author} {\bibfnamefont {J.}~\bibnamefont {Neugebauer}}, \bibinfo
				{author} {\bibfnamefont {C.}~\bibnamefont {Freysoldt}},\ and\ \bibinfo
				{author} {\bibfnamefont {M.}~\bibnamefont {Scheffler}},\ }\bibfield  {title}
			{\bibinfo {title} {{Combining GW calculations with exact-exchange
						density-functional theory: an analysis of valence-band photoemission for
						compound semiconductors}},\ }\href
			{https://doi.org/10.1088/1367-2630/7/1/126} {\bibfield  {journal} {\bibinfo
					{journal} {New Journal of Physics}\ }\textbf {\bibinfo {volume} {7}},\
				\bibinfo {pages} {126} (\bibinfo {year} {2005})}\BibitemShut {NoStop}%
			\bibitem [{\citenamefont {Atalla}\ \emph {et~al.}(2013)\citenamefont {Atalla},
				\citenamefont {Yoon}, \citenamefont {Caruso}, \citenamefont {Rinke},\ and\
				\citenamefont {Scheffler}}]{Atalla_2013}%
			\BibitemOpen
			\bibfield  {author} {\bibinfo {author} {\bibfnamefont {V.}~\bibnamefont
					{Atalla}}, \bibinfo {author} {\bibfnamefont {M.}~\bibnamefont {Yoon}},
				\bibinfo {author} {\bibfnamefont {F.}~\bibnamefont {Caruso}}, \bibinfo
				{author} {\bibfnamefont {P.}~\bibnamefont {Rinke}},\ and\ \bibinfo {author}
				{\bibfnamefont {M.}~\bibnamefont {Scheffler}},\ }\bibfield  {title} {\bibinfo
				{title} {Hybrid density functional theory meets quasiparticle calculations: A
					consistent electronic structure approach},\ }\href
			{https://doi.org/10.1103/PhysRevB.88.165122} {\bibfield  {journal} {\bibinfo
					{journal} {Phys.~Rev.~B}\ }\textbf {\bibinfo {volume} {88}},\ \bibinfo
				{pages} {165122} (\bibinfo {year} {2013})}\BibitemShut {NoStop}%
			\bibitem [{\citenamefont {Bruneval}\ and\ \citenamefont
				{Marques}(2013)}]{Bruneval_2013}%
			\BibitemOpen
			\bibfield  {author} {\bibinfo {author} {\bibfnamefont {F.}~\bibnamefont
					{Bruneval}}\ and\ \bibinfo {author} {\bibfnamefont {M.~A.}\ \bibnamefont
					{Marques}},\ }\bibfield  {title} {\bibinfo {title} {{Benchmarking the
						starting points of the $GW$ approximation for molecules}},\ }\href@noop {}
			{\bibfield  {journal} {\bibinfo  {journal} {Journal of chemical theory and
						computation}\ }\textbf {\bibinfo {volume} {9}},\ \bibinfo {pages} {324}
				(\bibinfo {year} {2013})}\BibitemShut {NoStop}%
			\bibitem [{\citenamefont {Marom}\ \emph {et~al.}(2012)\citenamefont {Marom},
				\citenamefont {Caruso}, \citenamefont {Ren}, \citenamefont {Hofmann},
				\citenamefont {K\"orzd\"orfer}, \citenamefont {Chelikowsky}, \citenamefont
				{Rubio}, \citenamefont {Scheffler},\ and\ \citenamefont
				{Rinke}}]{Marom_2012}%
			\BibitemOpen
			\bibfield  {author} {\bibinfo {author} {\bibfnamefont {N.}~\bibnamefont
					{Marom}}, \bibinfo {author} {\bibfnamefont {F.}~\bibnamefont {Caruso}},
				\bibinfo {author} {\bibfnamefont {X.}~\bibnamefont {Ren}}, \bibinfo {author}
				{\bibfnamefont {O.~T.}\ \bibnamefont {Hofmann}}, \bibinfo {author}
				{\bibfnamefont {T.}~\bibnamefont {K\"orzd\"orfer}}, \bibinfo {author}
				{\bibfnamefont {J.~R.}\ \bibnamefont {Chelikowsky}}, \bibinfo {author}
				{\bibfnamefont {A.}~\bibnamefont {Rubio}}, \bibinfo {author} {\bibfnamefont
					{M.}~\bibnamefont {Scheffler}},\ and\ \bibinfo {author} {\bibfnamefont
					{P.}~\bibnamefont {Rinke}},\ }\bibfield  {title} {\bibinfo {title}
				{{Benchmark of $GW$ methods for azabenzenes}},\ }\href
			{https://doi.org/10.1103/PhysRevB.86.245127} {\bibfield  {journal} {\bibinfo
					{journal} {Phys.~Rev.~B}\ }\textbf {\bibinfo {volume} {86}},\ \bibinfo
				{pages} {245127} (\bibinfo {year} {2012})}\BibitemShut {NoStop}%
			\bibitem [{\citenamefont {Ren}\ \emph {et~al.}(2009)\citenamefont {Ren},
				\citenamefont {Rinke},\ and\ \citenamefont {Scheffler}}]{Ren_2009}%
			\BibitemOpen
			\bibfield  {author} {\bibinfo {author} {\bibfnamefont {X.}~\bibnamefont
					{Ren}}, \bibinfo {author} {\bibfnamefont {P.}~\bibnamefont {Rinke}},\ and\
				\bibinfo {author} {\bibfnamefont {M.}~\bibnamefont {Scheffler}},\ }\bibfield
			{title} {\bibinfo {title} {{Exploring the random phase approximation:
						Application to CO adsorbed on Cu(111)}},\ }\href
			{https://doi.org/10.1103/PhysRevB.80.045402} {\bibfield  {journal} {\bibinfo
					{journal} {Phys. Rev. B}\ }\textbf {\bibinfo {volume} {80}},\ \bibinfo
				{pages} {045402} (\bibinfo {year} {2009})}\BibitemShut {NoStop}%
			\bibitem [{\citenamefont {Jiang}\ \emph {et~al.}(2013)\citenamefont {Jiang},
				\citenamefont {Gómez-Abal}, \citenamefont {Li}, \citenamefont
				{Meisenbichler}, \citenamefont {Ambrosch-Draxl},\ and\ \citenamefont
				{Scheffler}}]{Jiang_2013}%
			\BibitemOpen
			\bibfield  {author} {\bibinfo {author} {\bibfnamefont {H.}~\bibnamefont
					{Jiang}}, \bibinfo {author} {\bibfnamefont {R.~I.}\ \bibnamefont
					{Gómez-Abal}}, \bibinfo {author} {\bibfnamefont {X.-Z.}\ \bibnamefont {Li}},
				\bibinfo {author} {\bibfnamefont {C.}~\bibnamefont {Meisenbichler}}, \bibinfo
				{author} {\bibfnamefont {C.}~\bibnamefont {Ambrosch-Draxl}},\ and\ \bibinfo
				{author} {\bibfnamefont {M.}~\bibnamefont {Scheffler}},\ }\bibfield  {title}
			{\bibinfo {title} {{FHI-gap: A GW code based on the all-electron augmented
						plane wave method}},\ }\href {https://doi.org/10.1016/j.cpc.2012.09.018}
			{\bibfield  {journal} {\bibinfo  {journal} {Computer Physics Communications}\
				}\textbf {\bibinfo {volume} {184}},\ \bibinfo {pages} {348} (\bibinfo {year}
				{2013})}\BibitemShut {NoStop}%
			\bibitem [{\citenamefont {Aguilera}\ \emph {et~al.}(2013)\citenamefont
				{Aguilera}, \citenamefont {Friedrich},\ and\ \citenamefont
				{Bl\"ugel}}]{Aguilera_2013}%
			\BibitemOpen
			\bibfield  {author} {\bibinfo {author} {\bibfnamefont {I.}~\bibnamefont
					{Aguilera}}, \bibinfo {author} {\bibfnamefont {C.}~\bibnamefont
					{Friedrich}},\ and\ \bibinfo {author} {\bibfnamefont {S.}~\bibnamefont
					{Bl\"ugel}},\ }\bibfield  {title} {\bibinfo {title} {Spin-orbit coupling in
					quasiparticle studies of topological insulators},\ }\href
			{https://doi.org/10.1103/PhysRevB.88.165136} {\bibfield  {journal} {\bibinfo
					{journal} {Phys. Rev. B}\ }\textbf {\bibinfo {volume} {88}},\ \bibinfo
				{pages} {165136} (\bibinfo {year} {2013})}\BibitemShut {NoStop}%
			\bibitem [{\citenamefont {Singh}(1994)}]{singh_1994}%
			\BibitemOpen
			\bibfield  {author} {\bibinfo {author} {\bibfnamefont {D.~J.}\ \bibnamefont
					{Singh}},\ }\href@noop {} {\emph {\bibinfo {title} {Planes Waves,
						Pseudopotentials and the ...}}}\ (\bibinfo  {publisher} {Method, Kluwer
				Academic},\ \bibinfo {year} {1994})\BibitemShut {NoStop}%
			\bibitem [{\citenamefont {MacDonald}\ \emph {et~al.}(1980)\citenamefont
				{MacDonald}, \citenamefont {Picket},\ and\ \citenamefont
				{Koelling}}]{macdonald1980linearised}%
			\BibitemOpen
			\bibfield  {author} {\bibinfo {author} {\bibfnamefont {A.}~\bibnamefont
					{MacDonald}}, \bibinfo {author} {\bibfnamefont {W.}~\bibnamefont {Picket}},\
				and\ \bibinfo {author} {\bibfnamefont {D.}~\bibnamefont {Koelling}},\
			}\bibfield  {title} {\bibinfo {title} {A linearised relativistic
					augmented-plane-wave method utilising approximate pure spin basis
					functions},\ }\href@noop {} {\bibfield  {journal} {\bibinfo  {journal}
					{Journal of Physics C: Solid State Physics}\ }\textbf {\bibinfo {volume}
					{13}},\ \bibinfo {pages} {2675} (\bibinfo {year} {1980})}\BibitemShut
			{NoStop}%
			\bibitem [{\citenamefont {Li}\ \emph {et~al.}(1990)\citenamefont {Li},
				\citenamefont {Freeman}, \citenamefont {Jansen},\ and\ \citenamefont
				{Fu}}]{li1990magnetic}%
			\BibitemOpen
			\bibfield  {author} {\bibinfo {author} {\bibfnamefont {C.}~\bibnamefont
					{Li}}, \bibinfo {author} {\bibfnamefont {A.~J.}\ \bibnamefont {Freeman}},
				\bibinfo {author} {\bibfnamefont {H.}~\bibnamefont {Jansen}},\ and\ \bibinfo
				{author} {\bibfnamefont {C.}~\bibnamefont {Fu}},\ }\bibfield  {title}
			{\bibinfo {title} {{Magnetic anisotropy in low-dimensional ferromagnetic
						systems: Fe monolayers on Ag (001), Au (001), and Pd (001) substrates}},\
			}\href@noop {} {\bibfield  {journal} {\bibinfo  {journal} {Physical Review
						B}\ }\textbf {\bibinfo {volume} {42}},\ \bibinfo {pages} {5433} (\bibinfo
				{year} {1990})}\BibitemShut {NoStop}%
			\bibitem [{\citenamefont {Vona}\ \emph {et~al.}(2023)\citenamefont {Vona},
				\citenamefont {Lubeck}, \citenamefont {Kleine}, \citenamefont {Gulans},\ and\
				\citenamefont {Draxl}}]{vona_2023}%
			\BibitemOpen
			\bibfield  {author} {\bibinfo {author} {\bibfnamefont {C.}~\bibnamefont
					{Vona}}, \bibinfo {author} {\bibfnamefont {S.}~\bibnamefont {Lubeck}},
				\bibinfo {author} {\bibfnamefont {H.}~\bibnamefont {Kleine}}, \bibinfo
				{author} {\bibfnamefont {A.}~\bibnamefont {Gulans}},\ and\ \bibinfo {author}
				{\bibfnamefont {C.}~\bibnamefont {Draxl}},\ }\href@noop {} {\bibinfo {title}
				{Accurate and efficient treatment of spin-orbit coupling via second variation
					employing local orbitals}} (\bibinfo {year} {2023}),\ \Eprint
			{https://arxiv.org/abs/2306.02965} {arXiv:2306.02965 [cond-mat.mtrl-sci]}
			\BibitemShut {NoStop}%
			\bibitem [{\citenamefont {Lenthe}\ \emph {et~al.}(1993)\citenamefont {Lenthe},
				\citenamefont {Baerends},\ and\ \citenamefont {Snijders}}]{Lenthe_1993}%
			\BibitemOpen
			\bibfield  {author} {\bibinfo {author} {\bibfnamefont {E.~v.}\ \bibnamefont
					{Lenthe}}, \bibinfo {author} {\bibfnamefont {E.~J.}\ \bibnamefont
					{Baerends}},\ and\ \bibinfo {author} {\bibfnamefont {J.~G.}\ \bibnamefont
					{Snijders}},\ }\bibfield  {title} {\bibinfo {title} {{Relativistic regular
						two‐component Hamiltonians}},\ }\href {https://doi.org/10.1063/1.466059}
			{\bibfield  {journal} {\bibinfo  {journal} {The Journal of Chemical Physics}\
				}\textbf {\bibinfo {volume} {99}},\ \bibinfo {pages} {4597} (\bibinfo {year}
				{1993})}\BibitemShut {NoStop}%
			\bibitem [{\citenamefont {van Lenthe}\ \emph {et~al.}(1994)\citenamefont {van
					Lenthe}, \citenamefont {Baerends},\ and\ \citenamefont
				{Snijders}}]{Lenthe_1994}%
			\BibitemOpen
			\bibfield  {author} {\bibinfo {author} {\bibfnamefont {E.}~\bibnamefont {van
						Lenthe}}, \bibinfo {author} {\bibfnamefont {E.~J.}\ \bibnamefont
					{Baerends}},\ and\ \bibinfo {author} {\bibfnamefont {J.~G.}\ \bibnamefont
					{Snijders}},\ }\bibfield  {title} {\bibinfo {title} {{Relativistic total
						energy using regular approximations}},\ }\href
			{https://doi.org/10.1063/1.467943} {\bibfield  {journal} {\bibinfo  {journal}
					{The Journal of Chemical Physics}\ }\textbf {\bibinfo {volume} {101}},\
				\bibinfo {pages} {9783} (\bibinfo {year} {1994})}\BibitemShut {NoStop}%
			\bibitem [{\citenamefont {Betzinger}\ \emph {et~al.}(2010)\citenamefont
				{Betzinger}, \citenamefont {Friedrich},\ and\ \citenamefont
				{Bl{\"u}gel}}]{betzinger2010hybrid}%
			\BibitemOpen
			\bibfield  {author} {\bibinfo {author} {\bibfnamefont {M.}~\bibnamefont
					{Betzinger}}, \bibinfo {author} {\bibfnamefont {C.}~\bibnamefont
					{Friedrich}},\ and\ \bibinfo {author} {\bibfnamefont {S.}~\bibnamefont
					{Bl{\"u}gel}},\ }\bibfield  {title} {\bibinfo {title} {{Hybrid functionals
						within the all-electron FLAPW method: implementation and applications of
						PBE0}},\ }\href@noop {} {\bibfield  {journal} {\bibinfo  {journal} {Physical
						Review B}\ }\textbf {\bibinfo {volume} {81}},\ \bibinfo {pages} {195117}
				(\bibinfo {year} {2010})}\BibitemShut {NoStop}%
			\bibitem [{\citenamefont {Huhn}\ and\ \citenamefont {Blum}(2017)}]{Huhn_2017}%
			\BibitemOpen
			\bibfield  {author} {\bibinfo {author} {\bibfnamefont {W.~P.}\ \bibnamefont
					{Huhn}}\ and\ \bibinfo {author} {\bibfnamefont {V.}~\bibnamefont {Blum}},\
			}\bibfield  {title} {\bibinfo {title} {One-hundred-three compound
					band-structure benchmark of post-self-consistent spin-orbit coupling
					treatments in density functional theory},\ }\href
			{https://doi.org/10.1103/PhysRevMaterials.1.033803} {\bibfield  {journal}
				{\bibinfo  {journal} {Phys. Rev. Mater.}\ }\textbf {\bibinfo {volume} {1}},\
				\bibinfo {pages} {033803} (\bibinfo {year} {2017})}\BibitemShut {NoStop}%
			\bibitem [{\citenamefont {Wang}\ \emph {et~al.}(2018)\citenamefont {Wang},
				\citenamefont {Liu}, \citenamefont {Guo},\ and\ \citenamefont
				{Yao}}]{Wang_2018}%
			\BibitemOpen
			\bibfield  {author} {\bibinfo {author} {\bibfnamefont {M.}~\bibnamefont
					{Wang}}, \bibinfo {author} {\bibfnamefont {G.-B.}\ \bibnamefont {Liu}},
				\bibinfo {author} {\bibfnamefont {H.}~\bibnamefont {Guo}},\ and\ \bibinfo
				{author} {\bibfnamefont {Y.}~\bibnamefont {Yao}},\ }\bibfield  {title}
			{\bibinfo {title} {An efficient method for hybrid density functional
					calculation with spin–orbit coupling},\ }\href
			{https://doi.org/https://doi.org/10.1016/j.cpc.2017.11.010} {\bibfield
				{journal} {\bibinfo  {journal} {Computer Physics Communications}\ }\textbf
				{\bibinfo {volume} {224}},\ \bibinfo {pages} {90} (\bibinfo {year}
				{2018})}\BibitemShut {NoStop}%
			\bibitem [{\citenamefont {Vona}\ \emph {et~al.}(2022)\citenamefont {Vona},
				\citenamefont {Nabok},\ and\ \citenamefont {Draxl}}]{Vona_2022}%
			\BibitemOpen
			\bibfield  {author} {\bibinfo {author} {\bibfnamefont {C.}~\bibnamefont
					{Vona}}, \bibinfo {author} {\bibfnamefont {D.}~\bibnamefont {Nabok}},\ and\
				\bibinfo {author} {\bibfnamefont {C.}~\bibnamefont {Draxl}},\ }\bibfield
			{title} {\bibinfo {title} {Electronic structure of (organic-)inorganic metal
					halide perovskites: The dilemma of choosing the right functional},\ }\href
			{https://doi.org/https://doi.org/10.1002/adts.202100496} {\bibfield
				{journal} {\bibinfo  {journal} {Advanced Theory and Simulations}\ }\textbf
				{\bibinfo {volume} {5}},\ \bibinfo {pages} {2100496} (\bibinfo {year}
				{2022})},\ \Eprint
			{https://arxiv.org/abs/https://onlinelibrary.wiley.com/doi/pdf/10.1002/adts.202100496}
			{https://onlinelibrary.wiley.com/doi/pdf/10.1002/adts.202100496} \BibitemShut
			{NoStop}%
			\bibitem [{\citenamefont {Blum}\ \emph {et~al.}(2009)\citenamefont {Blum},
				\citenamefont {Gehrke}, \citenamefont {Hanke}, \citenamefont {Havu},
				\citenamefont {Havu}, \citenamefont {Ren}, \citenamefont {Reuter},\ and\
				\citenamefont {Scheffler}}]{Blum2009}%
			\BibitemOpen
			\bibfield  {author} {\bibinfo {author} {\bibfnamefont {V.}~\bibnamefont
					{Blum}}, \bibinfo {author} {\bibfnamefont {R.}~\bibnamefont {Gehrke}},
				\bibinfo {author} {\bibfnamefont {F.}~\bibnamefont {Hanke}}, \bibinfo
				{author} {\bibfnamefont {P.}~\bibnamefont {Havu}}, \bibinfo {author}
				{\bibfnamefont {V.}~\bibnamefont {Havu}}, \bibinfo {author} {\bibfnamefont
					{X.}~\bibnamefont {Ren}}, \bibinfo {author} {\bibfnamefont {K.}~\bibnamefont
					{Reuter}},\ and\ \bibinfo {author} {\bibfnamefont {M.}~\bibnamefont
					{Scheffler}},\ }\bibfield  {title} {\bibinfo {title} {Ab initio molecular
					simulations with numeric atom-centered orbitals},\ }\href
			{https://doi.org/10.1016/j.cpc.2009.06.022} {\bibfield  {journal} {\bibinfo
					{journal} {Computer Physics Communications}\ }\textbf {\bibinfo {volume}
					{180}},\ \bibinfo {pages} {2175} (\bibinfo {year} {2009})}\BibitemShut
			{NoStop}%
			\bibitem [{\citenamefont {Ren}\ \emph {et~al.}(2012)\citenamefont {Ren},
				\citenamefont {Rinke}, \citenamefont {Blum}, \citenamefont {Wieferink},
				\citenamefont {Tkatchenko}, \citenamefont {Sanfilippo}, \citenamefont
				{Reuter},\ and\ \citenamefont {Scheffler}}]{Ren2012}%
			\BibitemOpen
			\bibfield  {author} {\bibinfo {author} {\bibfnamefont {X.}~\bibnamefont
					{Ren}}, \bibinfo {author} {\bibfnamefont {P.}~\bibnamefont {Rinke}}, \bibinfo
				{author} {\bibfnamefont {V.}~\bibnamefont {Blum}}, \bibinfo {author}
				{\bibfnamefont {J.}~\bibnamefont {Wieferink}}, \bibinfo {author}
				{\bibfnamefont {A.}~\bibnamefont {Tkatchenko}}, \bibinfo {author}
				{\bibfnamefont {A.}~\bibnamefont {Sanfilippo}}, \bibinfo {author}
				{\bibfnamefont {K.}~\bibnamefont {Reuter}},\ and\ \bibinfo {author}
				{\bibfnamefont {M.}~\bibnamefont {Scheffler}},\ }\bibfield  {title} {\bibinfo
				{title} {{Resolution-of-identity approach to Hartree-Fock, hybrid density
						functionals, RPA, MP2 and GW with numeric atom-centered orbital basis
						functions}},\ }\href {https://doi.org/10.1088/1367-2630/14/5/053020}
			{\bibfield  {journal} {\bibinfo  {journal} {New Journal of Physics}\ }\textbf
				{\bibinfo {volume} {14}},\ \bibinfo {pages} {053020} (\bibinfo {year}
				{2012})}\BibitemShut {NoStop}%
			\bibitem [{\citenamefont {Ellis}\ \emph {et~al.}(2011)\citenamefont {Ellis},
				\citenamefont {Lucero},\ and\ \citenamefont {Scuseria}}]{Ellis_2011}%
			\BibitemOpen
			\bibfield  {author} {\bibinfo {author} {\bibfnamefont {J.~K.}\ \bibnamefont
					{Ellis}}, \bibinfo {author} {\bibfnamefont {M.~J.}\ \bibnamefont {Lucero}},\
				and\ \bibinfo {author} {\bibfnamefont {G.~E.}\ \bibnamefont {Scuseria}},\
			}\bibfield  {title} {\bibinfo {title} {{The indirect to direct band gap
						transition in multilayered MoS$_2$ as predicted by screened hybrid density
						functional theory}},\ }\href {https://doi.org/10.1063/1.3672219} {\bibfield
				{journal} {\bibinfo  {journal} {Appl.~Phys.~Lett.~}\ }\textbf {\bibinfo
					{volume} {99}},\ \bibinfo {pages} {261908} (\bibinfo {year}
				{2011})}\BibitemShut {NoStop}%
			\bibitem [{\citenamefont {Böker}\ \emph {et~al.}(2001)\citenamefont {Böker},
				\citenamefont {Severin}, \citenamefont {Müller}, \citenamefont {Janowitz},
				\citenamefont {Manzke}, \citenamefont {Voß}, \citenamefont {Krüger},
				\citenamefont {Mazur},\ and\ \citenamefont {Pollmann}}]{Boeker_2001}%
			\BibitemOpen
			\bibfield  {author} {\bibinfo {author} {\bibfnamefont {T.}~\bibnamefont
					{Böker}}, \bibinfo {author} {\bibfnamefont {R.}~\bibnamefont {Severin}},
				\bibinfo {author} {\bibfnamefont {A.}~\bibnamefont {Müller}}, \bibinfo
				{author} {\bibfnamefont {C.}~\bibnamefont {Janowitz}}, \bibinfo {author}
				{\bibfnamefont {R.}~\bibnamefont {Manzke}}, \bibinfo {author} {\bibfnamefont
					{D.}~\bibnamefont {Voß}}, \bibinfo {author} {\bibfnamefont {P.}~\bibnamefont
					{Krüger}}, \bibinfo {author} {\bibfnamefont {A.}~\bibnamefont {Mazur}},\
				and\ \bibinfo {author} {\bibfnamefont {J.}~\bibnamefont {Pollmann}},\
			}\bibfield  {title} {\bibinfo {title} {{Band structure of MoS$_2$, MoSe$_2$,
						and $\alpha$-MoTe$_2$: Angle-resolved photoelectron spectroscopy and ab
						initio calculations}},\ }\href {https://doi.org/10.1103/physrevb.64.235305}
			{\bibfield  {journal} {\bibinfo  {journal} {Phys.~Rev.~B}\ }\textbf {\bibinfo
					{volume} {64}},\ \bibinfo {pages} {235305} (\bibinfo {year} {2001})},\
			\Eprint {https://arxiv.org/abs/cond-mat/0107541} {cond-mat/0107541}
			\BibitemShut {NoStop}%
			\bibitem [{\citenamefont {Kadantsev}\ and\ \citenamefont
				{Hawrylak}(2012)}]{Kadantsev_2012}%
			\BibitemOpen
			\bibfield  {author} {\bibinfo {author} {\bibfnamefont {E.~S.}\ \bibnamefont
					{Kadantsev}}\ and\ \bibinfo {author} {\bibfnamefont {P.}~\bibnamefont
					{Hawrylak}},\ }\bibfield  {title} {\bibinfo {title} {{Electronic structure of
						a single MoS$_2$ monolayer}},\ }\href
			{https://doi.org/10.1016/j.ssc.2012.02.005} {\bibfield  {journal} {\bibinfo
					{journal} {Solid State Communications}\ }\textbf {\bibinfo {volume} {152}},\
				\bibinfo {pages} {909} (\bibinfo {year} {2012})}\BibitemShut {NoStop}%
			\bibitem [{\citenamefont {Kang}\ \emph {et~al.}(2013)\citenamefont {Kang},
				\citenamefont {Tongay}, \citenamefont {Zhou}, \citenamefont {Li},\ and\
				\citenamefont {Wu}}]{Kang2013}%
			\BibitemOpen
			\bibfield  {author} {\bibinfo {author} {\bibfnamefont {J.}~\bibnamefont
					{Kang}}, \bibinfo {author} {\bibfnamefont {S.}~\bibnamefont {Tongay}},
				\bibinfo {author} {\bibfnamefont {J.}~\bibnamefont {Zhou}}, \bibinfo {author}
				{\bibfnamefont {J.}~\bibnamefont {Li}},\ and\ \bibinfo {author}
				{\bibfnamefont {J.}~\bibnamefont {Wu}},\ }\bibfield  {title} {\bibinfo
				{title} {Band offsets and heterostructures of two-dimensional
					semiconductors},\ }\href {https://doi.org/10.1063/1.4774090} {\bibfield
				{journal} {\bibinfo  {journal} {Applied Physics Letters}\ }\textbf {\bibinfo
					{volume} {102}},\ \bibinfo {pages} {012111} (\bibinfo {year}
				{2013})}\BibitemShut {NoStop}%
			\bibitem [{\citenamefont {Molina-Sánchez}\ \emph {et~al.}(2016)\citenamefont
				{Molina-Sánchez}, \citenamefont {Palummo}, \citenamefont {Marini},\ and\
				\citenamefont {Wirtz}}]{MolinaSanchez_2016}%
			\BibitemOpen
			\bibfield  {author} {\bibinfo {author} {\bibfnamefont {A.}~\bibnamefont
					{Molina-Sánchez}}, \bibinfo {author} {\bibfnamefont {M.}~\bibnamefont
					{Palummo}}, \bibinfo {author} {\bibfnamefont {A.}~\bibnamefont {Marini}},\
				and\ \bibinfo {author} {\bibfnamefont {L.}~\bibnamefont {Wirtz}},\ }\bibfield
			{title} {\bibinfo {title} {{Temperature-dependent excitonic effects in the
						optical properties of single-layer MoS$_2$}},\ }\href
			{https://doi.org/10.1103/physrevb.93.155435} {\bibfield  {journal} {\bibinfo
					{journal} {Phys.~Rev.~B}\ }\textbf {\bibinfo {volume} {93}},\ \bibinfo
				{pages} {155435} (\bibinfo {year} {2016})},\ \Eprint
			{https://arxiv.org/abs/1604.00943} {1604.00943} \BibitemShut {NoStop}%
			\bibitem [{\citenamefont {Mansouri}\ \emph {et~al.}(2023)\citenamefont
				{Mansouri}, \citenamefont {Koval}, \citenamefont {Sharifzadeh},\ and\
				\citenamefont {S{\'a}nchez-Portal}}]{Mansouri_2023}%
			\BibitemOpen
			\bibfield  {author} {\bibinfo {author} {\bibfnamefont {M.}~\bibnamefont
					{Mansouri}}, \bibinfo {author} {\bibfnamefont {P.}~\bibnamefont {Koval}},
				\bibinfo {author} {\bibfnamefont {S.}~\bibnamefont {Sharifzadeh}},\ and\
				\bibinfo {author} {\bibfnamefont {D.}~\bibnamefont {S{\'a}nchez-Portal}},\
			}\bibfield  {title} {\bibinfo {title} {Molecular doping in the organic
					semiconductor diindenoperylene: Insights from many-body perturbation
					theory},\ }\href {https://doi.org/10.1021/acs.jpcc.3c03758} {\bibfield
				{journal} {\bibinfo  {journal} {The Journal of Physical Chemistry C}\
				}\textbf {\bibinfo {volume} {127}},\ \bibinfo {pages} {16668} (\bibinfo
				{year} {2023})}\BibitemShut {NoStop}%
			\bibitem [{\citenamefont {Sun}\ and\ \citenamefont {Ullrich}(2020)}]{Sun_2020}%
			\BibitemOpen
			\bibfield  {author} {\bibinfo {author} {\bibfnamefont {J.}~\bibnamefont
					{Sun}}\ and\ \bibinfo {author} {\bibfnamefont {C.~A.}\ \bibnamefont
					{Ullrich}},\ }\bibfield  {title} {\bibinfo {title} {{Optical properties of
						$\mathrm{Cs}{\mathrm{Cu}}_{2}{X}_{3}$ $(X=\mathrm{Cl}, \mathrm{Br},$ and I):
						A comparative study between hybrid time-dependent density-functional theory
						and the Bethe-Salpeter equation}},\ }\href
			{https://doi.org/10.1103/PhysRevMaterials.4.095402} {\bibfield  {journal}
				{\bibinfo  {journal} {Phys. Rev. Mater.}\ }\textbf {\bibinfo {volume} {4}},\
				\bibinfo {pages} {095402} (\bibinfo {year} {2020})}\BibitemShut {NoStop}%
			\bibitem [{\citenamefont {Laflamme~Janssen}\ \emph {et~al.}(2015)\citenamefont
				{Laflamme~Janssen}, \citenamefont {Rousseau},\ and\ \citenamefont
				{C\^ot\'e}}]{Janssen_2015}%
			\BibitemOpen
			\bibfield  {author} {\bibinfo {author} {\bibfnamefont {J.}~\bibnamefont
					{Laflamme~Janssen}}, \bibinfo {author} {\bibfnamefont {B.}~\bibnamefont
					{Rousseau}},\ and\ \bibinfo {author} {\bibfnamefont {M.}~\bibnamefont
					{C\^ot\'e}},\ }\bibfield  {title} {\bibinfo {title} {Efficient dielectric
					matrix calculations using the lanczos algorithm for fast many-body
					${G}_{0}{W}_{0}$ implementations},\ }\href
			{https://doi.org/10.1103/PhysRevB.91.125120} {\bibfield  {journal} {\bibinfo
					{journal} {Phys. Rev. B}\ }\textbf {\bibinfo {volume} {91}},\ \bibinfo
				{pages} {125120} (\bibinfo {year} {2015})}\BibitemShut {NoStop}%
			\bibitem [{\citenamefont {Klimeš}\ \emph {et~al.}(2014)\citenamefont
				{Klimeš}, \citenamefont {Kaltak},\ and\ \citenamefont
				{Kresse}}]{Klimes_2014}%
			\BibitemOpen
			\bibfield  {author} {\bibinfo {author} {\bibfnamefont {J.}~\bibnamefont
					{Klimeš}}, \bibinfo {author} {\bibfnamefont {M.}~\bibnamefont {Kaltak}},\
				and\ \bibinfo {author} {\bibfnamefont {G.}~\bibnamefont {Kresse}},\
			}\bibfield  {title} {\bibinfo {title} {{Predictive GW calculations using
						plane waves and pseudopotentials}},\ }\href
			{https://doi.org/10.1103/physrevb.90.075125} {\bibfield  {journal} {\bibinfo
					{journal} {Physical Review B}\ }\textbf {\bibinfo {volume} {90}},\ \bibinfo
				{pages} {075125} (\bibinfo {year} {2014})}\BibitemShut {NoStop}%
			\bibitem [{\citenamefont {Salehi}\ and\ \citenamefont
				{Saffarzadeh}(2018)}]{Salehi_2018}%
			\BibitemOpen
			\bibfield  {author} {\bibinfo {author} {\bibfnamefont {S.}~\bibnamefont
					{Salehi}}\ and\ \bibinfo {author} {\bibfnamefont {A.}~\bibnamefont
					{Saffarzadeh}},\ }\bibfield  {title} {\bibinfo {title} {{Optoelectronic
						properties of defective MoS$_2$ and WS$_2$ monolayers}},\ }\href
			{https://doi.org/10.1016/j.jpcs.2018.05.020} {\bibfield  {journal} {\bibinfo
					{journal} {Journal of Physics and Chemistry of Solids}\ }\textbf {\bibinfo
					{volume} {121}},\ \bibinfo {pages} {172} (\bibinfo {year} {2018})},\ \Eprint
			{https://arxiv.org/abs/1805.09532} {1805.09532} \BibitemShut {NoStop}%
			\bibitem [{\citenamefont {Kormányos}\ \emph {et~al.}(2015)\citenamefont
				{Kormányos}, \citenamefont {Burkard}, \citenamefont {Gmitra}, \citenamefont
				{Fabian}, \citenamefont {Zólyomi}, \citenamefont {Drummond},\ and\
				\citenamefont {Fal’ko}}]{Kormanyos_2015}%
			\BibitemOpen
			\bibfield  {author} {\bibinfo {author} {\bibfnamefont {A.}~\bibnamefont
					{Kormányos}}, \bibinfo {author} {\bibfnamefont {G.}~\bibnamefont {Burkard}},
				\bibinfo {author} {\bibfnamefont {M.}~\bibnamefont {Gmitra}}, \bibinfo
				{author} {\bibfnamefont {J.}~\bibnamefont {Fabian}}, \bibinfo {author}
				{\bibfnamefont {V.}~\bibnamefont {Zólyomi}}, \bibinfo {author}
				{\bibfnamefont {N.~D.}\ \bibnamefont {Drummond}},\ and\ \bibinfo {author}
				{\bibfnamefont {V.}~\bibnamefont {Fal’ko}},\ }\bibfield  {title} {\bibinfo
				{title} {{k·p theory for two-dimensional transition metal dichalcogenide
						semiconductors}},\ }\href {https://doi.org/10.1088/2053-1583/2/2/022001}
			{\bibfield  {journal} {\bibinfo  {journal} {2D Materials}\ }\textbf {\bibinfo
					{volume} {2}},\ \bibinfo {pages} {022001} (\bibinfo {year}
				{2015})}\BibitemShut {NoStop}%
			\bibitem [{\citenamefont {Dou}\ \emph {et~al.}(2016)\citenamefont {Dou},
				\citenamefont {Ding}, \citenamefont {Jiang}, \citenamefont {Fan},\ and\
				\citenamefont {Sun}}]{Dou_2016}%
			\BibitemOpen
			\bibfield  {author} {\bibinfo {author} {\bibfnamefont {X.}~\bibnamefont
					{Dou}}, \bibinfo {author} {\bibfnamefont {K.}~\bibnamefont {Ding}}, \bibinfo
				{author} {\bibfnamefont {D.}~\bibnamefont {Jiang}}, \bibinfo {author}
				{\bibfnamefont {X.}~\bibnamefont {Fan}},\ and\ \bibinfo {author}
				{\bibfnamefont {B.}~\bibnamefont {Sun}},\ }\bibfield  {title} {\bibinfo
				{title} {{Probing Spin–Orbit Coupling and Interlayer Coupling in Atomically
						Thin Molybdenum Disulfide Using Hydrostatic Pressure}},\ }\href
			{https://doi.org/10.1021/acsnano.5b07273} {\bibfield  {journal} {\bibinfo
					{journal} {ACS Nano}\ }\textbf {\bibinfo {volume} {10}},\ \bibinfo {pages}
				{1619} (\bibinfo {year} {2016})}\BibitemShut {NoStop}%
			\bibitem [{\citenamefont {Zhu}\ \emph {et~al.}(2011)\citenamefont {Zhu},
				\citenamefont {Cheng},\ and\ \citenamefont {Schwingenschlögl}}]{Zhu_2011}%
			\BibitemOpen
			\bibfield  {author} {\bibinfo {author} {\bibfnamefont {Z.~Y.}\ \bibnamefont
					{Zhu}}, \bibinfo {author} {\bibfnamefont {Y.~C.}\ \bibnamefont {Cheng}},\
				and\ \bibinfo {author} {\bibfnamefont {U.}~\bibnamefont
					{Schwingenschlögl}},\ }\bibfield  {title} {\bibinfo {title} {{Giant
						spin-orbit-induced spin splitting in two-dimensional transition-metal
						dichalcogenide semiconductors}},\ }\href
			{https://doi.org/10.1103/physrevb.84.153402} {\bibfield  {journal} {\bibinfo
					{journal} {Phys.~Rev.~B}\ }\textbf {\bibinfo {volume} {84}},\ \bibinfo
				{pages} {153402} (\bibinfo {year} {2011})}\BibitemShut {NoStop}%
			\bibitem [{\citenamefont {Miwa}\ \emph {et~al.}(2014)\citenamefont {Miwa},
				\citenamefont {Ulstrup}, \citenamefont {Sørensen}, \citenamefont {Dendzik},
				\citenamefont {Čabo}, \citenamefont {Bianchi}, \citenamefont {Lauritsen},\
				and\ \citenamefont {Hofmann}}]{Miwa_2014}%
			\BibitemOpen
			\bibfield  {author} {\bibinfo {author} {\bibfnamefont {J.~A.}\ \bibnamefont
					{Miwa}}, \bibinfo {author} {\bibfnamefont {S.}~\bibnamefont {Ulstrup}},
				\bibinfo {author} {\bibfnamefont {S.~G.}\ \bibnamefont {Sørensen}}, \bibinfo
				{author} {\bibfnamefont {M.}~\bibnamefont {Dendzik}}, \bibinfo {author}
				{\bibfnamefont {A.~G.}\ \bibnamefont {Čabo}}, \bibinfo {author}
				{\bibfnamefont {M.}~\bibnamefont {Bianchi}}, \bibinfo {author} {\bibfnamefont
					{J.~V.}\ \bibnamefont {Lauritsen}},\ and\ \bibinfo {author} {\bibfnamefont
					{P.}~\bibnamefont {Hofmann}},\ }\bibfield  {title} {\bibinfo {title}
				{{Electronic Structure of Epitaxial Single-Layer MoS$_2$}},\ }\href
			{https://doi.org/10.1103/physrevlett.114.046802} {\bibfield  {journal}
				{\bibinfo  {journal} {Phys.~Rev.~Lett.~}\ }\textbf {\bibinfo {volume}
					{114}},\ \bibinfo {pages} {046802} (\bibinfo {year} {2014})},\ \Eprint
			{https://arxiv.org/abs/1410.0615} {1410.0615} \BibitemShut {NoStop}%
			\bibitem [{\citenamefont {Zhang}\ \emph {et~al.}(2015)\citenamefont {Zhang},
				\citenamefont {Li}, \citenamefont {Wang}, \citenamefont {Liu}, \citenamefont
				{Zhang},\ and\ \citenamefont {Qiu}}]{Zhang_2015}%
			\BibitemOpen
			\bibfield  {author} {\bibinfo {author} {\bibfnamefont {Y.}~\bibnamefont
					{Zhang}}, \bibinfo {author} {\bibfnamefont {H.}~\bibnamefont {Li}}, \bibinfo
				{author} {\bibfnamefont {H.}~\bibnamefont {Wang}}, \bibinfo {author}
				{\bibfnamefont {R.}~\bibnamefont {Liu}}, \bibinfo {author} {\bibfnamefont
					{S.-L.}\ \bibnamefont {Zhang}},\ and\ \bibinfo {author} {\bibfnamefont
					{Z.-J.}\ \bibnamefont {Qiu}},\ }\bibfield  {title} {\bibinfo {title} {{On
						Valence-Band Splitting in Layered MoS$_2$}},\ }\href
			{https://doi.org/10.1021/acsnano.5b03505} {\bibfield  {journal} {\bibinfo
					{journal} {ACS Nano}\ }\textbf {\bibinfo {volume} {9}},\ \bibinfo {pages}
				{8514} (\bibinfo {year} {2015})}\BibitemShut {NoStop}%
			\bibitem [{\citenamefont {Splendiani}\ \emph {et~al.}(2010)\citenamefont
				{Splendiani}, \citenamefont {Sun}, \citenamefont {Zhang}, \citenamefont {Li},
				\citenamefont {Kim}, \citenamefont {Chim}, \citenamefont {Galli},\ and\
				\citenamefont {Wang}}]{Splendiani_2010}%
			\BibitemOpen
			\bibfield  {author} {\bibinfo {author} {\bibfnamefont {A.}~\bibnamefont
					{Splendiani}}, \bibinfo {author} {\bibfnamefont {L.}~\bibnamefont {Sun}},
				\bibinfo {author} {\bibfnamefont {Y.}~\bibnamefont {Zhang}}, \bibinfo
				{author} {\bibfnamefont {T.}~\bibnamefont {Li}}, \bibinfo {author}
				{\bibfnamefont {J.}~\bibnamefont {Kim}}, \bibinfo {author} {\bibfnamefont
					{C.-Y.}\ \bibnamefont {Chim}}, \bibinfo {author} {\bibfnamefont
					{G.}~\bibnamefont {Galli}},\ and\ \bibinfo {author} {\bibfnamefont
					{F.}~\bibnamefont {Wang}},\ }\bibfield  {title} {\bibinfo {title} {{Emerging
						Photoluminescence in Monolayer MoS$_2$}},\ }\bibfield  {journal} {\bibinfo
				{journal} {Nano Letters}\ }\textbf {\bibinfo {volume} {10}},\ \href
			{https://doi.org/10.1021/nl903868w} {10.1021/nl903868w} (\bibinfo {year}
			{2010})\BibitemShut {NoStop}%
			\bibitem [{\citenamefont {Li}\ \emph {et~al.}(2014)\citenamefont {Li},
				\citenamefont {Birdwell}, \citenamefont {Amani}, \citenamefont {Burke},
				\citenamefont {Ling}, \citenamefont {Lee}, \citenamefont {Liang},
				\citenamefont {Peng}, \citenamefont {Richter}, \citenamefont {Kong},
				\citenamefont {Gundlach},\ and\ \citenamefont {Nguyen}}]{Li_2014}%
			\BibitemOpen
			\bibfield  {author} {\bibinfo {author} {\bibfnamefont {W.}~\bibnamefont
					{Li}}, \bibinfo {author} {\bibfnamefont {A.~G.}\ \bibnamefont {Birdwell}},
				\bibinfo {author} {\bibfnamefont {M.}~\bibnamefont {Amani}}, \bibinfo
				{author} {\bibfnamefont {R.~A.}\ \bibnamefont {Burke}}, \bibinfo {author}
				{\bibfnamefont {X.}~\bibnamefont {Ling}}, \bibinfo {author} {\bibfnamefont
					{Y.-H.}\ \bibnamefont {Lee}}, \bibinfo {author} {\bibfnamefont
					{X.}~\bibnamefont {Liang}}, \bibinfo {author} {\bibfnamefont
					{L.}~\bibnamefont {Peng}}, \bibinfo {author} {\bibfnamefont {C.~A.}\
					\bibnamefont {Richter}}, \bibinfo {author} {\bibfnamefont {J.}~\bibnamefont
					{Kong}}, \bibinfo {author} {\bibfnamefont {D.~J.}\ \bibnamefont {Gundlach}},\
				and\ \bibinfo {author} {\bibfnamefont {N.~V.}\ \bibnamefont {Nguyen}},\
			}\bibfield  {title} {\bibinfo {title} {{Broadband optical properties of
						large-area monolayer CVD molybdenum disulfide}},\ }\href
			{https://doi.org/10.1103/PhysRevB.90.195434} {\bibfield  {journal} {\bibinfo
					{journal} {Phys.~Rev.~B}\ }\textbf {\bibinfo {volume} {90}},\ \bibinfo
				{pages} {195434} (\bibinfo {year} {2014})}\BibitemShut {NoStop}%
			\bibitem [{\citenamefont {Shen}\ \emph {et~al.}(2013)\citenamefont {Shen},
				\citenamefont {Hsu}, \citenamefont {Li},\ and\ \citenamefont
				{Liu}}]{Shen_2013}%
			\BibitemOpen
			\bibfield  {author} {\bibinfo {author} {\bibfnamefont {C.-C.}\ \bibnamefont
					{Shen}}, \bibinfo {author} {\bibfnamefont {Y.-T.}\ \bibnamefont {Hsu}},
				\bibinfo {author} {\bibfnamefont {L.-J.}\ \bibnamefont {Li}},\ and\ \bibinfo
				{author} {\bibfnamefont {H.-L.}\ \bibnamefont {Liu}},\ }\bibfield  {title}
			{\bibinfo {title} {{Charge Dynamics and Electronic Structures of Monolayer
						MoS$_2$ Films Grown by Chemical Vapor Deposition}},\ }\href
			{https://doi.org/10.7567/apex.6.125801} {\bibfield  {journal} {\bibinfo
					{journal} {Applied Physics Express}\ }\textbf {\bibinfo {volume} {6}},\
				\bibinfo {pages} {125801} (\bibinfo {year} {2013})}\BibitemShut {NoStop}%
			\bibitem [{\citenamefont {Schmidt}\ \emph {et~al.}(2015)\citenamefont
				{Schmidt}, \citenamefont {Yudhistira}, \citenamefont {Chu}, \citenamefont
				{Neto}, \citenamefont {Özyilmaz}, \citenamefont {Adam},\ and\ \citenamefont
				{Eda}}]{Schmidt_2015}%
			\BibitemOpen
			\bibfield  {author} {\bibinfo {author} {\bibfnamefont {H.}~\bibnamefont
					{Schmidt}}, \bibinfo {author} {\bibfnamefont {I.}~\bibnamefont {Yudhistira}},
				\bibinfo {author} {\bibfnamefont {L.}~\bibnamefont {Chu}}, \bibinfo {author}
				{\bibfnamefont {A.~H.~C.}\ \bibnamefont {Neto}}, \bibinfo {author}
				{\bibfnamefont {B.}~\bibnamefont {Özyilmaz}}, \bibinfo {author}
				{\bibfnamefont {S.}~\bibnamefont {Adam}},\ and\ \bibinfo {author}
				{\bibfnamefont {G.}~\bibnamefont {Eda}},\ }\bibfield  {title} {\bibinfo
				{title} {{Quantum Transport and Observation of Dyakonov-Perel Spin-Orbit
						Scattering in Monolayer MoS$_2$}},\ }\href
			{https://doi.org/10.1103/physrevlett.116.046803} {\bibfield  {journal}
				{\bibinfo  {journal} {Phys.~Rev.~Lett.~}\ }\textbf {\bibinfo {volume}
					{116}},\ \bibinfo {pages} {046803} (\bibinfo {year} {2015})},\ \Eprint
			{https://arxiv.org/abs/1503.00428} {1503.00428} \BibitemShut {NoStop}%
			\bibitem [{\citenamefont {Eknapakul}\ \emph {et~al.}(2014)\citenamefont
				{Eknapakul}, \citenamefont {King}, \citenamefont {Asakawa}, \citenamefont
				{Buaphet}, \citenamefont {He}, \citenamefont {Mo}, \citenamefont {Takagi},
				\citenamefont {Shen}, \citenamefont {Baumberger}, \citenamefont {Sasagawa},
				\citenamefont {Jungthawan},\ and\ \citenamefont
				{Meevasana}}]{Eknapakul_2014}%
			\BibitemOpen
			\bibfield  {author} {\bibinfo {author} {\bibfnamefont {T.}~\bibnamefont
					{Eknapakul}}, \bibinfo {author} {\bibfnamefont {P.~D.~C.}\ \bibnamefont
					{King}}, \bibinfo {author} {\bibfnamefont {M.}~\bibnamefont {Asakawa}},
				\bibinfo {author} {\bibfnamefont {P.}~\bibnamefont {Buaphet}}, \bibinfo
				{author} {\bibfnamefont {R.-H.}\ \bibnamefont {He}}, \bibinfo {author}
				{\bibfnamefont {S.-K.}\ \bibnamefont {Mo}}, \bibinfo {author} {\bibfnamefont
					{H.}~\bibnamefont {Takagi}}, \bibinfo {author} {\bibfnamefont {K.~M.}\
					\bibnamefont {Shen}}, \bibinfo {author} {\bibfnamefont {F.}~\bibnamefont
					{Baumberger}}, \bibinfo {author} {\bibfnamefont {T.}~\bibnamefont
					{Sasagawa}}, \bibinfo {author} {\bibfnamefont {S.}~\bibnamefont
					{Jungthawan}},\ and\ \bibinfo {author} {\bibfnamefont {W.}~\bibnamefont
					{Meevasana}},\ }\bibfield  {title} {\bibinfo {title} {{Electronic Structure
						of a Quasi-Freestanding MoS$_2$ Monolayer}},\ }\href
			{https://doi.org/10.1021/nl4042824} {\bibfield  {journal} {\bibinfo
					{journal} {Nano Letters}\ }\textbf {\bibinfo {volume} {14}},\ \bibinfo
				{pages} {1312} (\bibinfo {year} {2014})}\BibitemShut {NoStop}%
			\bibitem [{\citenamefont {Jin}\ \emph {et~al.}(2015)\citenamefont {Jin},
				\citenamefont {Yeh}, \citenamefont {Zaki}, \citenamefont {Zhang},
				\citenamefont {Liou}, \citenamefont {Sadowski}, \citenamefont {Barinov},
				\citenamefont {Yablonskikh}, \citenamefont {Dadap}, \citenamefont {Sutter}
				\emph {et~al.}}]{Jin_2015}%
			\BibitemOpen
			\bibfield  {author} {\bibinfo {author} {\bibfnamefont {W.}~\bibnamefont
					{Jin}}, \bibinfo {author} {\bibfnamefont {P.-C.}\ \bibnamefont {Yeh}},
				\bibinfo {author} {\bibfnamefont {N.}~\bibnamefont {Zaki}}, \bibinfo {author}
				{\bibfnamefont {D.}~\bibnamefont {Zhang}}, \bibinfo {author} {\bibfnamefont
					{J.~T.}\ \bibnamefont {Liou}}, \bibinfo {author} {\bibfnamefont {J.~T.}\
					\bibnamefont {Sadowski}}, \bibinfo {author} {\bibfnamefont {A.}~\bibnamefont
					{Barinov}}, \bibinfo {author} {\bibfnamefont {M.}~\bibnamefont
					{Yablonskikh}}, \bibinfo {author} {\bibfnamefont {J.~I.}\ \bibnamefont
					{Dadap}}, \bibinfo {author} {\bibfnamefont {P.}~\bibnamefont {Sutter}}, \emph
				{et~al.},\ }\bibfield  {title} {\bibinfo {title} {Substrate interactions with
					suspended and supported monolayer {MoS 2}: Angle-resolved photoemission
					spectroscopy},\ }\href@noop {} {\bibfield  {journal} {\bibinfo  {journal}
					{Phys.~Rev.~B}\ }\textbf {\bibinfo {volume} {91}},\ \bibinfo {pages} {121409}
				(\bibinfo {year} {2015})}\BibitemShut {NoStop}%
			\bibitem [{\citenamefont {Peelaers}\ and\ \citenamefont
				{Walle}(2012)}]{Peelaers_2012}%
			\BibitemOpen
			\bibfield  {author} {\bibinfo {author} {\bibfnamefont {H.}~\bibnamefont
					{Peelaers}}\ and\ \bibinfo {author} {\bibfnamefont {C.~G. V.~d.}\
					\bibnamefont {Walle}},\ }\bibfield  {title} {\bibinfo {title} {{Effects of
						strain on band structure and effective masses in MoS$_2$}},\ }\href
			{https://doi.org/10.1103/physrevb.86.241401} {\bibfield  {journal} {\bibinfo
					{journal} {Phys.~Rev.~B}\ }\textbf {\bibinfo {volume} {86}},\ \bibinfo
				{pages} {241401} (\bibinfo {year} {2012})}\BibitemShut {NoStop}%
			\bibitem [{\citenamefont {Pulkin}\ and\ \citenamefont
				{Chan}(2020)}]{Pulkin_2020}%
			\BibitemOpen
			\bibfield  {author} {\bibinfo {author} {\bibfnamefont {A.}~\bibnamefont
					{Pulkin}}\ and\ \bibinfo {author} {\bibfnamefont {G.~K.-L.}\ \bibnamefont
					{Chan}},\ }\bibfield  {title} {\bibinfo {title} {{First-principles coupled
						cluster theory of the electronic spectrum of transition metal
						dichalcogenides}},\ }\href {https://doi.org/10.1103/physrevb.101.241113}
			{\bibfield  {journal} {\bibinfo  {journal} {Physical Review B}\ }\textbf
				{\bibinfo {volume} {101}},\ \bibinfo {pages} {241113} (\bibinfo {year}
				{2020})}\BibitemShut {NoStop}%
			\bibitem [{\citenamefont {Jin}\ \emph {et~al.}(2013)\citenamefont {Jin},
				\citenamefont {Yeh}, \citenamefont {Zaki}, \citenamefont {Zhang},
				\citenamefont {Sadowski}, \citenamefont {Al-Mahboob}, \citenamefont {Zande},
				\citenamefont {Chenet}, \citenamefont {Dadap}, \citenamefont {Herman},
				\citenamefont {Sutter}, \citenamefont {Hone},\ and\ \citenamefont
				{Osgood}}]{Jin_2013}%
			\BibitemOpen
			\bibfield  {author} {\bibinfo {author} {\bibfnamefont {W.}~\bibnamefont
					{Jin}}, \bibinfo {author} {\bibfnamefont {P.-C.}\ \bibnamefont {Yeh}},
				\bibinfo {author} {\bibfnamefont {N.}~\bibnamefont {Zaki}}, \bibinfo {author}
				{\bibfnamefont {D.}~\bibnamefont {Zhang}}, \bibinfo {author} {\bibfnamefont
					{J.~T.}\ \bibnamefont {Sadowski}}, \bibinfo {author} {\bibfnamefont
					{A.}~\bibnamefont {Al-Mahboob}}, \bibinfo {author} {\bibfnamefont {A.~M.
						v.~d.}\ \bibnamefont {Zande}}, \bibinfo {author} {\bibfnamefont {D.~A.}\
					\bibnamefont {Chenet}}, \bibinfo {author} {\bibfnamefont {J.~I.}\
					\bibnamefont {Dadap}}, \bibinfo {author} {\bibfnamefont {I.~P.}\ \bibnamefont
					{Herman}}, \bibinfo {author} {\bibfnamefont {P.}~\bibnamefont {Sutter}},
				\bibinfo {author} {\bibfnamefont {J.}~\bibnamefont {Hone}},\ and\ \bibinfo
				{author} {\bibfnamefont {R.~M.}\ \bibnamefont {Osgood}},\ }\bibfield  {title}
			{\bibinfo {title} {{Direct Measurement of the Thickness-Dependent Electronic
						Band Structure of MoS$_2$ Using Angle-Resolved Photoemission Spectroscopy}},\
			}\href {https://doi.org/10.1103/physrevlett.111.106801} {\bibfield  {journal}
				{\bibinfo  {journal} {Phys.~Rev.~Lett.~}\ }\textbf {\bibinfo {volume}
					{111}},\ \bibinfo {pages} {106801} (\bibinfo {year} {2013})}\BibitemShut
			{NoStop}%
			\bibitem [{\citenamefont {Fuchs}\ \emph {et~al.}(2007)\citenamefont {Fuchs},
				\citenamefont {Furthm\"uller}, \citenamefont {Bechstedt}, \citenamefont
				{Shishkin},\ and\ \citenamefont {Kresse}}]{Fuchs_2007}%
			\BibitemOpen
			\bibfield  {author} {\bibinfo {author} {\bibfnamefont {F.}~\bibnamefont
					{Fuchs}}, \bibinfo {author} {\bibfnamefont {J.}~\bibnamefont
					{Furthm\"uller}}, \bibinfo {author} {\bibfnamefont {F.}~\bibnamefont
					{Bechstedt}}, \bibinfo {author} {\bibfnamefont {M.}~\bibnamefont
					{Shishkin}},\ and\ \bibinfo {author} {\bibfnamefont {G.}~\bibnamefont
					{Kresse}},\ }\bibfield  {title} {\bibinfo {title} {{Quasiparticle band
						structure based on a generalized Kohn-Sham scheme}},\ }\href
			{https://doi.org/10.1103/PhysRevB.76.115109} {\bibfield  {journal} {\bibinfo
					{journal} {Phys.~Rev.~B}\ }\textbf {\bibinfo {volume} {76}},\ \bibinfo
				{pages} {115109} (\bibinfo {year} {2007})}\BibitemShut {NoStop}%
			\bibitem [{\citenamefont {Yadav}\ and\ \citenamefont
				{Ramprasad}(2012)}]{Yadav_2012}%
			\BibitemOpen
			\bibfield  {author} {\bibinfo {author} {\bibfnamefont {S.}~\bibnamefont
					{Yadav}}\ and\ \bibinfo {author} {\bibfnamefont {R.}~\bibnamefont
					{Ramprasad}},\ }\bibfield  {title} {\bibinfo {title} {{Strain-assisted
						bandgap modulation in Zn based II-VI semiconductors}},\ }\href
			{https://doi.org/10.1063/1.4729153} {\bibfield  {journal} {\bibinfo
					{journal} {Applied Physics Letters}\ }\textbf {\bibinfo {volume} {100}},\
				\bibinfo {pages} {241903} (\bibinfo {year} {2012})}\BibitemShut {NoStop}%
			\bibitem [{\citenamefont {Camarasa-G\'omez}\ \emph {et~al.}(2023)\citenamefont
				{Camarasa-G\'omez}, \citenamefont {Ramasubramaniam}, \citenamefont {Neaton},\
				and\ \citenamefont {Kronik}}]{CamarasaGomez_2023}%
			\BibitemOpen
			\bibfield  {author} {\bibinfo {author} {\bibfnamefont {M.}~\bibnamefont
					{Camarasa-G\'omez}}, \bibinfo {author} {\bibfnamefont {A.}~\bibnamefont
					{Ramasubramaniam}}, \bibinfo {author} {\bibfnamefont {J.~B.}\ \bibnamefont
					{Neaton}},\ and\ \bibinfo {author} {\bibfnamefont {L.}~\bibnamefont
					{Kronik}},\ }\bibfield  {title} {\bibinfo {title} {Transferable screened
					range-separated hybrid functionals for electronic and optical properties of
					van der waals materials},\ }\href
			{https://doi.org/10.1103/PhysRevMaterials.7.104001} {\bibfield  {journal}
				{\bibinfo  {journal} {Phys. Rev. Mater.}\ }\textbf {\bibinfo {volume} {7}},\
				\bibinfo {pages} {104001} (\bibinfo {year} {2023})}\BibitemShut {NoStop}%
			\bibitem [{\citenamefont {C{\'a}rsky}\ \emph {et~al.}(2010)\citenamefont
				{C{\'a}rsky}, \citenamefont {Paldus},\ and\ \citenamefont
				{Pittner}}]{Carsky}%
			\BibitemOpen
			\bibfield  {author} {\bibinfo {author} {\bibfnamefont {P.}~\bibnamefont
					{C{\'a}rsky}}, \bibinfo {author} {\bibfnamefont {J.}~\bibnamefont {Paldus}},\
				and\ \bibinfo {author} {\bibfnamefont {J.}~\bibnamefont {Pittner}},\
			}\bibfield  {title} {\bibinfo {title} {Recent progress in coupled cluster
					methods: Theory and applications},\ }\href@noop {} {\bibfield  {journal}
				{\bibinfo  {journal} {Springer Science \& Business Media}\ } (\bibinfo {year}
				{2010})}\BibitemShut {NoStop}%
			\bibitem [{\citenamefont {Helgaker}\ \emph {et~al.}(2008)\citenamefont
				{Helgaker}, \citenamefont {Klopper},\ and\ \citenamefont
				{Tew}}]{Helgaker_2008}%
			\BibitemOpen
			\bibfield  {author} {\bibinfo {author} {\bibfnamefont {T.}~\bibnamefont
					{Helgaker}}, \bibinfo {author} {\bibfnamefont {W.}~\bibnamefont {Klopper}},\
				and\ \bibinfo {author} {\bibfnamefont {D.~P.}\ \bibnamefont {Tew}},\
			}\bibfield  {title} {\bibinfo {title} {Quantitative quantum chemistry},\
			}\href {https://doi.org/10.1080/00268970802258591} {\bibfield  {journal}
				{\bibinfo  {journal} {Mol. Phys.}\ }\textbf {\bibinfo {volume} {106}},\
				\bibinfo {pages} {2107} (\bibinfo {year} {2008})}\BibitemShut {NoStop}%
			\bibitem [{\citenamefont {Draxl}\ and\ \citenamefont
				{Scheffler}(2019)}]{Draxl2019}%
			\BibitemOpen
			\bibfield  {author} {\bibinfo {author} {\bibfnamefont {C.}~\bibnamefont
					{Draxl}}\ and\ \bibinfo {author} {\bibfnamefont {M.}~\bibnamefont
					{Scheffler}},\ }\bibfield  {title} {\bibinfo {title} {The nomad laboratory:
					From data sharing to artificial intelligence},\ }\href
			{https://doi.org/10.1088/2515-7639/ab13bb} {\bibfield  {journal} {\bibinfo
					{journal} {J.~Phys.~Mater.~}\ }\textbf {\bibinfo {volume} {2}},\ \bibinfo
				{pages} {036001} (\bibinfo {year} {2019})}\BibitemShut {NoStop}%
			\bibitem [{nom()}]{nomad-doi}%
			\BibitemOpen
			\href@noop {} {\bibinfo {title} {Nomad repository, dataset: Gw-mos2.}},\
			\bibinfo {howpublished}
			{\url{https://dx.doi.org/10.17172/NOMAD/2023.09.16-1}}\BibitemShut {NoStop}%
			\bibitem [{\citenamefont {Friedrich}\ \emph {et~al.}(2011)\citenamefont
				{Friedrich}, \citenamefont {Müller},\ and\ \citenamefont
				{Blügel}}]{Friedrich_2011}%
			\BibitemOpen
			\bibfield  {author} {\bibinfo {author} {\bibfnamefont {C.}~\bibnamefont
					{Friedrich}}, \bibinfo {author} {\bibfnamefont {M.~C.}\ \bibnamefont
					{Müller}},\ and\ \bibinfo {author} {\bibfnamefont {S.}~\bibnamefont
					{Blügel}},\ }\bibfield  {title} {\bibinfo {title} {{Band convergence and
						linearization error correction of all-electron GW calculations: The extreme
						case of zinc oxide}},\ }\href {https://doi.org/10.1103/physrevb.83.081101}
			{\bibfield  {journal} {\bibinfo  {journal} {Phys.~Rev.~B}\ }\textbf {\bibinfo
					{volume} {83}},\ \bibinfo {pages} {081101} (\bibinfo {year}
				{2011})}\BibitemShut {NoStop}%
			\bibitem [{\citenamefont {Friedrich}\ \emph {et~al.}(2012)\citenamefont
				{Friedrich}, \citenamefont {Betzinger}, \citenamefont {Schlipf},
				\citenamefont {Blügel},\ and\ \citenamefont {Schindlmayr}}]{Friedrich_2012}%
			\BibitemOpen
			\bibfield  {author} {\bibinfo {author} {\bibfnamefont {C.}~\bibnamefont
					{Friedrich}}, \bibinfo {author} {\bibfnamefont {M.}~\bibnamefont
					{Betzinger}}, \bibinfo {author} {\bibfnamefont {M.}~\bibnamefont {Schlipf}},
				\bibinfo {author} {\bibfnamefont {S.}~\bibnamefont {Blügel}},\ and\ \bibinfo
				{author} {\bibfnamefont {A.}~\bibnamefont {Schindlmayr}},\ }\bibfield
			{title} {\bibinfo {title} {{Hybrid functionals and GW approximation in the
						FLAPW method}},\ }\href {https://doi.org/10.1088/0953-8984/24/29/293201}
			{\bibfield  {journal} {\bibinfo  {journal} {J.~Phys.~Condens.~Matter.~}\
				}\textbf {\bibinfo {volume} {24}},\ \bibinfo {pages} {293201} (\bibinfo
				{year} {2012})}\BibitemShut {NoStop}%
		\end{thebibliography}
	\end{document}